\documentclass[10pt]{article}
\usepackage{amsfonts,amsmath,amssymb,amsthm}
\usepackage[pdftex]{graphicx}
\usepackage[lmargin = 1in, rmargin=1in,vmargin=1in]{geometry}
\usepackage{dsfont}

\usepackage{natbib} 
\usepackage{setspace}
\usepackage{enumerate}
\usepackage[toc,title,titletoc,header]{appendix}
\usepackage[colorlinks,citecolor=blue,hypertexnames=false]{hyperref}
\usepackage{etoolbox}
\usepackage{bbm}
\usepackage{array,longtable}
\usepackage[font=small]{caption}
\usepackage{tikz,pgfplots}
\newcolumntype{C}{>{$}c<{$}} % automatic math mode, centered
\allowdisplaybreaks

\usepackage{float}
\usepackage{multirow}
\usepackage{booktabs}
\usepackage{adjustbox}
\usepackage{nicematrix}
\usepackage{comment}
\usepackage{subcaption}
\usepackage{marginnote}

\def\qed{\rule{2mm}{2mm}}

\newtheorem{lem}{Lemma}[section]

\newtheorem{prop}{Proposition}[section]
\theoremstyle{definition}

\newtheorem{rem}{Remark}[section]
\newtheorem{assu}{Assumption}[section]

\usepackage[bb=dsserif]{mathalpha}
\usepackage{bm}

\AtEndEnvironment{rem}{~\qed}
\AtEndEnvironment{example}{~\qed}

\pgfplotsset{compat=1.18}
\begin{document}

\title{Inference After Ranking with Applications to Economic Mobility\thanks{This document greatly benefited from generous comments from Isaiah Andrews, Anna Mikusheva, and Alex Torgovitsky. We are grateful for comments from seminar discussant Sarah Moon, and for discussion from Lanier Benkard, Aditya Bhardwaj, Alex Lehner, Yueran Ma, Yucheng Shang, Jesse Shapiro, Steven Shi, and Jann Speiss, as well as for comments by participants at IAAE 2025 and the ISSI online seminar. Xun Huang provided excellent research assistance. The authors would also like to extend their gratitude to Victor Lima, Kotaro Yoshida and students in the economics undergraduate honors workshop at the University of Chicago and to the social sciences computing services at the University of Chicago. }}

\author{Andreas Petrou-Zeniou \\
Department of Economics \\
MIT \\
\url{apetrouz@mit.edu}
\and
Azeem M.\ Shaikh \\
Department of Economics \\
University of Chicago \\
\url{amshaikh@uchicago.edu}
}

\bigskip

\maketitle

% \vspace{-0.3in}

\begin{spacing}{1.2}

\begin{abstract}
This paper considers the problem of inference after ranking. In our setting, we are interested in any population whose rank according to some random quantity, such as an estimated treatment effect, a measure of value-added, or benefit (net of cost), falls in a pre-specified range of values. As such, this framework generalizes the inference on winners setting previously considered in \cite{andrewsetal2023}, in which a winner is understood to be the single population whose rank according to some random quantity is highest. We show that this richer setting accommodates a broad variety of empirically-relevant applications. We develop a two-step method for inference, which we compare to existing methods or their natural generalizations to this setting. We first show the finite-sample validity of this method in a normal location model and then develop asymptotic counterparts to these results by proving uniform validity over a large class of distributions satisfying a weak uniform integrability condition. Importantly, our results permit degeneracy in the covariance matrix of the limiting distribution, which arises naturally in many applications. In an application to the literature on economic mobility, we find that it is difficult to distinguish between high and low-mobility census tracts when correcting for selection. Finally, we demonstrate the practical relevance of our theoretical results through an extensive set of simulations.
\end{abstract}

\end{spacing}

\section{Introduction}
\label{sec:intro}
In this paper, we consider the problem of conducting inference on multiple rank-based selections. Here, we define a selection to be a population whose rank according to some random value, such as a measure of value-added or an estimated treatment effect, lies in a set specified by the analyst. We seek to provide joint confidence sets for parameters corresponding to these populations selected according to their ranks. Our framework generalizes the setting of inference on  winners considered in \cite{andrewsetal2022} and \cite{andrewsetal2023}, in that we target joint coverage of parameters among multiple selected populations. This generalization allows us to address several important applications, such as inference after cutoff-based selections, inference on quantiles, or inference on statistical significance. These settings arise frequently in applied work. For example, cutoff-based selections arise frequently in the decision-theoretic literature on optimal subset selection, as in \cite{GuKoenker2023}. Similarly, inference on quantiles, or equivalently inference on the top-$\tau$ winners, arises in the applied literature on economic mobility, as in \cite{Bergmanetal2023}. Inference on statistical significance is particularly relevant in the literature on publication bias, as in \cite{AndrewsKasy2019}. Our framework identifies a common structure in these empirically relevant problems.  

Motivated by this setting, we propose a novel, two-step approach to inference. In particular, we identify a key nuisance parameter that characterizes rank-based selection. In the first step of our procedure, we construct confidence bounds for this nuisance parameter. In the second step, we use these confidence bounds to bound the errors on selected units and construct critical values. We apply a Bonferroni-type correction to account for the possibility that our first step confidence region does not cover the nuisance parameter driving selection. In this way, our approach is most similar to that of \cite{Romanoetal2014}, who study inference in moment inequality models. We demonstrate the finite-sample validity of our methods in a normal location model. We then provide results on feasible inference when the data generating process lies in a nonparametric class of distributions. We show, in particular, that our procedure enjoys uniform asymptotic validity under a non-restrictive uniform integrability condition on the class of distributions generating the data. 
% Moreover, we show that for each distribution in a large class, our two-step approach to inference asymptotically dominates projection  and simultaneous inferences in the spirit of \cite{bachoc2017}, \cite{Kuchibhotlaetal2022}, and \cite{Berketal2013}.

Our approach lies in a broader literature on selective inference. After constructing our two-step approach to inference, we show that some well-known, existing methods for selective inference and their natural generalizations may be ill-suited for inference after selection on ranks. While projection-based methods (see, e.g., [\cite{bachoc2017}, \cite{Berketal2013}, \cite{Kuchibhotlaetal2022}]) are robust to arbitrary selection rules, we show that for each distribution in a large class, our two-step approach to inference asymptotically dominates such approaches. Tools due to \cite{andrewsetal2023} and \cite{Leeetal2016}  provide a polyhedral characterization of the rank-based selections considered in this paper. However, this characterization does not provide a computationally feasible inference procedure when joint coverage is desired.  Consequently, it is unclear how to generalize the conditional and hybrid approaches of \cite{andrewsetal2023} to settings where multiple selections are made, further motivating our approach. 

Within the selective inference literature, our approach is most similar in spirit to that of \cite{ZrnicFithian2024}, who also apply a Bonferroni-type correction to a general selective inference setting. Similar Bonferroni-type corrections appear in \cite{McCloskey2017} and \cite{Silvapulle1996}.  As explained further in section \ref{sec:two-step}, the approach in \cite{ZrnicFithian2024} differs from ours, in that \cite{ZrnicFithian2024} propose projection inference localized to a set of likely selections, whereas our method constructs worst-case critical values over a confidence set for a key nuisance parameter driving selection. In our normal location model, we also show that our critical values are, with high probability, smaller than those of \cite{ZrnicFithian2024} for a broad set of data generating processes. The relevance of this result is evident in our extensive simulations comparing the methods.  In those simulation results, we also include comparisons with the alternative methodology developed in \cite{ZrnicFithian2024WP}.  

We consider two applications. In our first application, we use as an illustrative example a replication failure from two studies on job retraining due to the JOBSTART demonstration of \cite{Caveetal1993} and \cite{Milleretal2005}. We build on the previous analysis of \cite{andrewsetal2023} to determine whether correcting for selection bias can explain the replication failure between the two studies. We find that, even when considering alternative selection rules based on statistical significance, a winners' curse fails to explain this replication failure. In our second application, we revisit the literature on economic mobility, and the problem of selecting high-opportunity census tracts in the spirit of the Creating Moves to Opportunity (CMTO) Program of \cite{Bergmanetal2023}. We consider the problem of ex-post inference on tract-level measures of economic mobility, as defined in \cite{Chettyetal2020}, for a subset of high-mobility tracts considered in the randomized trial of \cite{Bergmanetal2023}. Our analysis revisits the findings of \cite{Mogstadetal2023}, who suggest that the selection of high-opportunity tracts may reflect noise as opposed to signal. 
% We observe that the methods of \cite{Mogstadetal2023} may be conservative since they typically impose a notion of joint coverage over parameters for all tracts, motivating an application of our methods, which focus power on selected tracts alone. 
We find similar results to \cite{Mogstadetal2023}, in that we generally fail to reject the null that pairwise differences between arbitrary high and low opportunity tracts are zero. We obtain more positive results, however, in comparing commuting zones. 

Finally, we conduct an extensive simulation study comparing our methods to existing approaches in the literature, namely projection inference as defined in \cite{andrewsetal2023}, the locally simultaneous approach of \cite{ZrnicFithian2024}, and a recent approach due to \cite{ZrnicFithian2024WP}. Our methods are able to outperform these existing methods across a broad range of simulation designs, both in terms of reducing over-coverage and in reducing confidence set length. In particular, our methods reduce over-coverage error by up to 96\%, and reduce confidence set length by up to 27\% relative to projection inference. We also reduce over-coverage error by up to 71\% and length by up to 11\% relative to locally simultaneous inference. 

The paper is organized as follows: In section \ref{sec:Setup}, we formally introduce the problem of inference after ranking and present four empirically-relevant settings to which it applies. In section \ref{sec:two-step}, we develop the two-step approach to inference after ranking and discuss its extensions. In section \ref{sec:existing_approaches}, we provide a more detailed discussion of existing approaches to the problem of inference after ranking, outline some of their shortcomings, and provide some  results comparing our two-step approach to inference to these existing methods. In section \ref{sec:JOBSTART_Application_Section}, we revisit the JOBSTART demonstration and compare different approaches to inference. In section \ref{sec:Neighborhood_Effects_Application_Section}, we apply our two-step inferences to the CMTO program, evaluating
neighborhood effects in selected census tracts. Finally, in section \ref{sec:Simulation_Study_Main}, we present the results from a simulation study comparing the performance of the two-step and projection approaches to inference in a range of synthetic, simulation designs. In the supplemental material, we describe further approaches in the literature, provide proofs and supplemental results, and provide results from an expanded simulation study.

\section{Setup and Notation}
\label{sec:Setup}
In this section, we formalize the problem of inference after ranking.  For the time being, we confine our description to a normal location model, but we later consider nonparametric settings. First, let $p$ be some natural number. For each population $j \in J := \{1,\ldots,p\}$, denote by $X_j$ and $Y_j$ scalar characteristics of the $j$-th population of interest. For the $p$-dimensional random vectors $X := (X_j : j \in J)'$ and $Y := (Y_j : j \in J)'$, we assume that:
% Let $X$ and $Y$ each be $p$-dimensional random vectors, whose distribution is:
% We consider a finite set of indices or populations $J := \{1,\ldots,p\}$, the $p$-dimensional random vector $Y$, as well as a correlated $p$-dimensional random vector $X$, which are drawn jointly from a multivariate normal as below:
\begin{equation}
\label{eq:normality_DGP}
\begin{pmatrix}
X\\
Y
\end{pmatrix}
\sim
\mathcal{N}\left(
\begin{pmatrix}
\mu_{X}\\
\mu_Y
\end{pmatrix},
\begin{pmatrix}
\Sigma_{X} & \Sigma_{XY}\\
\Sigma_{YX} & \Sigma_Y
\end{pmatrix}
\right),~ \mu := \begin{pmatrix}
\mu_{X}\\
\mu_Y
\end{pmatrix},~ \Sigma := \begin{pmatrix}
\Sigma_{X} & \Sigma_{XY}\\
\Sigma_{YX} & \Sigma_Y
\end{pmatrix}
~,
\end{equation}
where $\mu \in \mathbb{R}^{2p}$ is unknown and the $2p\times 2p$ matrix $\Sigma$ is known. Let $\mu_{X,j}$ and $\mu_{Y,j}$, denote the $j$-th elements of $\mu_X$ and $\mu_Y$ respectively, and let $\Sigma_{X,jj'}$, $\Sigma_{XY,jj'}$ and $\Sigma_{Y,jj'}$ denote the element in the $j$-th row and $j'$-th column of $\Sigma_X$, $\Sigma_{XY}$ and $\Sigma_Y$ respectively. We will denote the joint distribution of $X$ and $Y$ by $P_{\mu,\Sigma}$. 
% For  $\xi_X$ and $\xi_Y$ as follows:
%\begin{equation}
%    \label{eq:xi_defn}
%    \xi_X := X-\mu_X,\ \xi_Y = Y-\mu_Y
%\end{equation}
% For $j \in J := \{1, \ldots, p\}$, denote by $X_j$ and $Y_j$ the $j$-th elements in $X$ and $Y$, respectively, and by $\xi_{X,j}$ and $\xi_{Y,j}$ the $j$-th elements in $\xi_X$ and $\xi_Y$, respectively.

Our goal is to construct, using $X$ and $Y$, a rectangular confidence set for the values of $\mu_Y$ at indices selected according to the ranking given by the realized values of $X$.  In order to describe this problem more formally, we require some further notation.  To this end, let $R \subseteq J$ be some fixed set of ranks of interest, and let $k := |R|$ be the cardinality of the set $R$. We define the rank of population $j$, according to $X$, as follows:
\begin{equation}
    r_j(X) := \sum_{j'\in J} \mathds{1}\left(
X_{j'} \leq X_j
    \right)~.
\end{equation}
We define the set of selected indices $\hat{J}_R(X)$ as follows:
\begin{equation}
    \hat{J}_R(X) := \{j : r_j(X)\in R\}~.
\end{equation}
If we take $R$ to simply be $J$, then $\hat{J}_R(X)$ would be the full index set $J$. Similarly, taking $R := \{p\}$, our selected indices would correspond to the largest values in $X$. Unless stated otherwise, we will denote $\hat{J}_R(X)$ by $\hat{J}_R$. We emphasize that, even when $R$ is a singleton, $\hat{J}_R$ need not be a singleton because of ties.\footnote{\cite{andrewsetal2023} provide a sufficient condition in their lemma 1 for $\hat{J}_R$ to be a singleton almost surely, for any $R$ singleton. Imposing that $\Sigma_X$ be full rank provides an alternative sufficient condition that is stronger than that stated in lemma 1 of \cite{andrewsetal2023}.}

In terms of this notation, our goal can be formally described as follows: to construct a random set $CS := (CS_{\hat\jmath})_{\hat\jmath \in \hat{J}_R}$, where each $CS_{\hat\jmath}$ is an interval, such that 
%Our goal is to conduct simultaneous inference on the means $\mu_{Y,\hat{\jmath}}$ for all $\hat{\jmath}$ in $\hat{J}_R$, and in particular, construct a rectangular confidence set $CS$ indexed by $\hat{J}_R$ such that:
\begin{equation}
\label{eq:validity_requirement}
P_{\mu,\Sigma}\left(
\mu_{Y,\hat{\jmath}} \in CS_{\hat{\jmath}} \quad \text{for all} \quad \hat{\jmath} \in \hat{J}_R
\right) \geq 1-\alpha
\end{equation}
for all $\mu$ and $\Sigma$. When $R=\{p\}$, this problem is equivalent to the inference on winners problem considered in \cite{andrewsetal2023}. For convenience, where $J_c$ is an arbitrary subset of $J$, we may write $(\mu_{Y,j})_{j\in J_c} \in CS$ as shorthand for $\mu_{Y,j} \in CS_{j}\text{ for all } j \in J_c$.

\begin{rem}
In addition to the case of $R = \{p\}$ considered in \cite{andrewsetal2023}, \cite{andrewsetal2022} consider the case where selection occurs with respect to an arbitrary rank, such that $R$ is an arbitrary singleton. However, \cite{andrewsetal2022} do not consider the case where multiple rank-based selections are made. 
\end{rem}

Finally, we provide some additional notation used throughout the remainder of the paper and the supplemental material. For natural numbers $z \in \mathbb{N}$, we denote by $[z]$ the set $\{1,\ldots,z\}$ and by $\mathds{1}_{z}$ the vector of ones in $\mathbb{R}^{z}$. We will denote by $I_{z}$ the $z\times z$ identity matrix. Finally, we will denote by $2^J$ the power set of $J$. Finally, for arbitrary real matrices or vectors $\ell$ and $u$ of the same dimension, we denote by $\ell \leq u$ element-wise inequality.
\subsection{Review of Applications}
Before proceeding, we describe a broad array of empirically-relevant settings in which our results are of interest. In addition to the same applications considered in \cite{andrewsetal2023}, the inference after ranking setting accommodates a range of novel applications.
\\

\textbf{Post-Selection Inference on Quantiles:} \quad Suppose we want to conduct inference on the components of the mean of $Y$ corresponding to the components of $X$ in the top $\gamma$-quantile of all components of $X$. We can take $\tau := \lceil\gamma p\rceil$ and take $R=\{p-\tau+1,\ldots,p\}$. In other words, we seek to construct confidence sets for the elements in $\mu_Y$ corresponding to the $\tau$-largest elements in $X$. This setting naturally arises in our neighborhood effects application in section \ref{sec:Neighborhood_Effects_Application_Section}, where we study tract-level outcomes for high opportunity tracts in the setting of \cite{Bergmanetal2023}. This specialized setting also arises in \cite{Haushoferetal2016}, who conduct a randomized controlled trial in the top 40\% of villages in Rarieda, Kenya, selected according to the proportion of houses with thatched roofs. 
\\

\textbf{Inference After Cutoff-Based Selections:} \quad We may want to conduct inference on the values $\mu_{Y,j}$ for $j\in[p]$ such that $X_j\geq c$ for some non-negative real number $c$, in a version of the file drawer problem. In order to accommodate this setting in our framework, define $X_c := \begin{pmatrix}X' & c\mathds{1}_{p}'\end{pmatrix}'$, and $Y_c := \begin{pmatrix} Y' & c\mathds{1}_{p}'\end{pmatrix}'$. We can take $J:=[2p]$ and $R := \{p+1,\ldots,2p\}$, and consider inference using $X_c$ and $Y_c$. The indices of the top $p$ elements in $X_c$ correspond exactly to those indices $j$ in $X$ such that $X_j \geq c$, as well as a residual set of indices corresponding (non-uniquely) to elements in the appended constant vector in $X_c$. For a concrete, empirical example, policymakers may observe the marginal value of public funds (MVPFs) of \cite{HendrenandSprung_Keyser2020} for a menu of policies. A MVPF exceeding one corresponds to a policy whose benefits, in dollar terms, exceeds its costs. Consequently, policymakers may choose to proceed only with policies whose MVPFs exceed one, generating an inference after ranking problem. Our analysis of cutoff-based selections differs from that of \cite{andrewsetal2023} in that we can accommodate multiple selections. In contrast, the setting of \cite{andrewsetal2023} can only accommodate the above when $p=1$.
\\

\textbf{Inference on Statistical Significance:}\quad We observe, just as in \cite{andrewsetal2023}, that we can normalize the $X_j$ by standard deviations $\sqrt{\Sigma_{X,jj}}$. By applying the previous two examples, we can accommodate inference on all units that are statistically significant at a given level according to $X$. Note that $X$ and $Y$ need not be the same. Such problems arise in the literature on publication bias, as in \cite{AndrewsKasy2019}.
\\

\textbf{Inference on Multiple Outcomes:}\quad We may be concerned with the means of $K$ different outcomes of interest among selected populations. For example, in the CMTO intervention of \cite{Bergmanetal2023}, we may be concerned with mobility effects among different groups (say effects by race or gender). Denote by $Y_1,\ldots,Y_K$  the different outcomes of interest.  In order to accommodate this setting in our framework, define $Y_r := \begin{pmatrix}
    Y_1' & \ldots & Y_K'
\end{pmatrix}'$ and $X_r := \begin{pmatrix}
    X' & \ldots & X'
\end{pmatrix}'$, where $X$ is repeated $K$ times. Further define a new index set $R_r := \{l\cdot q : l \in R,\ 1\leq q\leq K\}$. Using $Y_r$, $X_r$, and $R_r$ in lieu of the original $Y$, $X$, and $R$ characterize the inference after ranking problem with the desired estimands.
\begin{rem}
    In many of these examples of applications, $\Sigma$ need not be full rank. Our methods are valid in finite samples when imposing normality as in \eqref{eq:normality_DGP}. Moreover, when deriving the asymptotic properties of our methods, we demonstrate that we can provide uniformly valid inferences over a class of data generating processes which may include distributions with degenerate covariance matrices. We discuss asymptotic validity in subsection \ref{sec:Theoretical_Results_Main}.
\end{rem}

\section{A Two-Step Approach to Inference After Ranking}
\label{sec:two-step}
In this section, we provide a two-step approach to inference after ranking. We construct a confidence set that is conceptually similar to the inferential approach of \cite{Romanoetal2014} and most recently in the selective inference literature, to \cite{ZrnicFithian2024}. We emphasize, however, that our approach is meaningfully different from that of  \cite{ZrnicFithian2024}; see our remark \ref{rem:comp_to_ZF} for further discussion. In the first step, we specify some $\beta$ in $(0,\alpha)$ and construct level $1-\beta$ confidence bounds on moment differences. In the second step, we use these bounds to model the errors on selected units and derive critical values. As we will show in sections \ref{sec:existing_approaches} and \ref{sec:Simulation_Study_Main}, our approach to inference performs well relative to existing methods.
\subsection{Construction}
Our goal is to construct a rectangular confidence set indexed by $\hat\jmath\in\hat{J}_R$ satisfying \eqref{eq:validity_requirement}. It is convenient to introduce the following notation:
\begin{equation}
    \label{eq:xi_defn}
    \xi_X := X-\mu_X,\ \xi_Y := Y-\mu_Y~.
\end{equation}
We additionally denote by $\xi_{X,j}$ and $\xi_{Y,j}$ the $j$-th components of $\xi_X$ and $\xi_Y$, respectively, with $j\in J$. In addition, for $j,j' \in J$, let $\Delta_{jj'} := \mu_{X,j} - \mu_{X,j'}$ be an unobserved nuisance parameter, and let $\text{var}_{jj'} := \text{Var}(\xi_{X,j} - \xi_{X,j'})$. We note that $\text{var}_{jj'} = \Sigma_{X,jj} +\Sigma_{X,j'j'} - 2\Sigma_{X,jj'}$ is a known function of $\Sigma$.

We will first construct a confidence region for the $p\times p$ matrix $\Delta := (\Delta_{jj'})_{j,j'\in J}$ given by a lower bound $L$ and upper bound $U$, both $p\times p$ matrices, and will then use these confidence bounds to approximate the critical values of the $\xi_{Y,\hat\jmath}$. In particular, for our choice of $\beta$ in $(0,\alpha)$, we obtain $L$ and $U$ such that $P_{\mu,\Sigma}\left(L\leq\Delta\leq U\right) \geq 1-\beta$, where the inequality $L\leq \Delta\leq U$ is interpreted element-wise. We construct $L$ and $U$ by first taking $d_{1-\beta\mbox{}}(\Sigma)$ to be the $1-\beta\mbox{}$-quantile of the following:
\begin{equation}
\label{eq:d_quantile_of}
\max_{j,j'\in J, j\neq j',\text{var}_{jj'}\neq 0}\frac{|\xi_{X,j} - \xi_{X,j'}|}{\sqrt{\text{var}_{jj'}}}~.
\end{equation}
%Because $d$ is the critical value of a studentized statistics, we have that $d$ is homogeneous of degree zero in $\Sigma$. 
We define $L$ and $U$ as follows:
\begin{align}
\label{eq:L_definition}
L_{jj'} &:= X_{j}-X_{j'}-d_{1-\beta\mbox{}}(\Sigma)\sqrt{\text{var}_{jj'}}\\
\label{eq:U_definition}
U_{jj'} &:= X_{j}-X_{j'}+d_{1-\beta\mbox{}}(\Sigma)\sqrt{\text{var}_{jj'}}~.
\end{align}
It follows that $P_{\mu,\Sigma}\left(L\leq\Delta\leq U\right)\geq 1-\beta$. 

In order to apply our confidence bounds for $\Delta$, we first consider non-random $p\times p$ matrices $\ell$ and $u$, and define $f(\ell,u)$ as follows:
$$ \max_{j\in J} \frac{|\xi_{Y,j}|}{\sqrt{\Sigma_{Y,jj}}}\mathds{1}\left(\exists r : r\in\left[\sum_{j'\in J}\mathds{1}\left(
\xi_{X,j} \geq \xi_{X,j'} + u_{j'j}\right),
\sum_{j'\in J}\mathds{1}\left(
\xi_{X,j} \geq \xi_{X,j'} + \ell_{j'j}\right)
\right] \right)~.
$$
We observe that $f(\ell,u)$ is a random variable given by a function of the $\xi_X$ and $\xi_Y$. Consequently, we denote by $\rho_{1-\alpha+\beta}(L,U)$ the $1-\alpha+\beta$ quantile of the $f(\ell,u)$, evaluated at the random values of $L$ and $U$. When necessary, we may emphasize that $\rho_{1-\alpha+\beta}(L,U)$ depends on $\Sigma$ by writing $\rho_{1-\alpha+\beta}(L,U ;\Sigma)$. Whenever $\ell \leq \Delta \leq u$, we note that the following holds:
\begin{equation}
\label{eq:key_inequality_two_step}
\frac{|\xi_{Y,\hat{\jmath}}|}{\sqrt{\Sigma_{Y,\hat{\jmath}\hat{\jmath}}}} \leq f(\ell,u), ~ \text{for each}~ \hat{\jmath} \in \hat{J}_R~, 
\end{equation}
since we can write:
\begin{equation}
\label{eq:trivial_obs_one}
\frac{|\xi_{Y,\hat{\jmath}}|}{\sqrt{\Sigma_{Y,\hat{\jmath}\hat{\jmath}}}} \leq \max_{j\in J} \frac{|\xi_{Y,j}|}{\sqrt{\Sigma_{Y,jj}}} \mathds{1}\left(
\sum_{j'\in J}\mathds{1}(\xi_{X,j}\geq \xi_{X,j'} +\Delta_{j'j})\in R
\right)~. 
\end{equation}
As a result of \eqref{eq:trivial_obs_one}, our construction follows:
\begin{equation}
\label{eq:CS_TS_definition}
CS_{\hat\jmath}^{TS}(1-\alpha;\beta) := \left[
Y_{\hat\jmath} - \rho_{1-\alpha+\beta}(L,U)\sqrt{\Sigma_{Y,\hat{\jmath}\hat{\jmath}}},
Y_{\hat\jmath} + \rho_{1-\alpha+\beta}(L,U)\sqrt{\Sigma_{Y,\hat{\jmath}\hat{\jmath}}}
\right]~.
\end{equation}
We denote by $CS^{TS}(1-\alpha;\beta)$ the joint confidence set $(CS_{\hat\jmath}^{TS}(1-\alpha;\beta))_{\hat\jmath\in\hat{J}_R}$. We claim the following:
\begin{prop}
\label{prop:normal_validity}
$CS^{TS}(1-\alpha;\beta)$ is a valid confidence set at the $(1-\alpha)$-level, such that (\ref{eq:validity_requirement}) holds for all $\mu$, $\Sigma$. 
\end{prop}
We prove this result in section D of the supplemental material.

\begin{rem}
    \label{rem:comp_to_ZF}
    Within the literature on selective inference, our methods are most similar to the locally simultaneous approach of \cite{ZrnicFithian2024}, who suggest focusing power by localizing simultaneous inference to a set of likely selections. In particular, \cite{ZrnicFithian2024} consider, as a nuisance parameter, a non-random subset $\widetilde{J}_R$ of $J$ such that for all $\mu$ and $\Sigma$, $P_{\mu,\Sigma}(\hat{J}_R \subseteq \widetilde{J}_R)\geq 1-\beta$. For a particular choice of $\widetilde{J}_R$, \cite{ZrnicFithian2024} provide the key insight that an outer confidence region $\hat{J}_R^+$ for $\widetilde{J}_R$ can be constructed such that $\widetilde{J}_R \subseteq \hat{J}_R^+$ if and only if $\hat{J}_R \subseteq \widetilde{J}_R$. Consequently, $P_{\mu,\Sigma}(\hat{J}_R \subseteq \widetilde{J}_R \subseteq \hat{J}_R^+) \geq 1-\beta$ for all $\mu$ and $\Sigma$. Using this insight, \cite{ZrnicFithian2024} use a Bonferroni-type correction to localize simultaneous inference to $\hat{J}_R^+$ and obtain valid coverage. 
    
    To draw a connection between our approach and  that of \cite{ZrnicFithian2024}, we observe that we too construct confidence bounds for a nuisance parameter, derive worst-case critical values over this confidence set, and apply a union bound to recover valid coverage. However, our approach builds on the insight that, when considering ranked-based selections, we can model the errors on selected units more explicitly by taking pairwise differences in means as our nuisance parameter, as in (\ref{eq:trivial_obs_one}). This insight leads to several key advantages associated with our approach. In particular, we can extend our methods to provide asymmetric confidence sets as we discuss further in section \ref{subsec:extensions}. Moreover, because our critical values $\rho_{1-\alpha+\beta}(L,U)$ vary smoothly in the confidence bounds $L$ and $U$, we are able to achieve power gains relative to existing approaches, as demonstrated in section \ref{sec:Simulation_Study_Main}. We demonstrate this formally in proposition \ref{prop:ZF_vs_two_step}, where we show that our two-step critical values are strictly smaller than the corresponding locally-simultaneous critical values with probability at least $1-\beta$, for a broad range of data generating processes where $\max_{j,j'}|\Delta_{jj'}|$ is sufficiently small. We provide further discussion of the approach of \cite{ZrnicFithian2024} in section \ref{subsec:ZF_subsec}.
\end{rem}
\subsection{Feasible Inference}
\label{sec:Theoretical_Results_Main}
In this section, we provide uniform asymptotic guarantees for our methods over a nonparametric class. 

First, we provide notation for our asymptotic setting. We assume that we observe a sequence of $2p$-dimensional random vectors $\widetilde{W}_i \equiv \begin{pmatrix}\widetilde{X}_i^{\prime} & \widetilde{Y}_i^\prime\end{pmatrix}^\prime$ for $i=1,\ldots,n$ drawn i.i.d. from some distribution $P$ in a nonparametric class of distributions $\mathcal{P}$. $\widetilde{X}_i$ and $\widetilde{Y}_i$ are both $p$-dimensional random vectors. We denote the $n$-th sample mean of the above sequence of random vectors by $\widetilde{S}_{W}^{\,n} := 
\begin{pmatrix}
\widetilde{S}_{X}^{n\ \prime} & \widetilde{S}_{Y}^{n\ \prime}
\end{pmatrix}^\prime$. That is, $\widetilde{S}_{W}^{\,n} = \frac{1}{n}\sum_{i=1}^n \widetilde{W}_i$, $\widetilde{S}_{X}^{\,n} = \frac{1}{n}\sum_{i=1}^n \widetilde{X}_i$, and $\widetilde{S}_{Y}^{\,n} = \frac{1}{n}\sum_{i=1}^n \widetilde{Y}_i$. 

As usual, we index $\widetilde{X}_i$ and $\widetilde{Y}_i$ by $J\equiv[p]$, such that for $j\in J$, $\widetilde{X}_{i,j}$ and $\widetilde{Y}_{i,j}$ denote the $j$-th elements in $\widetilde{X}_i$ and $\widetilde{Y}_i$, respectively. Similarly, we denote by $\widetilde{S}_{X,j}^{n}$ and $\widetilde{S}_{Y,j}^{n}$ the $j$-th elements in $\widetilde{S}_X^n$ and $\widetilde{S}_Y^n$, respectively. Moreover, we can denote by $\widetilde{W}_{i,j}$ the $j$-th element in $\widetilde{W}_i$ for $j\in [2p]$. Finally, we take $R\subseteq J$ to be a set of ranks of interest. We take $\hat{J}_{R;n}$ to be the set of indices given by $\hat{J}_R(\widetilde{S}_{X}^{n})$. 

For all $P$ in $\mathcal{P}$, we define 
\begin{equation*}
    \mu_{X}(P) := \mathbb{E}_{P}\left(
\widetilde{X}_i
    \right)~,~\mu_Y(P) := \mathbb{E}_{P}\left(
\widetilde{Y}_i\right)
\end{equation*}
and $\mu_W(P) = \begin{pmatrix}\mu_X(P)'& \mu_Y(P)'\end{pmatrix}'$. For $j\in J$, let $\mu_{X,j}(P)$ and $\mu_{Y,j}(P)$ denote the $j$-th elements of $\mu_X$ and $\mu_Y$, respectively. For $j\in [2p]$, let $\mu_{W,j}(P)$ denote the $j$-th element of $\mu_W(P)$.

Let the $p$-dimensional random vector $\xi_{\widetilde{X}_i} := \widetilde{X}_i - \mu_X(P)$ denote the demeaned version of $\widetilde{X}_i$, and let us similarly denote the demeaned version of $\widetilde{Y}_i$ by the $p$-dimensional random vector $\xi_{\widetilde{Y}_i} := \widetilde{Y}_i - \mu_Y(P)$. Similarly, we denote by $\xi_{\widetilde{S}_X^n} := \widetilde{S}_X^n -\mu_X(P)$ and $\xi_{\widetilde{S}_Y^n} := \widetilde{S}_Y^n - \mu_Y(P)$, both $p$-dimensional, the demeaned versions of $\widetilde{S}_X^n$ and $\widetilde{S}_Y^n$, respectively.

We define the $2p\times 2p$ variance-covariance matrix $\Sigma_W(P)$ as follows:
\begin{equation}
\label{eq:variance_covariance_definition_sample}
    \Sigma_W(P) := \mathbb{E}_P\left(
\begin{pmatrix}{\xi_{\widetilde{X}_i}}\\ {\xi_{\widetilde{Y}_i}}\end{pmatrix}
\begin{pmatrix}{\xi_{\widetilde{X}_i}}' & {\xi_{\widetilde{Y}_i}}'\end{pmatrix}
    \right)~.
\end{equation}
Taking $\hat{P}_n$ denotes the empirical distribution of the data $\widetilde{W}_i$ for $i=1,\ldots,n$, we can obtain some sequence of estimators for $\Sigma_W(P)$, given by $\Sigma_W(\hat{P}_n)$. We define $n\hat{\Sigma}^n := \Sigma_W(\hat{P}_n)$, such that $\hat{\Sigma}^n$ is an estimator of the variance of the sample means. Let $\hat{\Sigma}^n_X$ denote the upper-left $p\times p$ block of $\hat{\Sigma}^n$, and let $\hat{\Sigma}^n_{Y}$ be the lower-right $p\times p$ block of $\hat{\Sigma}^n$.

To apply our two-step confidence sets to our data $\widetilde{S}_W^{n}$, we proceed as follows. Let $\widehat{\text{var}}^n_{jj'} = \hat{\Sigma}^n_{X,j'j'}+\hat{\Sigma}^n_{X,jj}-2\hat{\Sigma}^n_{X,jj'}$, which we may think of as an estimator of the variance of the difference $\xi_{\widetilde{S}_X^n,j} - \xi_{\widetilde{S}_X^n,j'}$. We define the $p\times p$ random matrices $L^n$ and $U^n$ as follows:
\begin{align*}
    L_{jj'}^n &= \widetilde{S}_{X,j}^n - \widetilde{S}_{X,j'}^n - d_{1-\beta\mbox{}}(\hat{\Sigma}^n)\sqrt{\widehat{\text{var}}^n_{jj'}}\\
    U_{jj'}^n &= \widetilde{S}_{X,j}^n - \widetilde{S}_{X,j'}^n + d_{1-\beta\mbox{}}(\hat{\Sigma}^n)\sqrt{\widehat{\text{var}}^n_{jj'}}
\end{align*}
In order to suppress dependence on $\hat{\Sigma}^n$, let us write $\rho^n_{1-\alpha+\beta}(L^n,U^n) := \rho_{1-\alpha+\beta}(L^n,U^n;\hat{\Sigma}^n)$.

We define the feasible version of our two-step confidence set as follows:
\begin{equation*}
CS_{\hat\jmath}^{TS}(1-\alpha;\beta,n) := \left[
\widetilde{S}_{Y,\hat\jmath}^n - \rho^n_{1-\alpha+\beta}(L^n,U^n)\sqrt{\Sigma_{Y,\hat{\jmath}\hat{\jmath}}},
\widetilde{S}_{Y,\hat\jmath}^n + \rho^n_{1-\alpha+\beta}(L^n,U^n)\sqrt{\hat{\Sigma}^n_{Y,\hat{\jmath}\hat{\jmath}}}
\right]~.
\end{equation*}
As before, we denote by $CS^{TS}(1-\alpha;\beta,n)$ the joint confidence set $(CS_{\hat\jmath}^{TS}(1-\alpha;\beta,n))_{\hat\jmath\in\hat{J}_{R;n}}$. In most of our asymptotic analyses, we impose the following uniform integrability condition (see, e.g., [\cite{lehmann_romano_2022_testing}, \cite{RomanoShaikh_2012}, \cite{Romanoetal2014}] for discussions):
\begin{assu}
\label{assu:Uniform_Asymptotic_Normality_Sufficient}
    For $j=1,\ldots,2p$, it holds that:
    \begin{equation}
    \label{eq:UI_Assumption}
        \limsup_{K\to\infty}\sup_{P\in\mathcal{P}}\mathbb{E}_P\left(
        \frac{\left|\widetilde{W}_{1,j}-\mu_{W,j}(P)\right|^2}{\Sigma_{W,jj}(P)}\mathds{1}
        \left(
 \frac{\left|\widetilde{W}_{1,j}-\mu_{W,j}(P)\right|}{\sqrt{ \Sigma_{W,jj}(P)}}> K
        \right)
        \right) = 0
    \end{equation}
    \end{assu}
In appendix B, we prove the following uniform asymptotic validity result:
\begin{prop}
\label{prop:Two_Step_Uniform_Asymptotic_Validity}
    Given assumption \ref{assu:Uniform_Asymptotic_Normality_Sufficient}, the two-step confidence set given by $CS^{TS}(1-\alpha;\beta,n)$ is uniformly, asymptotically valid such that (\ref{eq:TS_unif_validity}) holds:
    \begin{equation}
    \label{eq:TS_unif_validity}
    \liminf_{n\to\infty}\inf_{P\in\mathcal{P}} {\Pr}_P\left(
\left(\mu_{Y,\hat{\jmath}}(P)\right)_{\hat{\jmath}\in\hat{J}_{R;n}} \in CS^{TS}(1-\alpha;\beta,n)
        \right) \geq 1-\alpha.
\end{equation}
\end{prop}

\subsection{Extensions}
\label{subsec:extensions}

Finally, we describe a range of extensions to our two-step approach to inference that deliver power gains, and moreover demonstrate that our methods can be applied to problems distinct from that of inference after ranking. 
\\

\textbf{Intersecting with Projection:} In appendix A, we construct an improved version of our two-step confidence set that exploits a joint projection confidence set for both $\mu_X$ and $\mu_Y$. Notably, when $X=Y$ it follows that this improved two-step confidence set is a subset of a projection confidence set for $\mu_Y$. This case corresponds to the case where units are selected according to their ranks in $Y$, or equivalently, where $\hat{J}_R(X) = \hat{J}_R(Y)$ almost surely. 
\\

\textbf{Simplified Inference for Quantiles:} In the case where $R$ is of the form $\{p-\tau+1,\ldots,p\}$, we can simplify our two-step approach to inference. In particular, let us define $\widetilde{L}_{jj'} := X_{j}-X_{j'}-d_{1-\beta}(\Sigma)\sqrt{\text{var}_{jj'}}$, for $j$ and $j'$ in $J$. This defines a $p\times p$ matrix $\widetilde{L}$. For a non-random, $p\times p$ matrix $\ell$, we define $\widetilde{f}$ as follows:
    \begin{equation*}
        \widetilde f(\ell) = \max_{j\in J} \frac{|\xi_{Y,j}|}{\sqrt{\Sigma_{Y,jj}}}\mathds{1}\left(\sum_{j'\in J}\mathds{1}\left(
\xi_{X,j} \geq \xi_{X,j'} + \ell_{j'j}\right)
\geq p-\tau+1 \right)~.
    \end{equation*}
    Whenever $\ell \leq \Delta$,  we obtain:
    \begin{equation*}
        \frac{|\xi_{Y,\hat{\jmath}}|}{\sqrt{\Sigma_{Y,\hat{\jmath}\hat{\jmath}}}} \leq \widetilde f(\ell), ~ \text{for each}~ \hat{\jmath} \in \hat{J}_R~. 
    \end{equation*}
    We define $\widetilde{\rho}_{1-\alpha+\beta}\left(\widetilde{L}\right)$, to be the $1-\alpha+\beta$ quantile of $\widetilde{f}(\ell)$ evaluated at the random value $\widetilde{L}$.  In this case, we may write:
    \begin{equation*}
CS^{TS}_{\hat\jmath}(1-\alpha;\beta):=\left[
Y_{\hat\jmath} - \widetilde{\rho}_{1-\alpha+\beta}\left(\widetilde{L}\right)\sqrt{\Sigma_{Y,\hat{\jmath}\hat{\jmath}}},
Y_{\hat\jmath} + \widetilde{\rho}_{1-\alpha+\beta}\left(\widetilde{L}\right)\sqrt{\Sigma_{Y,\hat{\jmath}\hat{\jmath}}}
\right]
\end{equation*}
giving a simplified two-step approach to inference for the top-$\tau$ winners. Again, we denote $CS^{TS}(1-\alpha;\beta)$ by $(CS_{\hat\jmath}^{TS}(1-\alpha;\beta))_{\hat\jmath\in\hat{J}_R}$.
\\

\textbf{Asymmetric Two-Step Inferences:} The observation in \eqref{eq:trivial_obs_one} generalizes to the positive and negative components of the errors. This allows us to construct an asymmetric version of our two-step confidence set. In particular, we write for $j=1,\ldots,J$, $\xi_{Y,j}^+ := \max\{\xi_{Y,j},0\}$ and $\xi_{Y,j}^- := \max\{-\xi_{Y,j},0\}$. For non-random $p\times p$ matrices $\ell$ and $u$, we define $f^+(\ell, u)$:
    $$ \max_{j\in J} \frac{\xi_{Y,j}^+}{\sqrt{\Sigma_{Y,jj}}}\mathds{1}\left(\left[\sum_{j'\in J}\mathds{1}\left(
\xi_{X,j} \geq \xi_{X,j'} + u_{j'j}\right),
\sum_{j'\in J}\mathds{1}\left(
\xi_{X,j} \geq \xi_{X,j'} + \ell_{j'j}\right)
\right]\cap R\neq\emptyset \right)~,
$$
    and $f^-(\ell,u)$:
    $$ \max_{j\in J} \frac{\xi_{Y,j}^-}{\sqrt{\Sigma_{Y,jj}}}\mathds{1}\left(\left[\sum_{j'\in J}\mathds{1}\left(
\xi_{X,j} \geq \xi_{X,j'} + u_{j'j}\right),
\sum_{j'\in J}\mathds{1}\left(
\xi_{X,j} \geq \xi_{X,j'} + \ell_{j'j}\right)
\right]\cap R\neq\emptyset \right)~.
$$
We can consequently take $\rho_{1-\frac{\alpha-\beta}{2}}^+(L,U)$ and $\rho_{1-\frac{\alpha-\beta}{2}}^-(L,U)$ to be the $\left ( 1-\frac{\alpha-\beta}{2} \right )$-quantile of $f^+(\ell, u)$ and $f^-(\ell,u)$, respectively, evaluated at the random values of $L$ and $U$ as constructed in (\ref{eq:L_definition}) and (\ref{eq:U_definition}). We can consequently take the following, asymmetric two-step confidence set:
\begin{equation*}
CS_{\hat\jmath}^{ATS} (1-\alpha;\beta):= \left[
Y_{\hat{\jmath}} - \rho_{1-\frac{\alpha-\beta}{2}}^+(L,U)\sqrt{\Sigma_{Y,\hat{\jmath}\hat{\jmath}}},\
Y_{\hat{\jmath}} + \rho_{1-\frac{\alpha-\beta}{2}}^-(L,U)\sqrt{\Sigma_{Y,\hat{\jmath}\hat{\jmath}}}
\right]
\end{equation*}
As usual, we denote by $CS^{ATS} (1-\alpha;\beta)$ the collection $(CS_{\hat\jmath}^{ATS} (1-\alpha;\beta))_{\hat\jmath\in\hat{J}_R}$. This confidence set reflects the intuition in \cite{andrewsetal2023} that the observation corresponding to the winner is upward biased. In particular, for certain choices of $R$, such as when $R=\{p\}$, $\rho_{1-\frac{\alpha-\beta}{2}}^+(L,U)$ will generally be larger than $\rho_{1-\frac{\alpha-\beta}{2}}^-(L,U)$.
\\

\textbf{Inference After Polyhedral Selection:} Our approach to inference can be generalized to arbitrary polyhedral selections, in the spirit of \cite{Leeetal2016}. In order to address this, we will need to introduce some new notation. Let $\mathcal{I}$ be some finite index set. Let us partition the space of possible realizations of $X$, being $\mathbb{R}^p$, into a finite collection of polyhedra $\mathcal{O} = \{\mathcal{O}_i\}_{i\in\mathcal{I}}$. We write $\mathcal{O}_i := \{x:A_ix\leq b_i\}$ for $i\in\mathcal{I}$ for some collections of matrices $A_i$ and vectors $b_i$. We take a collection of $k\times p$ matrices $B_i$ such that, whenever we observe $X$ satisfying $X \in \mathcal{O}_i$, we select $B_i Y$. We denote the rows of $B_i$ by the $p$-dimensional vector $\eta_{i,j}'$ for $j=1,\ldots,k$. 
    
    In general, we say that we observe $\hat{B}Y$, where the $k\times p$ matrix $\hat{B}= B_i$ if and only if $X \in \mathcal{O}_i$. Denoting the rows of $\hat{B}$ by the $p$-dimensional vectors $\hat{\eta}_j'$ for $j=1,\ldots,k$, we have that
    \begin{equation}
    \label{eq:trivial_obs_one_polyhedra}
    \frac{\left|\hat{\eta}_j'\xi_Y\right|}{\sqrt{\hat\eta_j' \Sigma_Y\hat\eta_j}} \leq \max_{i\in\mathcal{I}}\frac{|\eta_{i,j}'\xi_Y|}{\sqrt{
    \eta_{i,j}'\Sigma_Y\eta_{i,j}}}\mathds{1}\left(
    A_i\xi_X + A_i\mu_X \leq b_i
    \right)~,
    \end{equation}
    providing an analogue to (\ref{eq:trivial_obs_one}) for polyhedral selections. Our nuisance parameter of interest is the collection of $k$-dimensional vectors $(A_i\mu_X)_{i\in\mathcal{I}}$. One can then proceed as above to obtain a joint two-step confidence region for the $\hat\eta_j' \mu_Y$ for $j=1,\ldots,k$. Thus, our approach can be used as a more general tool for unconditional post-selection inference.
\\

\textbf{Inference for Pairwise Differences and Selected Ranks:} Our two-step approach can also be straightforwardly applied to the problem of inference for pairwise differences. Consider the problem of comparing units ranked in $R$ with those ranked in $R'$. For $\hat\jmath \in \hat{J}_R$ and $\hat\imath \in \hat{J}_{R'}$, we may modify \eqref{eq:trivial_obs_one} as follows:
    \begin{align*}
    \frac{|\xi_{Y,\hat{\jmath}} - \xi_{Y,\hat\imath}|}{\sqrt{\text{var}_{\hat\jmath\hat\imath}}}
\leq \max_{i,j\in J, i\neq j}  \frac{|\xi_{Y,j}-\xi_{Y,i}|}{\sqrt{\text{var}_{ji}}}\mathds{1}(r_j(X) \in R, r_i(X) \in R')~,
\end{align*}
where we recall that $\mathds{1}(r_j(X) \in R, r_i(X) \in R')$ can be written as follows:
\[
\mathds{1}\left(\sum_{j'\in J}\mathds{1}\left(
\xi_{X,j} \geq \xi_{X,j'} + \Delta_{j'j}\right)
\in R,\ \sum_{i'\in J}\mathds{1}\left(
\xi_{X,i} \geq \xi_{X,i'} + \Delta_{i'i}\right)
\in R' \right).
\]
    This suggests that we can obtain an analogue of equation (\ref{eq:key_inequality_two_step}) for pairwise differences. Importantly, as \cite{Mogstadetal2023} demonstrate, confidence regions for pairwise differences in unit means can be used to construct confidence sets for the ranks of these units. Thus, our methods can be applied to conduct inference for the ranks of selected units among all units, for example, or the ranks of selected units only among selected units.

\section{Existing Approaches to Inference}
\label{sec:existing_approaches}
We now turn our attention to existing approaches to post-selection inference. We discuss two classes of methods, namely projection and locally simultaneous, in depth. We also provide a brief discussion of conditional and hybrid approaches to inference (see, e.g., [\cite{andrewsetal2023}, \cite{Leeetal2016}, \cite{markovic2018}, \cite{McCloskey2023}]). The projection approach involves conducting inference that is simultaneously valid for all selection methods, as in \cite{bachoc2017} and \cite{Berketal2013}. A related approach is the locally simultaneous method of \cite{ZrnicFithian2024}, who apply projection inferences to a set of likely selections. We provide an analytical result comparing locally simultaneous inference with our two-step approach, which shows that with high-probability, our two-step confidence sets are smaller than locally simultaneous confidence sets for certain data generating processes. In appendix A, we discuss further, alternative approaches to inference, namely an extension of our two-step method and approaches based on inverting the zoom test of \cite{ZrnicFithian2024WP}. We also discuss conditional and hybrid approaches in greater depth. In general, we find that, in simulation, our two-step approach outperforms existing approaches for a broad range of data generating processes, which we demonstrate in section \ref{sec:Simulation_Study_Main}. Moreover, as we will show in proposition \ref{prop:clear_winner_two_step_best}, our method reduces over-coverage asymptotically relative to projection, and as we will show in proposition \ref{prop:ZF_vs_two_step}, our two-step critical values are smaller than locally-simultaneous critical values when all populations are sufficiently close.

\subsection{The Projection Approach to Inference}
\label{subsec:projection_def}
The projection approach as described in \cite{andrewsetal2023} is an example of the simultaneous approach to post-selection inference (see, e.g., [\cite{bachoc2017}, \cite{Berketal2013}, \cite{Kuchibhotlaetal2022}]). Such simultaneous inference methods are appropriate for the setting of inference after ranking, albeit conservative, since they are robust to arbitrary selection rules. 

For any subset $J_c$ of $J$, we define $c_{1-\alpha}(J_c)$ to be the $1-\alpha$-quantile:
\begin{equation}
\label{eq:c_quantile_def}
    \max_{j\in J_c}\frac{|\xi_{Y,j}|}{\sqrt{\Sigma_{Y,jj}}}
\end{equation}
noting, trivially, that, for all $\hat{\jmath} \in \hat{J}$:
\begin{equation*}
    \frac{|\xi_{Y,\hat{\jmath}}|}{\sqrt{\Sigma_{Y,\hat{\jmath}\hat{\jmath}}}} \leq 
    \max_{j\in J}\frac{|\xi_{Y,j}|}{\sqrt{\Sigma_{Y,jj}}}
\end{equation*}
Unless stated otherwise, we will denote $c_{1-\alpha}(J)$ by $c_{1-\alpha}$. Consequently, we can define the following rectangular confidence set:
\begin{equation}
    CS_{\hat\jmath}^P({1-\alpha}) := \left[Y_{\hat{\jmath}} - c_{1-\alpha}\sqrt{\Sigma_{Y,\hat{\jmath}\hat{\jmath}}},
Y_{\hat{\jmath}} + c_{1-\alpha}\sqrt{\Sigma_{Y,\hat{\jmath}\hat{\jmath}}}
    \right]~.
\end{equation}
$CS^P({1-\alpha}) := (CS_{\hat\jmath}^P({1-\alpha}))_{\hat\jmath\in\hat{J}_R}$ is a valid confidence set, satisfying \eqref{eq:validity_requirement}. This modified, projection confidence set can be easily constructed and has very simple statistical properties. However, it is generally quite conservative, particularly in cases where there may exist clear selections.\footnote{For further discussion, see e.g. [\cite{andrewsetal2023}, \cite{ZrnicFithian2024}].} 
% Nevertheless, the projection confidence set may be appropriate in settings where tight confidence intervals are not entirely necessary, or in cases where $k$ is large.

We formalize this in the normal location model developed in section \ref{sec:Setup}. In particular, we show that for a fixed $\mu$, and for some sequence of covariance matrices, our two-step methods reduce over-coverage relative to projection. In appendix C, we prove an asymptotic counterpart to this result for data drawn i.i.d. from some distribution in a nonparametric class as in section \ref{sec:Theoretical_Results_Main}. 

In our normal location model, we will consider fixed $\Sigma\neq 0$ and $c\Sigma$ for constants $c$. Moreover, for the $2p$-dimensional mean vector $\mu$, we define $J_R(\mu)$ to be the fixed set $\left\{j:\sum_{j'\in J}\mathds{1}\left(
    \mu_{X,j'} \leq \mu_{X,j}
    \right) \in R\right\}$. We can equivalently write $J_R(\mu) := \hat{J}_R(\mu_X)$. Crucially, for $c$ sufficiently small, $\hat J_R$ will equal $J_R(\mu)$ with high probability. This intuition underlies the following result:
    \begin{prop}
    \label{prop:clear_winner_two_step_best}
        Suppose that, for $\mu$, $J_R(\mu)$ is a strict subset of $J$ and that $\Sigma\neq 0$ is full rank. Then, for any $\beta$ sufficiently small, we have:
        \begin{equation}
            \lim_{c\downarrow 0} P_{\mu, c\Sigma}\left(
            (\mu_{Y,\hat\jmath})_{\hat\jmath\in\hat{J}_R} \in CS^{TS}(1-\alpha;\beta)
            \right)\leq \lim_{c\downarrow 0}P_{\mu,c\Sigma}\left(
            (\mu_{Y,\hat\jmath})_{\hat\jmath\in\hat{J}_R} \in CS^{P}(1-\alpha)
            \right)
        \end{equation}
    \end{prop}

We provide the proof of this proposition, along with an asymptotic counterpart, in appendix C.

\subsection{Locally Simultaneous Inference}
\label{subsec:ZF_subsec}
We now adapt the locally simultaneous approach of \cite{ZrnicFithian2024} to the problem of inference after ranking. As before, \cite{ZrnicFithian2024} consider some unknown set of likely selections $\widetilde{J}_R$ and conduct simultaneous inference restricted to an outer confidence set for $\widetilde{J}_R$, which we denote by $\hat{J}_R^+$. In general, \cite{ZrnicFithian2024} construct this $\hat{J}_R^+$ by first constructing a $1-\beta$ confidence region for the data generating process $P$. In our case, we can equivalently construct a $1-\beta$ confidence region for $\mu$, or some other carefully-chosen parameter of interest. Subsequently, for each $P$ in the confidence region described above, \cite{ZrnicFithian2024} derive a $(1-\beta)$-forecast set for the observations generated by this $P$, and the selections these observations imply. Taking the union of these forecast sets over all $P$ in the aforementioned confidence set yields a confidence set for $\widetilde J_R$ which is valid at level $1-\beta$. This forecast set will provide the $\hat{J}_R^+$ described above.

First, let $\bar{d}_{1-\beta\mbox{}}(\Sigma)$ denotes the $1-\beta$ quantile of:
\begin{equation*}
    \max_{j,j'\in J, j\neq j'}|\xi_{X,j} - \xi_{X,j'}|~.
\end{equation*}
We may notice that $\bar{d}_{1-\beta\mbox{}}(\Sigma)$ is a version of $d_{1-\beta\mbox{}}(\Sigma)$ without studentization. To apply the approach of \cite{ZrnicFithian2024} to the problem of inference after ranking, we take:
\begin{equation}
\label{eq:J_c_for_LS}
    \hat{J}_R^+ := \left\{j:X_j \in \bigcup_{\hat{\jmath}\in\hat{J}_R} \left[
    X_{\hat{\jmath}} - 2\bar{d}_{1-\beta\mbox{}}(\Sigma),\ X_{\hat{\jmath}}+2\bar{d}_{1-\beta\mbox{}}(\Sigma)
    \right]\right\}
\end{equation}
Finally, we may define the following confidence set:
\begin{equation}
    CS^{LS}_{\hat\jmath}({1-\alpha;\beta}) := \left[Y_{\hat{\jmath}} - c_{1-\alpha+\beta}\left(\hat{J}_R^+\right)\sqrt{\Sigma_{Y,\hat{\jmath}\hat{\jmath}}},\
Y_{\hat{\jmath}} + c_{1-\alpha+\beta}\left(\hat{J}_R^+\right)\sqrt{\Sigma_{Y,\hat{\jmath}\hat{\jmath}}}
    \right]
\end{equation}
We denote by $CS^{LS}({1-\alpha;\beta})$ the usual family $(CS^{LS}_{\hat\jmath}({1-\alpha;\beta}))_{\hat\jmath\in\hat{J}_R}$. It is worth noting that if $R$ is of the form $\{p-\tau+1,\ldots,p\}$, we can replace the two-sided interval in (\ref{eq:J_c_for_LS}) with a one-sided interval, as discussed in section \ref{subsec:extensions}. As before, we have the following proposition:
\begin{prop}
\label{prop:normal_validity_LS}
$CS^{LS}(1-\alpha;\beta)$ is a valid confidence set at the $1-\alpha$-level, such that (\ref{eq:validity_requirement}) holds for all $\mu$ and $\Sigma$.
\end{prop}
A proof of this proposition is provided in appendix D.\footnote{We also note that theorem 2 of \cite{ZrnicFithian2024} gives that $CS^{LS}({1-\alpha;\beta})$ may be intersected with a version of the level $1-\alpha$ projection confidence set, delivering a finite-sample, non-inferiority result for their methods. As we discuss in appendix A, and have already discussed in section \ref{subsec:extensions}, a similar result holds for our methods.} Comparing the non-intersected versions of two-step and locally-simultaneous inferences yields the following result, where we define:
\begin{align*}
\bar{L}_{jj'} &:= X_{j}-X_{j'}-\bar d_{1-\beta\mbox{}}(\Sigma)\\
\bar{U}_{jj'} &:= X_{j}-X_{j'}+\bar d_{1-\beta\mbox{}}(\Sigma)~.
\end{align*}
\begin{prop}
    \label{prop:ZF_vs_two_step}
    Let $\Sigma$ be full rank and let $R$ be a proper subset of $J$. Let $\mu$ be any mean vector such that:
    \begin{equation}
    \label{eq:suff_cond_for_ts_bet_ZF}
        \max_{j,j'\in J} |\mu_{X,j}-\mu_{X,j'}| \leq \bar{d}_{1-\beta}(\Sigma)
    \end{equation}
    It follows that, with probability at least $1-\beta$, $\rho_{1-\alpha+\beta}(\bar L, \bar U) < c_{1-\alpha+\beta}\left(
    \hat{J}_{R}^+
    \right)$.
\end{prop}
We provide a proof of this result in appendix C. It follows that, even when intersecting with projection as in proposition A.1 in the supplemental material and as in theorem 2 of \cite{ZrnicFithian2024}, our two-step confidence sets are contained by the corresponding locally-simultaneous confidence sets with probability at least $1-\beta$ whenever the condition (\ref{eq:suff_cond_for_ts_bet_ZF}) is satisfied. Our simulations suggest that our two-step critical values are smaller than locally simultaneous critical values for a broad range of designs, including those where (\ref{eq:suff_cond_for_ts_bet_ZF}) may not hold.

\subsection{Conditional and Hybrid Approaches to Inference}
Apart from the projection and locally-simultaneous approaches studied above, conditional (see, e.g., [\cite{andrewsetal2022}, \cite{andrewsetal2023}, \cite{Leeetal2016}]) and hybrid (see, e.g., [\cite{andrewsetal2023}, \cite{McCloskey2023}]) approaches have been applied to problems in selective inference, including to versions of the inference after ranking problem. In particular, both approaches involve characterizing the distribution of the $\begin{pmatrix}
    X' & Y'
\end{pmatrix}'$ conditional on the selection event $\hat{J}_R = J_c \in 2^J$ and a sufficient statistic for nuisance parameters corresponding to units not selected. While, whenever $R$ is a singleton, we can obtain a characterization of this distribution as a univariate truncated normal (see, e.g., [\cite{andrewsetal2022}, \cite{andrewsetal2023}]), we do not obtain such a characterization in the case where $R$ is not a singleton. Instead, using arguments due to \cite{Leeetal2016}, we demonstrate in lemma A.1 of the supplemental material that this conditional distribution becomes a multivariate normal truncated to a union of polyhedra whenever $R$ is not a singleton. As a result, conditional and hybrid inferences do not provide practical approaches to inference in the inferential setting described in this paper, an observation also made in \cite{ZrnicFithian2024}. We provide further discussion in appendix A.

\subsection{Further Approaches to Inference}
Our review of existing approaches has thus far been limited to projection, locally simultaneous, conditional and hybrid procedures. A recent approach due to \cite{ZrnicFithian2024WP} provides an unconditional approach to inference that recovers, or nearly recovers, uncorrected inference when the set of selections is clear. In particular, \cite{ZrnicFithian2024WP} construct tests of point hypotheses of the form $H_0:\mathbb{E}(Y)=\mu_Y,\ \mathbb{E}(X)=\mu_X$. \cite{ZrnicFithian2024WP} provide an acceptance region that is largest for populations that are unlikely to be selected. \cite{ZrnicFithian2024WP} use this acceptance region to define the zoom test, which they invert to provide confidence sets. Henceforth, we will refer to the inferential approach in \cite{ZrnicFithian2024WP} as the zoom approach. We describe this approach formally in appendix A, and include this approach in our extended simulation study in appendix E, alongside the methods discussed in the main text. Finally, in appendix A, we also provide some further approaches that extend our two-step approach to inference, or that apply our observation in (\ref{eq:trivial_obs_one}) to construct confidence sets in the spirit of \cite{ZrnicFithian2024}.

\section{Application: the JOBSTART Demonstration}
\label{sec:JOBSTART_Application_Section}
In this section, we revisit the JOBSTART demonstration due to \cite{Caveetal1993} and subsequent replication failure in \cite{Milleretal2005}, which has been previously studied in the selective inference literature by \cite{andrewsetal2023}. The JOBSTART demonstration was a randomized trial taking place between 1985 and 1988 across 13 sites, with the intention of studying the effects of a vocational training program on the employment outcomes of young, low-skilled high-school dropouts. The treatment group was given access to a suite of JOBSTART services which were inaccessible to those in the control group. Among the sites included in the JOBSTART study, only one site, the Center for Employment Training (CET) in San Jose, saw a large and statistically significant estimate of the effect on earnings. \cite{Caveetal1993} note that one cannot attribute the unique success of CET to a particular feature of the program, but suggest that the CET's strong connections with San Jose employers, or their robust placement efforts, may explain some of its value-add. Motivated by the success of the CET program, \cite{Milleretal2005} replicate the intervention at 12 sites. \cite{Milleretal2005} find that, even in replication sites deemed to have high fidelity to the original CET program of \cite{Caveetal1993}, the estimated effect of the program's services on enrollees' earnings was not statistically significant. 

\cite{andrewsetal2023} study the possibility that this replication failure is due to a winner's curse. In particular, \cite{andrewsetal2023} consider the possibility that the estimates of the effect of CET on earnings in the original JOBSTART demonstration are upwardly biased by virtue of CET being the site with the largest estimated effect. We consider a complementary thought experiment. In particular, since the exact mechanism by which \cite{Milleretal2005} select a program for replication is unknown, we consider the possibility that the replication failure between the two studies can be explained by an alternative selection rule in which the sites selected by \cite{Milleretal2005} are chosen according to a statistical significance cutoff. Under this selection rule, it is impossible for the analyst to know ex-ante how many sites will be selected for replication. When multiple selections are made, the conditional and hybrid approaches of \cite{andrewsetal2023} are difficult to apply, as discussed in section \ref{sec:existing_approaches} and in appendix A.\footnote{It is not difficult to provide conditional and hybrid confidence sets that apply in the event that only one selection is made via a particular choice of conditioning set. Such concerns are discussed in appendix C of \cite{andrewsetal2023}, but are beyond the scope of our analysis here.} 

Empirically, we show that our two-step confidence regions for the effect of the CET program on earnings exclude zero. This finding suggests that a winner's curse does not fully explain the replication failure between the studies of \cite{Caveetal1993} and \cite{Milleretal2005}, even under an alternative selection rule. To highlight the relevance of our approach in this setting, we illustrate the frequency with which multiple sites may be selected for replication under a statistical significance cutoff rule. In simulations calibrated to the JOBSTART demonstration, we find that multiple sites can be selected as statistically significant with probability ranging from 3.0\% to 99.9\%, depending on our choice of significance level by which to select sites for replication.

\subsection{JOBSTART: Empirical Findings}
In table \ref{tab:JOBSTART_TEs_SEs}, we report the estimated average treatment effects (ATEs) of the JOBSTART intervention on earnings at each of these 13 sites, with point estimates due to \cite{Caveetal1993} and standard errors due to \cite{andrewsetal2023}:

{
\begin{table}[H]
    \centering
    \caption{Estimated program treatment effects from the JOBSTART demonstration, as reported in \cite{Caveetal1993}. The reported standard errors are those derived in \cite{andrewsetal2023}.}
    \begin{tabular}{c|cc}
Intervention & Treatment Effect & Standard Error\\
\hline\hline
    Atlanta Job Corps & 2093 & 2288.40 \\
CET/San Jose & 6547$^{***}$ & 1496.17 \\
Chicago Commons & -1417 & 2168.21 \\
Connelley (Pittsburgh) & 785 & 1681.92 \\
East LA Skills Center & 1343 & 1735.51 \\
EGOS (Denver) & 401 & 1329.05 \\
Phoenix Job Corps & -1325 & 1598.03 \\
SET/Corpus Christi & 485 & 971.05 \\
El Centro (Dallas) & 336 & 1523.33 \\
LA Job Corps & -121 & 1409.79 \\
Allentown (Buffalo) & 904 & 1814.10 \\
BSA (New York City) & 1424 & 1768.44 \\
CREC (Hartford) & -1370 & 1860.45 \\
\hline\hline
\end{tabular}
    \label{tab:JOBSTART_TEs_SEs}
\end{table}
}

In what follows, we denote the estimated ATEs for the thirteen interventions by $Y$, and index $Y$ by $j$ in the set of interventions $J$. Of all the interventions from the JOBSTART demonstration, only the CET program had a statistically significant treatment effect on earnings at the $1\%$ significance level.\footnote{Or, similarly, at the $5\%$ level.} We provide confidence sets for the true treatment effects at selected sites in the thought experiment where \cite{Milleretal2005} select statistically significant programs for replication rather than the program with the largest effect. We present our empirical findings using both the symmetric and asymmetric versions of our novel two-step method in table \ref{tab:JOBSTART_CIs}. We also provide a conditional confidence set under the winner selection rule from \cite{andrewsetal2023}. We emphasize, however, that the winner selection rule leads to a different notion of coverage, so conditional inference is not directly comparable in this setting. We compute confidence sets for $\alpha = 0.05$ and $\beta = 0.005$, 

\begin{table}[H]
    \centering
    \caption{Confidence sets for the CET program treatment effect, correcting for cutoff-based selection. Confidence sets are presented for a 5\% significance cutoff. The conditional confidence set presented assumes the original, best-treatment selection rule considered in \cite{andrewsetal2023}.}
    \begin{tabular}{c|c}
Method & CS: 5\% Significance Cutoff \\
\hline\hline
    Two-Step & [\$2476, \$10618]  \\
    Two-Step (Asymmetric) & [\$2191, \$10114] \\
Original Conditional & [\$3485, \$9478] \\
% Revised Conditional & [\$2777, \$9477] \\
\hline\hline
\end{tabular}
    \label{tab:JOBSTART_CIs}
\end{table}

We now present results from simulations calibrated to the JOBSTART demonstration. We demonstrate that the probability of making multiple selections when using a cutoff rule is non-negligible. In such instances, conditional and hybrid confidence sets do not readily apply, as explained in section \ref{sec:existing_approaches}. In the simulation presented below, we draw $\mu_Y^b$ from a gaussian with mean $Y$ and with a diagonal covariance matrix using the standard errors computed in table \ref{tab:JOBSTART_TEs_SEs}, for $b=1,\ldots,1000$. For each draw of $\mu_Y^b$, we compute the probability of making multiple selections, conditional on $\mu_Y^b$. We find that in 90\% of our simulations, when choosing a 1\% significance cutoff rule for selection, the probability of making multiple selections lies between 0.153 and 0.828. When choosing a 5\% significance cutoff, this range becomes the interval $[0.473, 0.977]$. We present a histogram of these probabilities, over a confidence region for the means $\mu$ given our observed data, in figure \ref{fig:multiple_selections_prob} below:
\begin{figure}[H]
    \centering
    
    \includegraphics[width=0.45\linewidth]{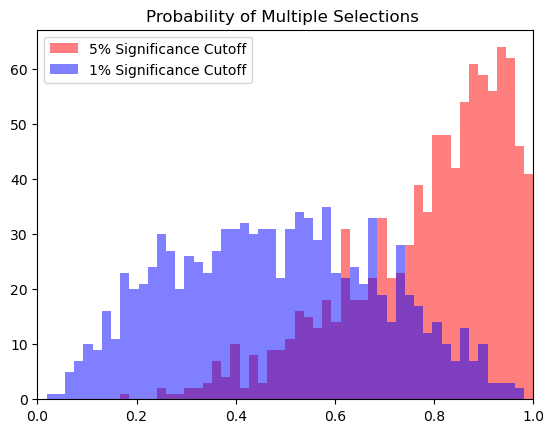}
    \caption{Probability of making multiple selections for different significance-based selection criteria. Probabilities are computed over 1000 simulation draws.}
    \label{fig:multiple_selections_prob}
\end{figure}

\section{Application: Neighborhood Effects Revisited}
\label{sec:Neighborhood_Effects_Application_Section}
In this section, we revisit the studies of \cite{Bergmanetal2023}, \cite{Chettyetal2020}, \cite{ChettyandHendren2018} and \cite{Chettyetal2014} on the geography of economic mobility. These studies have garnered substantial interest in the selective inference and multiple testing literatures, notably in \cite{andrewsetal2023} and \cite{Mogstadetal2023}. \cite{Chettyetal2020} construct the Opportunity Atlas, a dataset containing correlational estimates of the effect of childhood neighborhood in adulthood. In particular, \cite{Chettyetal2020} provide a measure of economic mobility by reporting a child's expected earnings in adulthood (as a percentile) conditional on growing up in a given census tract with parents whose earnings fall in a particular income percentile. \cite{Chettyetal2020} also report analogous economic mobility estimates at the commuting zone (CZ) level. In a study motivated by the findings of \cite{Chettyetal2020}, \cite{ChettyandHendren2018} and \cite{Chettyetal2014}, \cite{Bergmanetal2023} report the effects of an informational intervention on the residential decisions of low-income housing voucher recipients in Seattle via the Creating Moves to Opportunity (CMTO) program. In particular, \cite{Bergmanetal2023} use the economic mobility estimates of \cite{Chettyetal2020} to identify a set of high-opportunity tracts which \cite{Bergmanetal2023} subsequently advertised to treated households in the CMTO program. We seek to study, in a set of exercises related to \cite{andrewsetal2023} and \cite{Mogstadetal2023}, whether the CMTO can be expected to provide positive, long-run effects on the earnings of children in the treatment group, and relatedly whether the selection of high-opportunity neighborhoods reflects noise as opposed to signal. 

We seek to provide insight on both questions. To study the former question, we provide confidence regions for the economic mobility gains of the average, housing voucher recipient moving to an arbitrary, high-opportunity tract. We find that we fail to reject the possibility of null gains in the majority of urban Seattle tracts selected by the CMTO program. We replicate our analysis in the top fifty CZs in the US by population, and find heterogeneity in our findings. In some CZs, we are able to reject the possibility of null effects in the vast majority of selected tracts. Motivated by this finding, we present analyses focused on studying the latter question. In particular, we study pairwise comparisons of high and low-opportunity census tracts in urban Seattle. We find that, for the majority of high-low opportunity tract pairs, we cannot reject the possibility of a null effect. We replicate this analysis for pairwise comparisons of high and low-mobility commuting zones, and find the opposite. Indeed, our methods can provide informative inferences at the commuting zone level. 

Our analyses are closely related to a larger literature in selective inference and multiple testing studying the mobility estimates of \cite{Bergmanetal2023}, \cite{Chettyetal2020}, \cite{ChettyandHendren2018} and \cite{Chettyetal2014}. \cite{andrewsetal2023} study the CMTO program of \cite{Bergmanetal2023}, showing that there exists a statistically significant, positive difference in the average mobility of high-opportunity tracts and the mobility of the average housing voucher recipient. \cite{Mogstadetal2023} show that one can say little about the relative ranks of census tracts or commuting zones according to economic mobility. Our analysis imposes a more strict coverage criterion than that of \cite{andrewsetal2023}, but a less strict coverage criterion than that of \cite{Mogstadetal2023}. 
\subsection{Empirical Findings}

In this section, we provide an in-depth discussion of the empirical findings described above. First, we discuss tract-level effects in the CMTO program of \cite{Bergmanetal2023}. We then provide discussion of mobility effects at the commuting zone level. 

\subsubsection{Tract-level Effects in the CMTO Program}

We build on the work of \cite{Mogstadetal2023}, and seek to provide further insights on tract-level effects. In particular, \cite{Mogstadetal2023} study the problem of ranking tracts according to their true economic mobility effects. In a particularly stark finding, \cite{Mogstadetal2023} find that one cannot reject the possibility that the bottom-ranked tract in Seattle, according to estimated economic mobility, lies in the top third of tracts according to true economic mobility. As a result, \cite{Mogstadetal2023} conclude that:

\begin{quote}
    
The classification of a given tract as a high upward-mobility neighbourhood may simply reflect
statistical uncertainty (noise) rather than particularly high mobility (signal).
\end{quote}

\noindent This finding of \cite{Mogstadetal2023} is indeed surprising. As \cite{Chettyetal2020} remark, the methods of \cite{Mogstadetal2023} suggest that some of the poorest tracts in Los Angeles may be ranked above some of the wealthiest according to economic mobility. \cite{Chettyetal2020} suggest that the methods of \cite{Mogstadetal2023} may be too conservative. Indeed, per \cite{Chettyetal2020}:

\begin{quotation}
    This method is conservative because it assumes that the analyst is comparing all tracts in LA county (whereas in practice we focused on Watts given its well-known history of poverty and violence) and because it controls the family wise error rate (i.e., it requires that the probability that one or more of the millions of pairwise comparisons is wrong is less than 5\%).
\end{quotation}

In our analysis of tract-level effects, we focus power on selected tracts to address the former point. In particular, we seek to study tract-level effects for certain tracts of interest alone, namely high-opportunity tracts. However, we find that even when focusing inference on tracts of interest using our two-step approach to inference, the evidence on tract-level effects remains murky. Some aggregation of effects across tracts or loosening of the simultaneous coverage requirement may be necessary for informative inference. 

In the CMTO program, \cite{Bergmanetal2023} designate a subset of Seattle neighborhoods as high-opportunity according to their ranks. In particular, \cite{Bergmanetal2023} label the top 20\% and top 40\% of urban and non-urban tracts, respectively, in the Seattle commuting zone as high-opportunity. This roughly corresponds to the top third of all tracts in the commuting zone. 

In our first analysis of the CMTO program, we compare the economic mobility estimates in each selected high-opportunity tract with an estimate of the average economic mobility of housing voucher recipients. To be precise, we allow $X_j$ to be the economic mobility estimate for tract $j$ in the set of all census tracts $J$ in the Seattle commuting zone, or alternatively the set of all census tracts in urban Seattle. For each tract $j$, we also observe the number of housing voucher recipients residing in $j$, which we denote by $c_j$. We define $Y_j$ as follows:
\begin{equation}
    Y_j := X_j - \frac{\sum_{i\in J} X_i c_i}{\sum_{i\in J} c_i}
\end{equation}
$Y_j$ is an estimate of the expected gain or loss from moving the average housing voucher recipient to tract $j$.

In our first exercise, we take $J$ to be the set of all urban Seattle tracts. We let $p=132$ be the number of tracts in urban Seattle, we let $\tau = \lfloor p/5\rfloor = 26$ and we take $R$ to be $\{p-\tau + 1,\ldots, p\}$. We are concerned with inference for the means $\mu_{Y,\hat{\jmath}}$ for $\hat{\jmath}$ in $\hat{J}_R$. Here, $\hat J_R$ denotes the top quintile of tracts by economic mobility in urban Seattle. In figure \ref{fig:Neighborhood_Effects} below, we plot lower and upper confidence bounds for the mobility gains in selected Seattle tracts. We also apply the same exercise to Cleveland to demonstrate heterogeneity in findings between urban areas.
\begin{figure}[H]
    \centering
    \begin{tabular}[b]{cc}
        \begin{tabular}[b]{c}
            \begin{subfigure}[b]{0.4\columnwidth}
                \includegraphics[width=\textwidth]{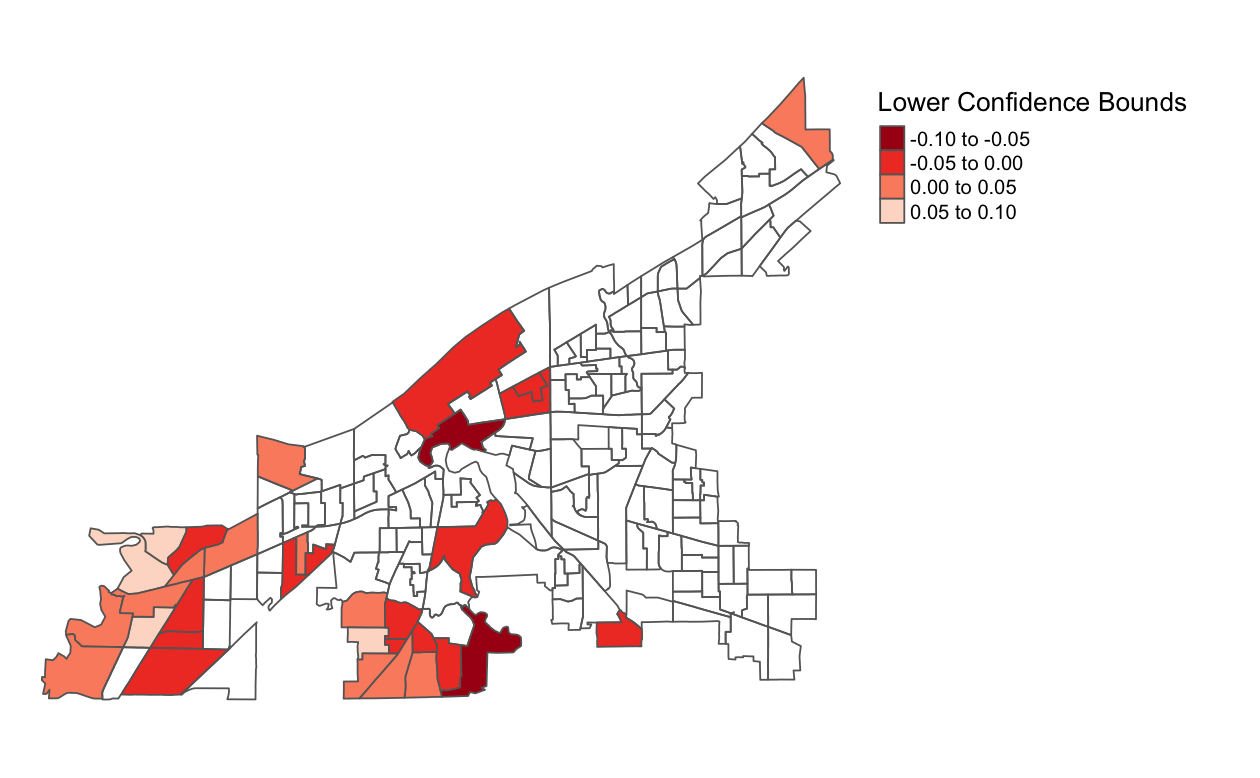}
                \caption{}
            \end{subfigure}\\
            \begin{subfigure}[b]{0.4\columnwidth}
                \includegraphics[width=\textwidth]{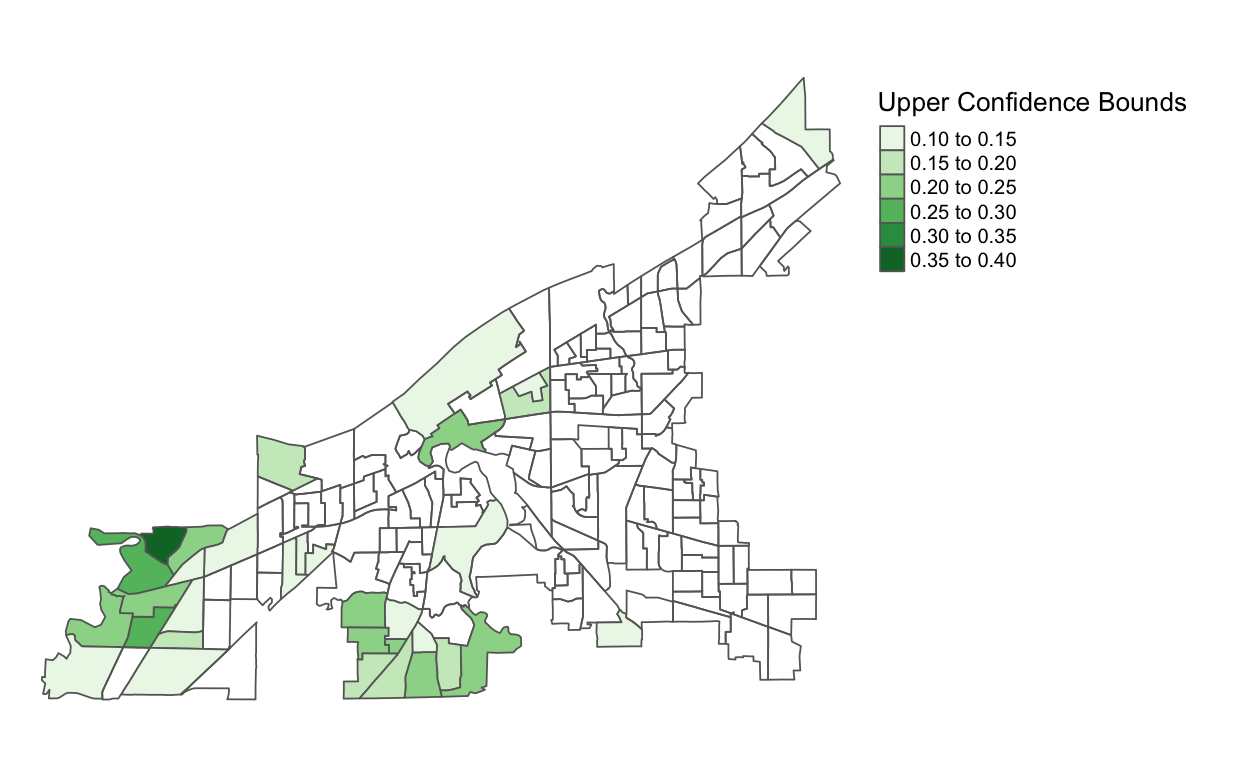}
                \caption{}
            \end{subfigure}
        \end{tabular}
        &
        \begin{tabular}[b]{c}
            \begin{subfigure}[b]{0.4\columnwidth}
                \includegraphics[width=\textwidth]{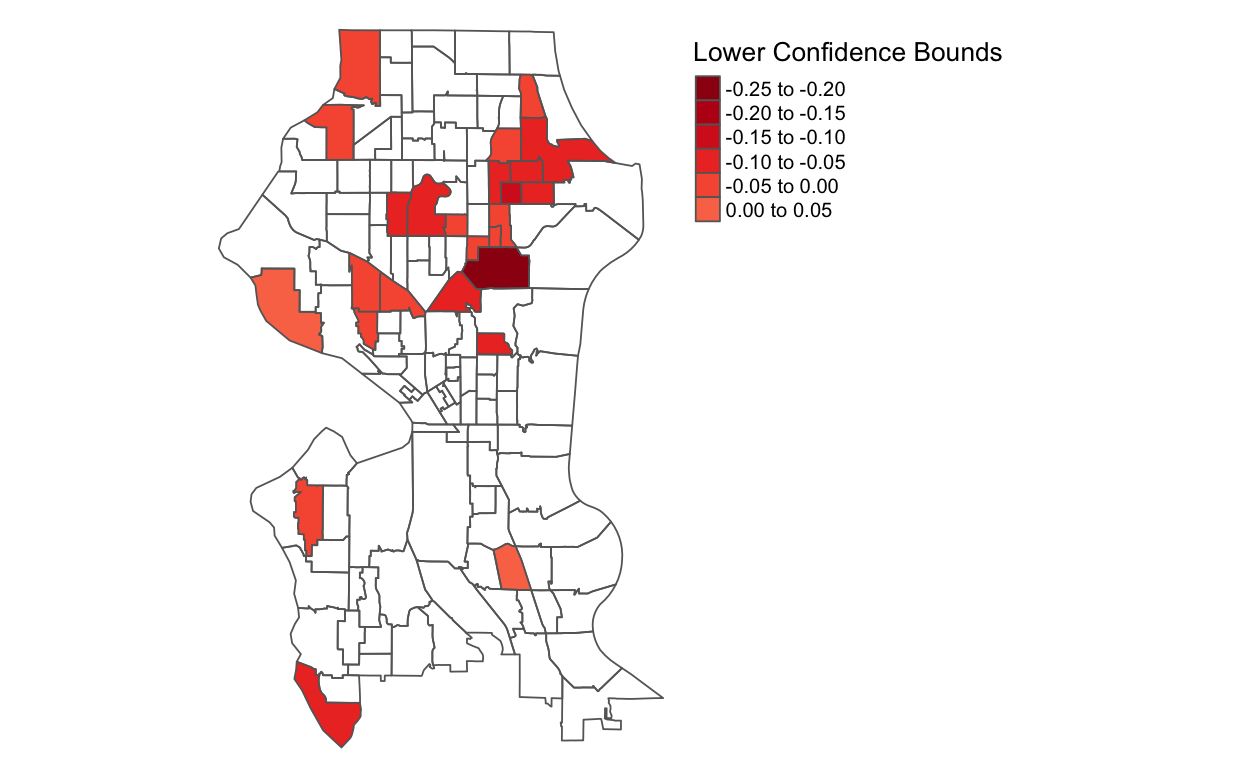}
                \caption{}
            \end{subfigure}\\
            \begin{subfigure}[b]{0.4\columnwidth}
                \includegraphics[width=\textwidth]{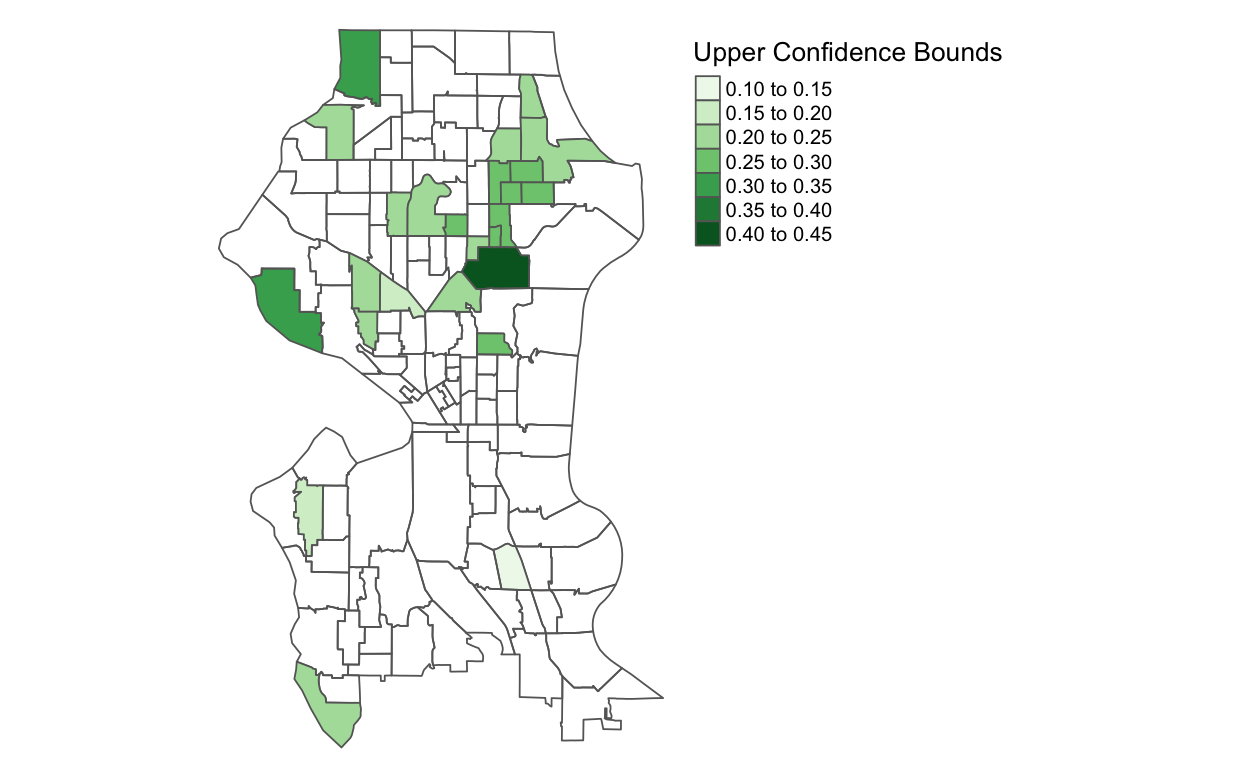}
                \caption{}
            \end{subfigure}
        \end{tabular}
    \end{tabular}
    \caption{Lower and upper confidence bounds for economic mobility gains in selected urban tracts. Subfigures (a) and (b) display lower and upper confidence bounds on neighborhood effects in Cleveland, respectively. Subfigures (c) and (d) display lower and upper confidence bounds on neighborhood effects in Seattle, respectively.\label{fig:Neighborhood_Effects}}
\end{figure}
In urban Seattle, we fail to reject the possibility of null effects at 92\% of selected tracts. In urban Cleveland, we fail to reject the possibility of null effects in a comparatively small 35\% of selected tracts. 
\begin{rem}
    \cite{andrewsetal2023} study the problem of inference for the mean:
    \begin{equation*}
        \bar{\mu}_{Y,\hat{J}_R} := \frac{1}{\lfloor p / 5\rfloor} \sum_{\hat{\jmath}\in\hat{J}_R} \mu_{Y,\hat{\jmath}}
    \end{equation*}
    Their confidence region for $\bar{\mu}_{Y,\hat{J}_R}$ lies above zero, implying that one can reject the possibility of a null effect on economic mobility on aggregate. Our analysis differs from that of \cite{andrewsetal2023} in that we are concerned with constructing a confidence region satisfying simultaneous coverage of the $\mu_{Y,\hat{\jmath}}$. We thereby provide insight into which of the selected tracts most credibly drive the positive aggregate effects.
\end{rem} 

\begin{figure}[H]
    \centering
    \includegraphics[width=0.5\linewidth, trim=0 0 80 0 0, clip]{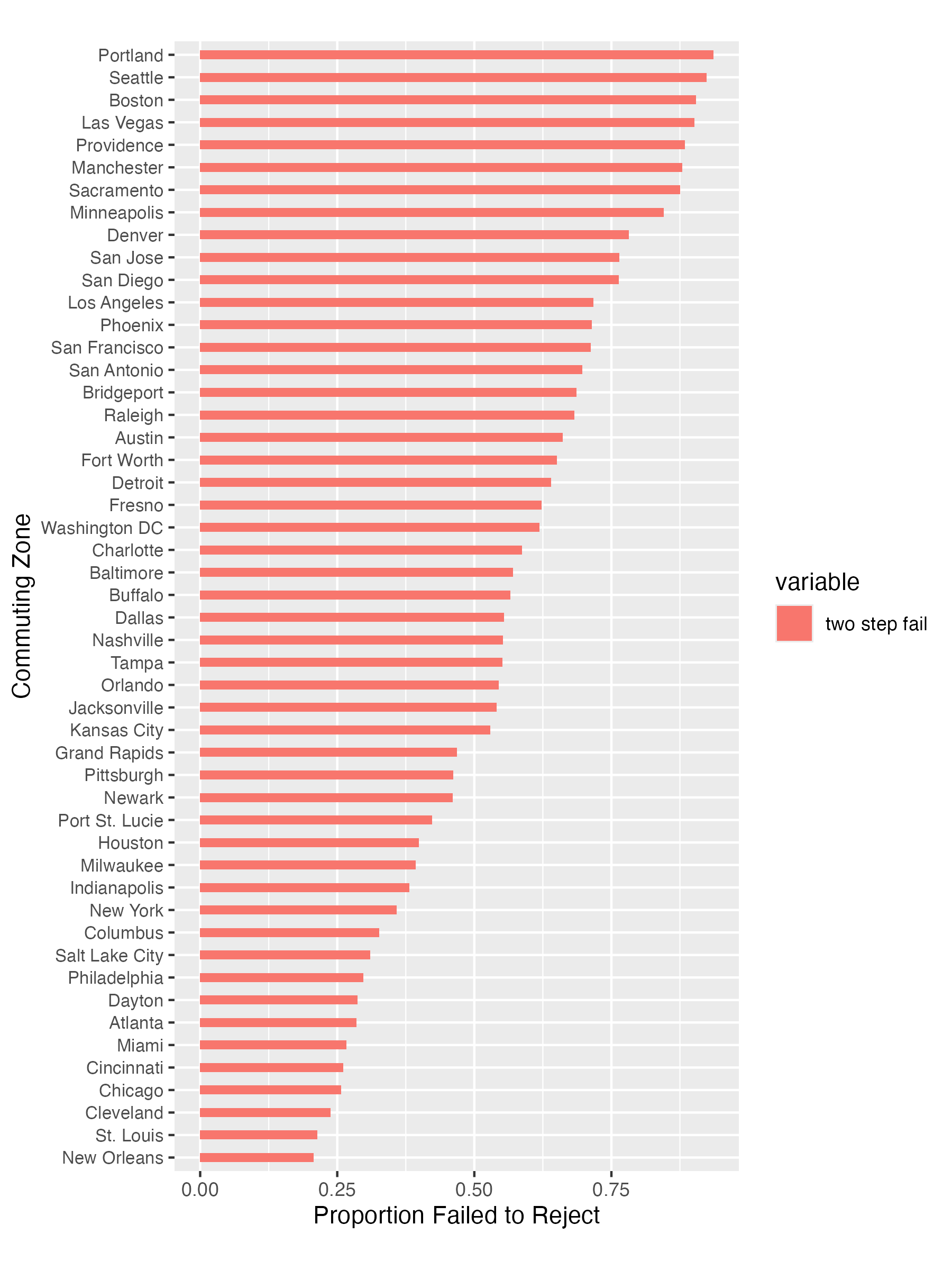}
    \caption{Proportion of selected tracts for which we fail to reject the possibility of a null effect on economic mobility relative to the average housing voucher recipient, by commuting zone. We provide results for all top-50 CZs by population. }
    \label{fig:failure_to_reject_by_CZ}
\end{figure}

Seattle is a fairly extreme example of this phenomenon. Repeating this analysis across the top 50 commuting zones by population in the US yields heterogeneous results. In certain commuting zones, our two-step approach to inference is reasonably powerful against the alternative of positive neighborhood effects. In this exercise, we focus on the economic mobility gains in the top-third of all tracts in a given commuting zone instead of focusing on urban tracts. We find that we fail to reject the null hypothesis of null effects in as few as 20.7\% of selected tracts (in New Orleans) and as many as 93.6\% of selected tracts (in Portland). Figure \ref{fig:failure_to_reject_by_CZ} plots the proportion of tract for which we fail to reject null effects in each of the top 50 commuting zones by population.

In our final exercise on tract-level effects, we study pairwise comparisons of low and high-opportunity tracts. In particular, we consider the thought experiment of a household moving from an arbitrary low-opportunity to an arbitrary high-opportunity tract, and seek to study how often we can reject the possibility of a null effect on economic mobility associated with this move. To be precise, we will consider comparisons of the $\tau$-best with the $\kappa$-worst tracts, for $\tau$ and $\kappa$ in $J$. As discussed in \cite{Mogstadetal2023}, this problem is closely related to the problem of inference for ranks. In their application to the economic mobility literature, \cite{Mogstadetal2023} consider the problem of inference on all pairwise comparisons in order to be agnostic on movement patterns among CMTO enrollees. Our analysis is notably different from that of \cite{Mogstadetal2023}, since we restrict attention to pairwise comparisons of selected tracts. This allows us to be agnostic on movement patterns only among CMTO enrollees living in low-ranked tracts prior to treatment. 

In particular, we consider the top and bottom fifths of urban tracts in Seattle, and for the sake of demonstrating heterogeneity between urban areas, Cleveland. In Seattle, we consider 26 high and 26 low-opportunity tracts of interest. We compare each of the top fifth and bottom fifth tracts, leading to 676 pairwise comparisons. This corresponds to comparisons of the $\tau=26$-best with $\kappa=26$-worst tracts. We also consider pairwise comparisons of each top-fifth tract with the bottom-ranked tract, and a comparison of the top-ranked and bottom-ranked tracts. We present our results in table \ref{tab:top_vs_bottom_CMTO} below.

\begin{table}[H]
    \centering
    \caption{Inference on the economic mobility gains associated with moving from low to high mobility urban tracts in Seattle and Cleveland.}
    \begin{tabular}{c|cc|ccc}
    \hline
         \multirow{2}{*}{CZ} & \multirow{2}{*}{$\tau$-best} &  \multirow{2}{*}{$\kappa$-worst} &  \% Fail to Reject&  Lowest LCB&  Highest LCB TS  \\
         & & & two-step & two-step & two-step \\
         \hline
        \multirow{3}{*}{Seattle} & 26 & 26 & 87\% & [-0.33, 0.59] &  [0.077, 0.50] \\
        & 26 & 1 & 100\% & [-0.19, 0.60] &  [-0.055, 0.66] \\
         & 1 & 1 & 100\% & [-0.055, 0.66] &  [-0.055, 0.66] \\
         \hline
        \multirow{3}{*}{Cleveland} & 34 & 34 & 31\% & [-0.076, 0.59] &  [0.11, 0.49] \\
        & 34 & 1 & 9\% & [-0.056, 0.38] &  [0.11, 0.49] \\
         & 1 & 1 & 0\% & [0.12, 0.48] &  [0.12, 0.48] \\
         \hline\hline
    \end{tabular}
    \label{tab:top_vs_bottom_CMTO}
\end{table}

We find that we fail to reject the possibility of a null effect associated with moving from many arbitrary low to arbitrary high-opportunity tracts. This finding matches those of \cite{Mogstadetal2023}, who suggest that one cannot reject the possibility of the bottom-ranked tract in the Seattle CZ according to estimated economic mobility lying in the top-third of tracts according to true economic mobility. Indeed, in Seattle, there remains little we can say about tract-level effects, even when attempting to focus the power of an inference procedure on certain tracts of interest. 

\subsubsection{Effects at the Commuting Zone Level}

Our previous analyses sought to study tract-level effects on economic mobility in the context of the CMTO program. However, policymakers concerned with designing national level policies may be more concerned with targeting interventions according to commuting zone level estimates of economic mobility. We apply our methods to revisit the studies of \cite{ChettyandHendren2018} and \cite{Chettyetal2014} on the geography of economic mobility at the commuting zone level. While the analysis of \cite{Mogstadetal2023} suggests that it is difficult to construct a ranking of all commuting zones in the U.S. according to economic mobility, we consider a complementary exercise where we compare high and low-mobility commuting zones.

In table \ref{tab:top_vs_bottom_CZ_level} below, we replicate our analysis of the mobility effects of moving from low to high-opportunity areas at the commuting zone level. Our results on mobility effects at the commuting zone level are more conclusive than our findings at the tract-zone level. This is in part due to the simple fact that the standard errors on CZ level effects are smaller than those on tract level effects. We find that, for the majority of high and low-opportunity commuting zone pairs, we can indeed conclude that there exists a non-zero difference in the true mobilities of both CZs.

\begin{table}[H]
    \centering
    \caption{Inference on the economic mobility gains associated with moving from low to high mobility commuting zones. We compare all commuting zones in the U.S.}
    \begin{tabular}{cc|ccc}
    \hline
         \multirow{2}{*}{Top} &  \multirow{2}{*}{Bottom} &  \% Fail to Reject&  Lowest LCB&  Highest LCB  \\
         & & two-step & two-step & two-step \\
         \hline
         50\% & 50\% & 18.5\% & [-0.28, 0.35] &  [0.33, 0.44] \\
        33\% & 33\% & 8.2\% & [-0.25, 0.50] &  [0.33, 0.44] \\
         20\% & 20\% & 4.7\% & [-0.19, 0.52] &  [0.34, 0.44] \\
         10\% & 10\% & 3.5\% & [-0.14, 0.55] & [0.34, 0.44] \\
         \hline
         33\% & 50\% & 12.0\% & [0.28, 0.35] & [0.33, 0.44] \\
         20\% & 50\% & 8.3\% & [-0.22, 0.38] & [0.33, 0.44] \\
         10\% & 50\% & 7.1\% & [-0.18, 0.41] & [0.34, 0.44] \\
         \hline 
         33\% & 67\% & 19.9\% & [-0.28, 0.35] & [0.33, 0.44] \\
         20\% & 80\% & 19.1\% & [-0.25, 0.29] & [0.33, 0.44] \\
         10\% & 90\% & 17.6\% & [-0.22, 0.31] & [0.34, 0.44] \\
         \hline\hline
    \end{tabular}
    \label{tab:top_vs_bottom_CZ_level}
\end{table}

Our results are particularly compelling in the context of \cite{Mogstadetal2023}, who comment that:
\begin{quote}
    It is often not possible to tell apart with
95\% confidence the CZs where children have opportunities to succeed from those without such
opportunities.
\end{quote}

\noindent By focusing on selected CZs, however, we are able to distinguish high and low-opportunity commuting zones with high frequency. 

\section{Simulation Study}
\label{sec:Simulation_Study_Main}
In this section, we conduct an extended simulation study comparing our two-step approach to inference after ranking with existing methods. We demonstrate that the two-step approach to inference after ranking outperforms the projection approach almost uniformly across a broad range of simulation designs. This outperformance is most pronounced in instances where the set of asymptotic selections is a proper subset of the set of all possible selections. In the simulation results presented in this section, we demonstrate that our methods perform favorably relative to standard approaches to inference such as projection, as well as relative to the state of the art of \cite{ZrnicFithian2024} and \cite{ZrnicFithian2024WP}.

We consider simulations where $J = [p]$, for some natural number $p$. In total, we consider 28 distinct simulation designs. We present four selected simulation studies in this section, and present the results from all simulation studies in appendix E.
We consider the following designs:
\begin{itemize}
    \item \textbf{Design A} $p = 5$, $R=\{5\}$, $\mu_Y = 0$, $\mu_{X} = \arctan (j-3)$
    \item \textbf{Design B} $p = 10$, $R=\{10\}$, $\mu_Y = 0$, $\mu_{X} = \arctan(j-5.5)$
    \item \textbf{Design C} $p = 5$, $R=\{5\}$, $\mu_Y = 0$, $\mu_{X} = 0$
    \item \textbf{Design D} $p = 5$, $R=\{5\}$, $\mu_Y = 0$, $\mu_{X} = \mathds{1}(j \in \{1,2\})$
\end{itemize}
In all of our simulations, we consider four distinct covariance cases. In particular, we have a simple covariance case where $X$ and $Y$ are perfectly correlated but the $X_j$ are independent, a low covariance case where all units are weakly correlated, a medium covariance case, and a high covariance case. In these four cases, we denote the variance covariance matrices by $\Sigma_{simple},\Sigma_{low}, \Sigma_{medium}$, or $\Sigma_{high}$. We provide explicit formulae for these variance covariance matrices in appendix E. In the simulation results presented in this section, we take $\Sigma = \Sigma_{simple}$. We present results from all other cases in appendix E. We scale $\Sigma$ by $1/n$ for sample size $n$ equal to $100,\ 1000,\ \text{and } 10000$. We emphasize that in all simulations, we treat $\Sigma$ as known for computational simplicity.

Table \ref{tab:sims_cov_prob} below includes results on coverage probability from selected designs, namely the simple correlation case of designs A, B, C, and D which demonstrate the two-step approach's ``clear-winner" property (see designs A and B).

\begin{table}[H]
    \centering
\caption{Coverage Probability in a Small Scale Simulation Study}
\begin{tabular}{cc|ccc}
\multirow{2}{*}{Design} & \multirow{2}{*}{Confidence Set} & \multicolumn{3}{c}{Sample Size} \\
& & 100 & 1000 & 10000 \\
\hline\hline
\multirow{4}{*}{ A } & Projection & 0.991 & 0.991 & 0.992 \\
& two-step & 0.966 & 0.955 & 0.955 \\
& zoom & 0.977 & 0.951 & 0.950\\
& Locally Simultaneous & 0.971 & 0.955 & 0.955 \\
\hline
\multirow{4}{*}{ B } & Projection & 0.993 & 0.996 & 0.995 \\
& two-step & 0.969 & 0.970 & 0.960 \\
& zoom & 0.985 & 0.985 & 0.971\\
& Locally Simultaneous & 0.979 & 0.976 & 0.962 \\
\hline
\multirow{4}{*}{ C } & Projection & 0.976 & 0.976 & 0.976 \\
& two-step & 0.962 & 0.962 & 0.962 \\
& zoom & 0.977 & 0.977 & 0.977 \\
& Locally Simultaneous & 0.979 & 0.979 & 0.979 \\
\hline
\multirow{4}{*}{ D } & Projection & 0.991 & 0.991 & 0.991 \\
& two-step & 0.970 & 0.970 & 0.969 \\
& zoom & 0.979 & 0.976 & 0.976 \\
& Locally Simultaneous & 0.978 & 0.978 & 0.979 \\
\hline\hline
\end{tabular}

    \label{tab:sims_cov_prob}
\end{table}
We find that, between the models described above, the two-step approach substantially reduces overcoverage relative to the projection approach when the set of selections $J_R(P)$ is clear, and specifically a proper subset of $J$ such that $J\setminus J_R(P)$ is large. Moreover, in intermediate cases when the set of selections is moderately clear, as in designs A and B for low $n$, the two-step approach outperforms the approach of \cite{ZrnicFithian2024WP}, which is based on the zoom test.\footnote{Our implementation of the zoom test is based on the step-down procedure outlined in section 4 of \cite{ZrnicFithian2024WP}.} In general, the two-step approach also outperforms the approach of \cite{ZrnicFithian2024}. Quantitatively, we have that in the four simulations presented above, the two-step approach to inference can reduce absolute overcoverage error by up to 88\% relative to projection inference, 50\% relative to locally simultaneous inference, and 56\% relative to the zoom test. Across all simulations, including those in appendix E, the two-step approach to inference reduces overcoverage error by up to 96\% relative to projection inference, up to 71\% relative to locally simultaneous inference, and up to 67\% relative to the zoom test.

Table \ref{tab:sims_length} below demonstrates that, over a wide range of data generating processes, the two-step approach to inference provides tighter confidence regions than the projection, zoom, and locally simultaneous approaches to inference. Indeed, the interval lengths of the two-step approach to inference may be up to 27\% shorter than projection inference. Moreover, two-step inference may be up to 11\% shorter than inversions of the zoom test and up to 8\% shorter than locally simultaneous inference.

\begin{table}[H]
    \centering
    \caption{Confidence Interval Length in a Small Scale Simulation Study, as a Fraction of Projection Interval Length}
\begin{tabular}{cc|ccc}
\multirow{2}{*}{Design} & \multirow{2}{*}{Confidence Set} & \multicolumn{3}{c}{Sample Size} \\
& & 100 & 1000 & 10000 \\
\hline\hline
\multirow{3}{*}{ A } & two-step & 0.835 & 0.780 & 0.780 \\
& zoom & 0.872 & 0.786 & 0.763\\
& Locally Simultaneous & 0.879 & 0.781 & 0.780 \\
\hline
\multirow{3}{*}{ B } & two-step & 0.824 & 0.784 & 0.730 \\
& zoom & 0.900 & 0.842 & 0.758\\
& Locally Simultaneous & 0.888 & 0.827 & 0.731 \\
\hline
\multirow{3}{*}{ C } & two-step & 0.937 & 0.937 & 0.937 \\
& zoom & 1.003 & 1.003 & 1.003 \\
& Locally Simultaneous & 1.014 & 1.014 & 1.014 \\
\hline
\multirow{3}{*}{ D } & two-step & 0.846 & 0.846 & 0.846 \\
& zoom & 0.914 & 0.873 & 0.873 \\
& Locally Simultaneous & 0.887 & 0.886 & 0.886 \\
\hline\hline
\end{tabular}
    \label{tab:sims_length}
\end{table}

\begin{rem}
In general, we recommend choosing $\beta = \alpha / 10$. As our simulation results demonstrate, the two-step approach to inference performs quite well under such a choice of $\beta$. This choice of $\beta$ is the same as in \cite{andrewsetal2023} and \cite{Romanoetal2014}.
\end{rem}

\bibliography{refs}    

\begin{appendix}
    \section{Alternative Approaches to Inference After Ranking}
\label{app:alternative_approaches}
In this section, we discuss four alternative approaches to the inference after ranking problem in the normal location model introduced in section 2. We first show that our two-step approach to inference can be intersected with a projection confidence set in the spirit of theorem 2 of \cite{ZrnicFithian2024}. We then discuss the approach to inference presented in \cite{ZrnicFithian2024WP}. Finally, we discuss conditional and hybrid inferences in depth (see e.g., [\cite{andrewsetal2022}, \cite{andrewsetal2023}, \cite{Leeetal2016}, \cite{McCloskey2023}]), and demonstrate that they do not easily generalize to settings where multiple selections are made.

Before proceeding, we will introduce some additional notation used throughout the remainder of the supplemental material. For real numbers $a$ and $b$, we define $a\wedge b := \min\{a,b\}$ and $a\vee b := \max\{a,b\}$. For natural numbers $z_1, z_2 \in \mathbb{N}$, we denote by $\boldsymbol{0}_{z_1\times z_2}$ the $z_1\times z_2$ zero matrix, and for $i\in [z_1]$ we denote by $e_i$ the $i$-th canonical basis vector in $\mathbb{R}^{z_1}$.
\subsection{A weakly improved two-step approach}
We provide a non-inferiority result in the spirit of theorem 2 of \cite{ZrnicFithian2024}. To do so, we need to introduce some new notation. For any subset $J_c$ of $J$, we define $\widetilde{c}_{1-\alpha}(J_c)$ to be the $(1-\alpha)$-quantile of:
\begin{equation}
\label{eq:c_bar_quantile_def}
    \max_{j\in J_c}\left[\frac{|\xi_{Y,j}|}{\sqrt{\Sigma_{Y,jj}}}\vee \frac{|\xi_{X,j}|}{\sqrt{\Sigma_{X,jj}}}\right]~.
\end{equation}
We will denote by $\widetilde{c}_{1-\alpha}$ the critical value $\widetilde{c}_{1-\alpha}(J)$. Under some mild conditions relating $\widetilde{c}_{1-\alpha}$ and $d_{1-\beta}(\Sigma)$ we can provide a weak improvement to two-step inference. In particular, we define the following confidence region:
\begin{equation*}
CS_{\hat\jmath}^{TS2}(1-\alpha;\beta) := \left[
Y_j - \left(\rho_{1-\alpha+\beta}(L,U)\wedge\widetilde{c}_{1-\alpha}\right)\sqrt{\Sigma_{Y,\hat{\jmath}\hat{\jmath}}},
Y_j + \left(\rho_{1-\alpha+\beta}(L,U)\wedge\widetilde{c}_{1-\alpha}\right)\sqrt{\Sigma_{Y,\hat{\jmath}\hat{\jmath}}}
\right]~.
\end{equation*} 
As usual, we will denote by $CS^{TS2}(1-\alpha;\beta)$ the collection $(CS_{\hat\jmath}^{TS2}(1-\alpha;\beta))_{\hat\jmath\in\hat{J}_R}$. 
\begin{prop}
\label{prop:normal_validity_non_inferior}
Suppose $\beta$ and $\Sigma$ satisfy:
\begin{equation*}
    \widetilde{c}_{1-\alpha}(J)(\sqrt{\Sigma_{X,jj}}+\sqrt{\Sigma_{X,j'j'}})\leq d_{1-\beta}(\Sigma)\sqrt{\text{var}_{jj'}}
\end{equation*}
for all $j\neq j'$. Then $CS^{TS2}(1-\alpha;\beta)$ is a valid confidence set at the $1-\alpha$-level, such that (4) holds for all $\mu$ and $\Sigma$.
\end{prop}
We prove this result in appendix \ref{app:Proofs_of_other_theoretical_results_appendix}. It follows from the above proposition that, whenever $X=Y$, $CS_{\hat\jmath}^{TS2}(1-\alpha;\beta)\subseteq CS_{\hat\jmath}^{P}({1-\alpha})$ for any $\hat\jmath\in\hat{J}_R$, providing a finite-sample non-inferiority result. 

\subsection{A Zoom Test for Inference} \label{app:ZF}
We now discuss an approach to inference after ranking based on the zoom test of \cite{ZrnicFithian2024WP}. The approach in \cite{ZrnicFithian2024WP} suggests allocating the error budget to near-winners by inverting a test based on an acceptance region which is increasing in the population suboptimality - the difference between a candidate's mean and the population selection's mean. While it is unclear how to precisely generalize their approach to the exact inference after ranking problem we discuss in this paper, \cite{ZrnicFithian2024WP} provide guidance on test inversion in the case where $R = \{p-\tau + 1, \ldots, p\}$ for some $\tau \geq 1$. 

Recall that, in section 4.1, we defined $J_R(\mu) := \hat{J}_R(\mu_X)$. Let the suboptimality $D_j := \min_{j' \not\in J_R(\mu)} |\mu_{X,j} - \mu_{X,j'}|$ for any $j\in J$. Clearly, $D_j \geq 0$ for all $j \in J$, with equality for all $j\not\in J_R(\mu)$. As in \cite{ZrnicFithian2024WP}, let us choose $r_\alpha$ to be the $1-\alpha$-quantile of:
\begin{equation*}
\max_{j\in J}|\xi_{Y,j}|\mathds{1}\left\{
|\xi_{Y,j}|>\frac{D_j}{2}
\right\}~.
\end{equation*}
With this choice of $r_\alpha$, we obtain that the following is a valid level $1-\alpha$ acceptance region:
\begin{equation}
\label{eq:acceptance_region_zoom}
    A_\alpha(\mu_Y,\mu_X) := \left\{y:y_j \in\left[
\mu_{Y,j} \pm \left(
r_\alpha \vee\frac{D_j}{2}
\right)
    \right] \quad \text{for all} \quad j\in J\right\}~.
\end{equation}
Under the point hypothesis $H_{0,(\mu_Y, \mu_{X})}:\mathbb{E}(Y) = \mu_Y,\ \mathbb{E}(X) = \mu_{X}$, the probability that $Y$ lies in $A_\alpha(\mu_Y,\mu_X)$ exceeds $1-\alpha$. We define the following, joint confidence set for $(\mu_Y, \mu_X)$ based on an inversion of the zoom test based on the acceptance region $A_\alpha$:
\begin{equation}
    \label{eq:joint_CS_zoom}
    CS(1-\alpha) = \left\{
\mu_Y,\mu_X : Y \in A_\alpha(\mu_Y, \mu_{X})
    \right\}
\end{equation}
For any $\hat{\jmath} \in \hat{J}_R$, we define the marginal confidence set:
\begin{equation*}
    CS_{\hat{\jmath}}^{zoom}(1-\alpha) := \left\{
c : \exists\ \widetilde{\mu}_Y,\ \widetilde{\mu}_X \in CS(1-\alpha)\quad\text{s.t.}\quad \widetilde{\mu}_{Y,\hat{\jmath}} = c
    \right\}
\end{equation*}
Consequently, we take $CS^{zoom}(1-\alpha)$ in the usual manner. It is not clear how to efficiently invert the test based on the acceptance region $A_\alpha$ for completely general $R$. \cite{ZrnicFithian2024WP} provide a parsimonious characterization of the confidence set based on inversion of the zoom test under certain cases, namely when $R$ is of the form $\{p-\tau + 1,\ldots,p\}$ for some $\tau\in [p]$. Computationally, \cite{ZrnicFithian2024WP} provide a step-wise implementation of their methods which we implement in all simulations included in this paper.

\subsection{The Conditional Approach to Inference}
\label{sec:Conditional_Inference}
The conditional inference approach outlined in \cite{andrewsetal2023} is an example of conditional selective inference, notably studied in \cite{Leeetal2016}. We outline a generalization of the conditional approach from \cite{andrewsetal2023} which accounts for multiple selections. We find that the exact distribution of the collection of the $(Y_{\hat{\jmath}})_{\hat{\jmath}\in\hat{J}_R}$, conditional on the selection event for the $\hat{J}_R$, is a multivariate normal truncated to a union of convex polyhedra. Unfortunately, due to computational challenges, this characterization does not lead to a practical inference procedure for the problem of inference after ranking.

In our analysis of conditional and hybrid inference, we will neglect ties. Formally, we assume the following:
\begin{assu}
\label{assu:selection_uniqueness}
    For any $r$ in $[p]$, $\hat{J}_{\{r\}}$ is a singleton, almost surely.
\end{assu}
We will let $j_l$ be the unique indices in $J$ such that $r_{j_l}(X) = l$, and condition on the event $j_l = i_l$ for $l$ in $R$, with $(i_l)_{l\in R} \subseteq J$ being some collection of indices. We additionally condition on a sufficient statistic $Z$ for the nuisance parameters associated with the elements of $\mu$ not corresponding to the $Y_{i_l}$. Conditional inferences allow us to obtain joint confidence sets $CS({1-\alpha})$ in $\mathbb{R}^k$ such that:
\begin{equation}
\label{eq:conditional_validity_CS}
    P_{\mu,\Sigma}\left(
(\mu_{Y,j_l})_{l\in R} \in CS({1-\alpha})~ \big|~ j_l = i_l  ~\text{for all}~ l\in R,\ Z=z
    \right) \geq 1-\alpha
\end{equation}
for all $\mu$ and $\Sigma$. By the law of iterated expectations, such a confidence set satisfies (4) as well. In order to obtain some confidence set satisfying \eqref{eq:acceptance_region_zoom}, we first derive an extension of the polyhedral selection lemma in \cite{Leeetal2016}; see, in particular, lemma 5.1 therein. In order to describe our generalization of this result, we require some further notation. Let $(e_{j}' - e_i')_{i \neq j}$ denotes the matrix obtained from stacking the vectors $e_{i_1}' - e_i'$ for all $i\neq j$. For ease of exposition, let $R := \{p-\tau+1,\ldots,p\}$.\footnote{When $R$ is not of the form $\{p-\tau+1,\ldots,p\}$, we can take the selection event to be a union of polyhedra. Details on such unions of polyhedral selection events for ranked objects are contained in \cite{andrewsetal2022}. However, in our setting, it is unclear how to derive a parsimonious characterization of the union of polyhedra in the spirit of algorithm 1 in \cite{andrewsetal2022}. It is worth noting that conditional inference is especially challenging for such $R$, since the number of polyhedra over which we take a union can be very large, leading to well-known numerical integration issues, as per the discussion in \cite{Leeetal2016}.} Define:
\begin{equation*}
    B' := \begin{pNiceArray}{cc}
    \Block{4-1}<\Large>{\mathbf{0}_{\tau\times p}} & e_{i_{p}}'\\
    & e_{i_{p-1}}'\\
    & \vdots\\
    & e_{i_{p-\tau+1}}'
\end{pNiceArray}~,\quad 
        A:= \begin{pmatrix}
            (e_{i_p}' - e_i')_{i \neq i_1} & \mathbf{0}_{p-1\times p}\\
            (e_{i_{p-1}}' - e_i')_{i \neq i_1,i_2} & \mathbf{0}_{p-2\times p}\\
            \vdots & \vdots \\
            (e_{i_{p-\tau +1}}' - e_i')_{i\neq i_1,\ldots,i_\tau} & \mathbf{0}_{p-\tau\times p}
        \end{pmatrix} ~,
\end{equation*}
and $c := \Sigma B (B'\Sigma B)^{-1}$. By assumption \ref{assu:selection_uniqueness}, the selection event $j_l = i_l$ for $l$ in $R$ is equivalent to the following:
    \begin{equation*}
        A \begin{pmatrix}
            X\\
            Y
        \end{pmatrix} \geq 0~.
    \end{equation*}
 Finally, we define the random variable $Z := (I_{2p} - cB')Y$. We will condition on $Z=z$. Using this notation, we have the following result:
    \begin{lem}
    \label{lem:Polyhedral_lemma}
        The collection $(Y_{\hat{\jmath}})_{\hat{\jmath} \in \hat{J}_R} = (Y_{j_l})_{l\in R}$, conditional on $Z=z$ and $j_l = i_l$ for $l$ in $R$, is distributed according to a multivariate normal with mean $B'\mu$ and variance-covariance $B'\Sigma B$ truncated to the polyhedron $\mathcal{Y}((i_l)_{l\in R},z) := \{x : (Ac)x \geq -Az\}$. We write:
        \begin{equation}
        (Y_{j_l})_{l\in R} | Z=z, j_l = i_l\ \text{for}\ l\in R \sim TN_{\mathcal{Y}((i_l)_{l\in R},z)}(B'\mu, B'\Sigma B)
        \end{equation}
    \end{lem}
Notice that $B'\mu = (\mu_{Y,i_l})_{l\in R}$. Given this lemma, we may consider the problem of testing the null, for $m_Y \in \mathbb{R}^k$ indexed by $i_l$ for $l$ in $R$:
\begin{equation}
\label{eq:null_for_cond_test_inv}
H_{0,m}: (\mu_{Y,i_l})_{l\in R} = (m_{Y,i_l})_{l\in R}
\end{equation}
conditional on $Z=z, j_l = i_l\ \text{for all}\ l\in R$. Let us denote by $c_{TN,\alpha}((i_l)_{l\in R}, z, (m_{Y,i_l})_{l\in R})$ the critical following value:
\begin{equation*}
     \inf\left\{x : P_{m,B'\Sigma B} \left(\left\|\left(Y_{i_l} - m_{Y,i_l}\right)_{l \in R}\right\| \leq x | Z=z, j_l = i_l\ \text{for all}\ l\in R \right) \geq 1 - \alpha\right\}
\end{equation*}
under the conditional distribution calculated using the distribution in lemma \ref{lem:Polyhedral_lemma}. We suggest the following level-$\alpha$ test of the null in \eqref{eq:null_for_cond_test_inv}:
\begin{align*}
&\phi\left((Y_{i_l})_{l\in R};(i_l)_{l\in R}, z, (m_{Y,i_l})_{l\in R}\right) \\
& = \begin{cases}
    1 & \text{if} \quad c_{TN,\alpha}((i_l)_{l\in R}, z, (m_{Y,i_l})_{l\in R}) < \left\|\left(Y_{i_l} - m_{Y,i_l}\right)_{l \in R}\right\|\\ 
    0 & \text{otherwise}
\end{cases}
~.
\end{align*}
We therefore construct a confidence set for the $\left(\mu_{Y,i_l}\right)_{l\in R}$ by inverting $\phi$. In particular, let:
\begin{equation*}
    CS^c(1-\alpha) := \left\{(m_{Y,i_l})_{l\in R} :
\phi\left((Y_{i_l})_{l\in R};(i_l)_{l\in R}, z, (m_{Y,i_l})_{l\in R}\right) = 0
    \right\}
\end{equation*}
For the resulting confidence set $CS^c({1-\alpha})$, the following proposition holds:
\begin{prop}
\label{prop:conditional_validity}
    For $\alpha$ in $(0,1)$, under assumption \ref{assu:selection_uniqueness}, $CS^c({1-\alpha})$ is a valid conditional confidence set at level $\alpha$ such that (\ref{eq:conditional_validity_CS}) holds for all $\mu$ and $\Sigma$. 
\end{prop}
We provide a proof in appendix \ref{app:Proofs_of_other_theoretical_results_appendix}. Now, we provide the following remark on the feasibility of conditional inference:
\begin{rem}
\cite{andrewsetal2023} show that, in the $R = \{p\}$ case, $Y_{\hat{\jmath}}$ is distributed according to a univariate normal truncated to an interval for $\hat{\jmath}$ in $\hat{J}_{\{p\}}$. Given this, the test inversion procedure described above is quite tractable, since the test statistic used to compute $\phi$ can be easily computed using the cumulative distribution function of a univariate truncated normal. However, in the multivariate case, computing these test statistics over a high-dimensional grid for the values of $(\mu_{Y,i_l})_{l\in R}$ is a challenging numerical integration problem.\footnote{Motivated by such concerns, \cite{liu2023} applies the separation of variables technique from \cite{Genz1992} to problems of polyhedral selection. \cite{liu2023} achieves performance gains in the numerical integration of multivariate Gaussians over polyhedra, but the issue of test-inversion on a potentially high-dimensional grid remains. } 
\end{rem}

\subsection{The Hybrid Approach to Inference}
\cite{andrewsetal2023} and \cite{McCloskey2023} suggest an approach to selective inference related to both approaches described above, in particular the conditional and projection approaches. Again, ignoring ties, we assume that $\hat{J}_R$ can be written as $(j_l)_{l\in R}$, such that $r_{j_l}(X) = l$. In particular,  \cite{andrewsetal2023} and \cite{McCloskey2023} suggest conditioning not only on the selection event $j_l = i_l$ for $l$ in $R$, but on the event that $(\mu_{Y,j})_{j\in J} \in CS^P({1-\beta})$ for $\beta$ in $(0,\alpha)$. In particular, in our setting, one potential approach to hybrid inference would involve inverting tests of the null \eqref{eq:null_for_cond_test_inv} conditional on $Z=z,~ j_l = i_l,~ m_{Y,i_l} \in CS_{i_l}^P(1-\beta)$. Let us denote by $c_{TN,\frac{1-\alpha}{1-\beta}}((i_l)_{l\in R}, z, (m_{Y,i_l})_{l\in R}, CS^P({1-\beta}))$ the infimum of the set of $x$ satisfying:
\begin{equation*}
\resizebox{0.9\textwidth}{!}{$
P_{m,B'\Sigma B} \left(\left\|\left(Y_{j_l} - m_{Y,j_l}\right)_{l \in R}\right\| \leq x | Z=z, j_l = i_l,\ m_{Y,i_l} \in CS_{i_l}^P(1-\beta)\ \text{for all}\ l\in R \right) \geq 
\frac{1-\alpha}{1-\beta}~.
$}
\end{equation*}
As before, we notice that our polyhedral lemma \ref{lem:Polyhedral_lemma} applies to hybrid inference as well as to conditional inference, providing an exact characterization of the above conditional distribution. We generalize the hybrid approaches of \cite{andrewsetal2023} and \cite{McCloskey2023} by using a test of the following form:
\begin{align*}
&\phi\left((y_{i_l})_{l\in R};(i_l)_{l\in R}, z, (m_{Y,i_l})_{l\in R}\right) \\
& = \begin{cases}
    1 & \text{if} \quad c_{TN,\frac{1-\alpha}{1-\beta}}((i_l)_{l\in R}, z, (m_{Y,i_l})_{l\in R}, CS^P({1-\beta})) \geq \left\|\left(y_{i_l} - m_{Y,i_l}\right)_{l \in R}\right\| \\
    0 & \text{otherwise}
\end{cases}
~.
\end{align*}
Inverting this test provides the hybrid confidence set $CS^H({1-\alpha;\beta})$, such that the following proposition holds.
\begin{prop}
\label{prop:hybrid_validity}
    For $\alpha$ in $(0,1)$ and $\beta$ in $(0,\alpha)$, under assumption \ref{assu:selection_uniqueness}, $CS^H({1-\alpha;\beta})$ is an unconditionally valid confidence set at level $\alpha$, such that:
    \begin{equation*}
        P_{\mu,\Sigma}\left(
(\mu_{Y,\hat{\jmath}})_{\hat{\jmath}\in\hat{J}_R} \in CS^H({1-\alpha;\beta})
    \right) \geq 1-\alpha
    \end{equation*}
\end{prop}
To see that validity holds, it suffices to notice that:
\begin{align*}
    & P_{\mu,\Sigma}\left(
(\mu_{Y,\hat{\jmath}})_{\hat{\jmath}\in\hat{J}_R} \in CS^H({1-\alpha;\beta})
    \right)\\
    & \geq P_{\mu,\Sigma}\left(
(\mu_{Y,\hat{\jmath}})_{\hat{\jmath}\in\hat{J}_R} \in CS^H({1-\alpha;\beta}) | (\mu_{Y,\hat{\jmath}})_{\hat{\jmath}\in\hat{J}_R} \in CS^P ({1-\beta})
    \right)\\
    & \quad \cdot P_{\mu,\Sigma}\left(
(\mu_{Y,\hat{\jmath}})_{\hat{\jmath}\in\hat{J}_R} \in CS^P ({1-\beta })
    \right)\\
    & \geq \frac{1-\alpha}{1-\beta} (1-\beta) = 1-\alpha
\end{align*}
Because hybrid inference involves conditioning in the construction of a test $\phi$, it is subject to the same computational concerns as conditional inference in our generalized setting. In cases where hybrid inference easily applies, as in the setting of \cite{andrewsetal2023}, we observe that any confidence set satisfying valid unconditional coverage of the $(\mu_{Y,\hat{\jmath}})_{\hat{\jmath}\in\hat{J}_R}$ can be used in lieu of projection, including our proposed two-step approach to inference.\footnote{This point has also been noted in \cite{ZrnicFithian2024WP}.} 

\subsection{Simulations Comparing Different Approaches to Inference}
We now provide results from a small simulation study comparing our two-step approach to inference with the approaches above. We provide results from a more extensive simulation study in appendix \ref{app:all_sims}. In the simulations below, we consider $\Sigma_X = \Sigma_Y = \Sigma_{XY} = I_2$, $\mu_X = \begin{pmatrix}
    \mu_1 & 0
\end{pmatrix}'$ and $\mu_Y = \begin{pmatrix}
    0 & 0
\end{pmatrix}'$, and vary $\mu_1$. Our two-step approach to inference substantially outperforms the approaches of \cite{ZrnicFithian2024} and \cite{ZrnicFithian2024WP} for intermediate values of $\mu_1$. In particular, our approach reduces over-coverage error by as much as 68.6\% relative to these approaches. Our simulation results are plotted below:
\begin{figure}[H]
    \centering
    \begin{tabular}[b]{cc}
        \begin{tabular}[b]{c}
            \begin{subfigure}[b]{0.45\columnwidth}
                \includegraphics[width=\textwidth]{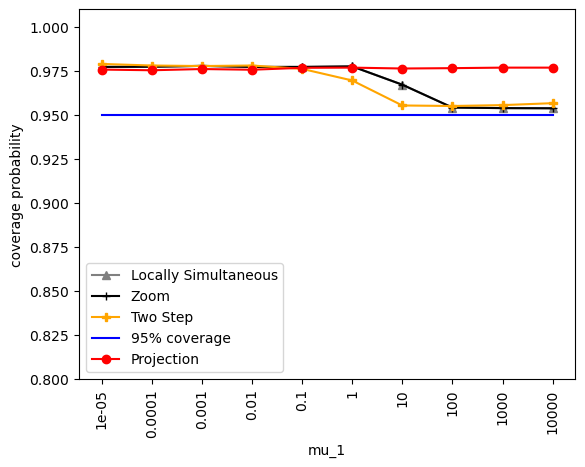}
                \caption{}
            \end{subfigure}
        \end{tabular}
        &
        \begin{tabular}[b]{c}
            \begin{subfigure}[b]{0.45\columnwidth}
                \includegraphics[width=\textwidth]{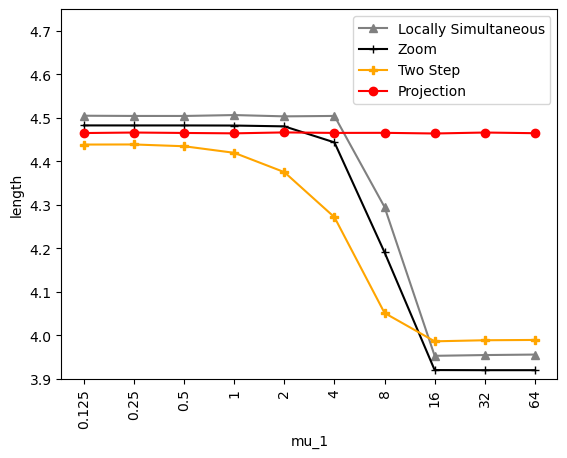}
                \caption{}
            \end{subfigure}
        \end{tabular}
    \end{tabular}
    \caption{Confidence set coverage (a) and width (b) as $\mu_1$ varies. Results are plotted for projection and two-step inference, as well as the locally simultaneous approach of \cite{ZrnicFithian2024}, and the zoom test of \cite{ZrnicFithian2024WP}}
\end{figure}

\section{Uniform Asymptotic Validity and Proofs}
\label{app:Two_Step_Uniform_Asymptotic_Validity_Appendix}
In this section, we restate the assumptions stated in subsection 3.2 and introduce several lemmas supporting the proof of proposition 2. The results and proofs in this section take place in the asymptotic framework of section 3.2. First, we provide the following uniform integrability assumption, which is sufficient for uniform convergence in distribution, per lemma \ref{lem:Uniform_Convergence_in_Distribution}.
\begin{assu}
\label{assu:Uniform_Asymptotic_Normality_Sufficient}
    For $j=1,\ldots,2p$, it holds that:
    \begin{equation}
    \label{eq:UI_Assumption}
        \limsup_{K\to\infty}\sup_{P\in\mathcal{P}}\mathbb{E}_P\left(
        \frac{\left|\widetilde{W}_{1,j}-\mu_{W,j}(P)\right|^2}{\Sigma_{W,jj}(P)}\mathds{1}
        \left(
 \frac{\left|\widetilde{W}_{1,j}-\mu_{W,j}(P)\right|}{\sqrt{ \Sigma_{W,jj}(P)}}> K
        \right)
        \right) = 0
    \end{equation}
    \end{assu}
% assu:Uniform_Asymptotic_Consistency_Covariance

Before stating lemma \ref{lem:Uniform_Convergence_in_Distribution}, we introduce some new notation that we will use throughout the remainder of the supplemental material.

For $j\in J$, we denote by $\xi_{\widetilde{X}_i,j}$ and $\xi_{\widetilde{Y}_i,j}$ the $j$-th elements of $\xi_{\widetilde{X}_i}$ and $\xi_{\widetilde{Y}_i}$, respectively. Similarly, we denote by $\xi_{\widetilde{S}_X^n,j}$ and $\xi_{\widetilde{S}_Y^n,j}$ the $j$-th elements of $\xi_{\widetilde{S}_X^n}$ and $\xi_{\widetilde{S}_Y^n}$, respectively. Let $\text{var}_{jj'}(P) = \Sigma_{X,j'j'}(P) + \Sigma_{X,jj}(P) - 2\Sigma_{X,jj'}(P)$. We define $\Delta_{jj'}(P) := \mu_{X,j}(P) - \mu_{X,j'}(P)$. This defines the $p\times p$ matrix $\Delta(P)$. In order to streamline notation, we may denote $\Delta(P)$ by $\Delta$.

Let $\Phi_{V}$ denote the law of a random variable distributed according to a multivariate Gaussian with mean zero and variance-covariance $V$. For multivariate Gaussians not mean zero, we may denote by $\Phi_{\mu,V}$ the law of a random variable distributed according to a multivariate Gaussian with mean $\mu$ and variance-covariance $V$. Moreover, let $\mathcal{C}$ denote the set of convex subsets $S$ of $\mathbb{R}^{2p}$ satisfying $\Phi_V(\partial S)=0$ for all p.s.d. covariance matrices $V$ with diagonal elements $V_{jj}=1$. 

The following lemma simply restates lemma 3.1 of \cite{RomanoandShaikh2008}.
\begin{lem}
\label{lem:Uniform_Convergence_in_Distribution}
Under assumption \ref{assu:Uniform_Asymptotic_Normality_Sufficient}, we obtain uniform convergence in distribution such that $\mu_W(P)$ satisfy:
\begin{equation*}
\label{eq:Romano_and_Shaikh_Result}
\lim_{n\to\infty}\sup_{P\in\mathcal{P}}\sup_{S\in\mathcal{C}}\left|{\Pr}_P\left(
\sqrt{n}\left(
\widetilde{S}_W^n - \mu_{W}(P)
\right)\in S
    \right)-\Phi_{\Sigma(P)}(S)\right|=0~.
\end{equation*}
\end{lem}
\begin{proof}
The result is an immediate consequence of lemma 3.1 in \cite{RomanoandShaikh2008}.
\end{proof}
We also provide the following technical lemma. Let $R_n(\widetilde{S}_W^n,P_n)$ be some function of $\widetilde{S}_W^n$ and of $P_n \in \mathcal{P}$, with the cumulative distribution function $J_n(x,P_n)$ under $P_n \in \mathcal{P}$. Let $(\widetilde{P}_n)_{n\in\mathbb{N}}$ be some sequence of distributions in $\mathcal{P}$. The following lemma holds:
\begin{lem}
\label{lem:CDF_cv_implies_uniform_cv}
    Suppose, for any $\varepsilon >0$, that the sequence $(\widetilde{P}_n)_{n\in\mathbb{N}}$ satisfies the following:
    \begin{equation}
        \liminf_{n\to\infty}\inf_{P\in\mathcal{P}}{\Pr}_P\left(
        \sup_{x\in\mathbb{R}}\left|
J_n(x,\widetilde{P}_n)-J_n(x,P)
         \right|\leq\varepsilon
        \right) = 1
    \end{equation}
    Then, for any $0\leq\alpha_1$, $0\leq\alpha_2$ such that $0\leq\alpha_1+\alpha_2 < 1$, the following holds:
    \begin{equation}
    \label{eq:uniform_validity_of_critical_values}
        \liminf_{n\to\infty}\inf_{P\in\mathcal{P}}{\Pr}_P\left(
        J_n^{-1}(\alpha_1,\widetilde{P}_n) \leq R_n(\widetilde{S}_W^n,P_n) \leq J_n^{-1}(1-\alpha_2,\widetilde{P}_n)
        \right) \geq 1-\alpha_1-\alpha_2
    \end{equation}
\end{lem}
\begin{proof}
    Let us fix arbitrary $\eta > 0$. For any $n$ sufficiently large, we have:
    \begin{equation*}
        1-\frac{\eta}{2} \leq \inf_{P\in\mathcal{P}}{\Pr}_P\left(
    \sup_{x\in\mathbb{R}}\left|
J_n(x,\widetilde{P}_n) - J_n(x,P)
    \right|\leq \frac{\eta}{2}
        \right)
    \end{equation*}
    This implies, by part viii of lemma A.1 of \cite{RomanoShaikh_2012}, that:
    \begin{equation*}
        \inf_{P\in\mathcal{P}}{\Pr}_P\left(
J_n^{-1}(\alpha_1,\widetilde{P}_n) \leq R_n(\widetilde{S}_W^n,P_n) \leq J_n^{-1}(1-\alpha_2,\widetilde{P}_n)
        \right) \geq 1-\alpha_1-\alpha_2-\eta
    \end{equation*}
    Finally, because $\eta$ was arbitrary, we find that (\ref{eq:uniform_validity_of_critical_values}) holds, thus proving the lemma.
\end{proof}

The following lemma concerns the asymptotic properties of $L^n$ and $U^n$, and namely, whether $L^n$ and $U^n$ jointly satisfy a uniform, asymptotic validity condition as bounds of $\Delta(P)$. 
\begin{lem}
\label{lem:L_asymptotic_validity}
    Let us assume that assumption \ref{assu:Uniform_Asymptotic_Normality_Sufficient} holds. Under these conditions, it follows that $L^n$ and $U^n$ satisfy:
    \begin{equation}
    \label{eq:L_U_uniformly_valid}
        \liminf_{n\to\infty}\inf_{P\in\mathcal{P}}{\Pr}_P\left(
L^n \leq \Delta(P) \leq U^n
\right) \geq 1 - \beta
    \end{equation}
    where inequality is interpreted elementwise.
\end{lem}
\begin{proof}
We first notice that, by lemma s.6.1 and lemma s.7.1 of \cite{RomanoShaikh_2012}, and the continuous mapping theorem, we obtain following uniform consistency condition on the $\widehat{\text{var}}_{jj'}^n$:\footnote{We note that $\widehat{\text{var}}_{jj'} = 0$ if and only if $\text{var}_{jj'}(P)=0$.}
    \begin{equation}
    \label{eq:uniform_consistency_var_jk} 
    \limsup_{n\to\infty}\sup_{P\in\mathcal{P}}\sup_{j,j'\in J : \text{var}_{jj'}(P) > 0}{\Pr}_P\left(
    \left|
    \frac{n\widehat{\text{var}}_{jj'}^n}{\text{var}_{jj'}(P)}-1
    \right|>\varepsilon
    \right) = 0\quad \text{for $\varepsilon > 0$}
    \end{equation}
    The remainder of our proof will use the results of lemma \ref{lem:CDF_cv_implies_uniform_cv}. In particular, we define $R_n(\widetilde{S}_W^n, P)$ as follows:
\begin{equation}
\label{eq:R_n_def_d_1_b}
    \max_{j,j'\in J, \text{var}_{jj'}(P)\neq 0}\frac{\sqrt{n}\left|(\widetilde{S}_{X,j}^n - \mu_{X,j}(P)) - (\widetilde{S}_{X,j'}^n - \mu_{X,j'}(P))\right|}{\sqrt{\text{var}_{jj'}(P)}}
\end{equation}
We denote by $J_n(x,P)$ the cumulative distribution function of $R_n(\widetilde{S}_W^n,P)$ when the $\Tilde{W}_i$ are i.i.d. from $P$. Let $x>0$. We note that the event that $R_n(\widetilde{S}_W^n, P) \leq x$ holds if and only if $\sqrt{n}(\widetilde{S}_W^n - \mu_{W}(P))$ lies in a convex set $S$ contained in $\mathcal{C}$. To see this, notice that $R_n(\widetilde{S}_W^n, P) \leq x$ if and only if for all $j, j'$ in $J$ such that $\text{var}_{jj'}(P) \neq 0$:
\begin{equation*}
    \sqrt{n}\left|(\widetilde{S}_{X,j}^n - \mu_{X,j}(P)) - (\widetilde{S}_{X,j'}^n - \mu_{X,j'}(P))\right| \leq x \sqrt{\text{var}_{jj'}(P)}
\end{equation*}
It suffices to show that for any $j,j'$ such that $\text{var}_{jj'}(P) \neq 0$, the event:
\begin{equation*}
    \sqrt{n}\left|(\widetilde{S}_{X,j}^n - \mu_{X,j}(P)) - (\widetilde{S}_{X,j'}^n - \mu_{X,j'}(P))\right| = x \sqrt{\text{var}_{jj'}(P)}
\end{equation*}
is probability zero according to $\Phi_V$, where $V$ is such that $V_{jj} = 1$ for $j\in [2p]$. This holds since we imposed that $\text{var}_{jj'}(P)\neq 0$ and $x>0$.
Consequently, lemma \ref{lem:Uniform_Convergence_in_Distribution} gives us that:
\begin{equation*}
    \sup_{P\in\mathcal{P}}\sup_{x>0}\left|
    J_n(x,P) - J_n(x,\Phi_{{\Sigma(P)}})
    \right| = o(1)
\end{equation*}
Moreover, because cumulative distribution functions are right-continuous, we obtain:
\begin{equation*}
    \sup_{P\in\mathcal{P}}\sup_{x\geq 0}\left|
    J_n(x,P) - J_n(x,\Phi_{{\Sigma(P)}})
    \right| = o(1)
\end{equation*}
The continuous mapping theorem, the uniform consistency result in (\ref{eq:uniform_consistency_var_jk}), lemmas s.6.1 and s.7.1 in \cite{RomanoandShaikh2008}, and Polya's theorem provide that, for any $\varepsilon>0$:
\begin{equation*}
\liminf_{n\to\infty}\inf_{P\in\mathcal{P}}{\Pr}_P\left(\sup_{x\geq 0}\left|
    J_n(x,\Phi_{{n\hat{\Sigma}^n}}) - J_n(x,\Phi_{{\Sigma(P)}})
    \right|\leq \varepsilon \right)=1~.
\end{equation*}
Thus, we can verify:
\begin{equation*}
\liminf_{n\to\infty}\inf_{P\in\mathcal{P}}{\Pr}_P\left(\sup_{x\geq 0}\left|
    J_n(x,\Phi_{{n\hat{\Sigma}^n}}) - J_n(x,P)
    \right|\leq \varepsilon \right)=1
\end{equation*}
Applying lemma \ref{lem:CDF_cv_implies_uniform_cv} gives us that:
\begin{equation*}
    \liminf_{n\to\infty}\inf_{P\in\mathcal{P}}{\Pr}_P\left(
J_n^{-1}(0, \Phi_{{n\hat{\Sigma}^n}}) \leq R_n(\widetilde{S}_W^n,P) \leq J_n^{-1}\left(1-\beta\mbox{}, \Phi_{{n\hat{\Sigma}^n}}\right)
    \right) \geq 1-\beta\mbox{}
\end{equation*}
We can note that $J_n^{-1}(0, \Phi_{{n\hat{\Sigma}^n}}) = 0$, while $J_n^{-1}(1-\beta\mbox{}, \Phi_{{n\hat{\Sigma}^n}}) = d_{1-\beta\mbox{}}(\hat{\Sigma}^n)$.
    Consequently, algebraic manipulation of $R_n(\widetilde{S}_W^n,P)$ as defined in (\ref{eq:R_n_def_d_1_b}) and an application of the uniform consistency result in (\ref{eq:uniform_consistency_var_jk}) jointly imply that (\ref{eq:L_U_uniformly_valid}) holds, proving the lemma. \end{proof}
In addition, the following proposition demonstrates that $\rho_{1-\gamma}^n(\Delta,\Delta)$ is an asymptotically valid critical value for the maximum, studentized deviation between all $\hat{\jmath}$ in $\hat{J}_R$, where $\gamma := \alpha-\beta$. Formally, we have the following lemma.
\begin{lem}
\label{lem:rho_asymptotic_validity}
    Under assumption \ref{assu:Uniform_Asymptotic_Normality_Sufficient}, $\rho^n_{1-\gamma}$ satisfies:
    \begin{equation}
    \label{eq:rho_asymptotic_validity_key_eqn}
\limsup_{n\to\infty} \sup_{P\in\mathcal{P}}{\Pr}_P\left(
\max_{j\in J}\left\{\frac{|\xi_{\widetilde{S}_Y^n,j
}|}{\sqrt{\hat{\Sigma}_{Y,jj}^n}}\mathds{1}\left(\sum_{j'\in J}\mathds{1}\left(
\xi_{\widetilde{S}_X^n,j} \geq \xi_{\widetilde{S}_X^n,j'} + \Delta_{j'j}\right)
\in R \right)
\nonumber\right\}
>\rho_{1-\gamma}^n(\Delta,\Delta)
\right) \leq \gamma~.
    \end{equation}
\end{lem}
\begin{proof}
Our proof proceeds much as the proof of lemma \ref{lem:L_asymptotic_validity}. Indeed, we define $R_n(\widetilde{S}_W^n,P)$ as follows:
\begin{equation}
\resizebox{0.85\textwidth}{!}{$
    \max_{j \in J}\frac{\left|\sqrt{n}\left(\widetilde{S}_{Y,j}^n - \mu_{Y,j}(P)\right)
\right|}{\sqrt{\Sigma_{Y,jj}(P)}}\mathds{1}\left(\sum_{j'\in J}\mathds{1}\left(
\sqrt{n}\left(\widetilde{S}_{X,j}^n - \mu_{X,j}(P)\right) - \sqrt{n}\left(\widetilde{S}_{X,j'}^n - \mu_{X,j'}(P)\right) \geq  \sqrt{n}\Delta_{j'j}(P)\right)
\in R \right)$}~.
\end{equation}
Similarly, let us denote by $\hat{R}_n(\widetilde{S}_W^n,P)$ the following:
\begin{equation}
\resizebox{0.85\textwidth}{!}{$
    \max_{j \in J}\frac{\left|\sqrt{n}\left(\widetilde{S}_{Y,j}^n - \mu_{Y,j}(P)\right)
\right|}{\sqrt{n\hat\Sigma_{Y,jj}(P)}}\mathds{1}\left(\sum_{j'\in J}\mathds{1}\left(
\sqrt{n}\left(\widetilde{S}_{X,j}^n - \mu_{X,j}(P)\right) - \sqrt{n}\left(\widetilde{S}_{X,j'}^n - \mu_{X,j'}(P)\right) \geq  \sqrt{n}\Delta_{j'j}(P)\right)
\in R \right)$}~.
\end{equation}
The indicator in the above expression is equivalent to the indicator $\mathds{1}(j \in \hat{J}_{R;n})$, where we recall that $\hat{J}_{R;n} := \hat{J}_R(\widetilde{S}_X^n)$. Thus, we seek to show that the event:
\begin{equation}
\label{eq:event_need_to_lie_in_C}
    \max_{j \in \hat{J}_{R;n}}\frac{\left|\sqrt{n}\left(\widetilde{S}_{Y,j}^n - \mu_{Y,j}(P)\right)
\right|}{\sqrt{\Sigma_{Y,jj}(P)}} \leq x
\end{equation}
lies in $\mathcal{C}$. The above event can be rewritten as:
\begin{equation*}
    A(x) := \bigcup_{J_c \in 2^J}\left(
\left\{
\max_{j\in J_c}\frac{\left|\sqrt{n}\left(\widetilde{S}_{Y,j}^n - \mu_{Y,j}(P)\right)
\right|}{\sqrt{\Sigma_{Y,jj}(P)}} \leq x
\right\}\bigcap \left\{
\hat{J}_{R;n} = J_c
\right\}
    \right)
\end{equation*}
We may notice that:
\begin{equation*}
    \partial A(x) \subseteq \bigcup_{J_c \in 2^J}
\left\{
\max_{j\in J_c}\frac{\left|\sqrt{n}\left(\widetilde{S}_{Y,j}^n - \mu_{Y,j}(P)\right)
\right|}{\sqrt{\Sigma_{Y,jj}(P)}} = x
\right\} =: S
\end{equation*}
The above union is a subset of the a union of boundaries of hyperrectangles, and thus is such that $\Phi_V(S) = 0$ for any $2p\times 2p$ covariance matrix $V$ with $V_{jj} = 1$ for all $j\in [2p]$. Thus, $A(x)$ lies in $\mathcal{C}$. Let $J_n(x,P)$ denote the cumulative distribution function of $R_n(\widetilde{S}_W^n,P)$ when the $\Tilde{W}_i$ are i.i.d. from $P$. Let $\widetilde{\Phi}_n$ denote the distribution given by $\Phi_{\mu_W(P),n\hat\Sigma^n}$. Now, the following hold:
\begin{align}
    \sup_{P\in\mathcal{P}}\sup_{x\in\mathbb{R}}\left|
    J_n(x,P) - J_n(x,\Phi_{{\Sigma(P)}})
    \right| &= o(1)\nonumber\\
\liminf_{n\to\infty}\inf_{P\in\mathcal{P}}{\Pr}_P\left(\sup_{x\in\mathbb{R}}\left|
    J_n(x,\widetilde{\Phi}_n) - J_n(x,\Phi_{{\Sigma(P)}})
    \right|\leq \varepsilon \right) &= 1\nonumber\\
    \liminf_{n\to\infty}\inf_{P\in\mathcal{P}}{\Pr}_P\left(\sup_{x\in\mathbb{R}}\left|
    J_n(x,\widetilde{\Phi}_n) - J_n(x,P)
    \right|\leq \varepsilon \right) &=1\nonumber\\
        \liminf_{n\to\infty}\inf_{P\in\mathcal{P}}{\Pr}_P\left(
J_n^{-1}(0, \widetilde{\Phi}_n) \leq R_n(\widetilde{S}_W^n,P) \leq J_n^{-1}\left(1-\gamma, \widetilde{\Phi}_n\right)
    \right) &\geq 1-\gamma~. \label{eq:final_step_for_rho_validity}
\end{align}
The first equality holds by an application of lemma \ref{lem:Uniform_Convergence_in_Distribution}. The second equality holds by lemmas s.6.1 and s.7.1 in \cite{RomanoShaikh_2012}, and Polya's theorem. The third equality is a consequence of the first two equalities, and the final equality is a consequence of lemma \ref{lem:CDF_cv_implies_uniform_cv}. Finally, we may notice that $J_n^{-1}(0, \widetilde{\Phi}_n) = 0$, and $J_n^{-1}(1-\gamma,\widetilde{\Phi}_n) = \rho_{1-\gamma}^n(\Delta,\Delta)$ by construction. We also notice that assumption \ref{assu:Uniform_Asymptotic_Normality_Sufficient} implies that for any $\varepsilon>0$:
\begin{equation*}
    \limsup_{n\to\infty}\sup_{P\in\mathcal{P}}{\Pr}_P\left(
\left|
\frac{n\hat{\Sigma}_{Y,jj}^n}{\Sigma_{Y,jj}(P)}-1
\right|>\varepsilon
    \right) = 0
\end{equation*}
by lemma s.6.1 of \cite{RomanoShaikh_2012}. This result, along with (\ref{eq:final_step_for_rho_validity}) gives that the following holds:
\begin{equation*}
    \liminf_{n\to\infty}\inf_{P\in\mathcal{P}}{\Pr}_P\left(
J_n^{-1}(0, \widetilde{\Phi}_n) \leq \hat{R}_n(\widetilde{S}_W^n,P) \leq J_n^{-1}(1-\gamma, \widetilde{\Phi}_n)
    \right) \geq 1-\gamma
\end{equation*}
thus proving the lemma.
\end{proof}

Finally, we can prove proposition 2:
\begin{proof}
\textbf{Proof of Proposition 2.}
First, we note that, by lemma \ref{lem:L_asymptotic_validity}:
\begin{equation}
\label{eq:lem:L_asymptotic_validity}
\liminf_{n\to\infty}\inf_{P\in\mathcal{P}}{\Pr}_P\left(
L^n \leq \Delta(P) \leq U^n
\right) \geq 1 - \beta
\end{equation}
We can define the event $B^n := \{L^n \leq \Delta \leq U^n\}$, with the inequality interpreted elementwise. In addition, we note that, for any sequence of events $\{A^n(P)\}_{n=1}^\infty$: 
\begin{equation*}
    \liminf_{n\to\infty} \inf_{P\in\mathcal{P}} {\Pr}_P(A^n(P)) \geq 1-\alpha \quad \text{if and only if} \quad \limsup_{n\to\infty} \sup_{P\in\mathcal{P}}{\Pr}_P(A^n(P)^c) \leq \alpha
\end{equation*}
Thus, we have that, by (\ref{eq:lem:L_asymptotic_validity}):
\begin{equation*}
    \limsup_{n\to\infty} \sup_{P\in\mathcal{P}}{\Pr}_P\left((B^n)^c\right) \leq \beta
\end{equation*}
and we seek to show that:
\begin{equation*}
     \limsup_{n\to\infty} \sup_{P\in\mathcal{P}}{\Pr}_P\left(
\left(
\mu_{Y,\hat{\jmath}}(P)\right)_{\hat{\jmath}\in\hat{J}_R} \not\in CS^{TS}({1-\alpha;\beta,n})
    \right) \leq \alpha
\end{equation*}
Indeed:
\begin{align}
   & \limsup_{n\to\infty} \sup_{P\in\mathcal{P}}{\Pr}_P\left(
\left(
\mu_{Y,\hat{\jmath}}(P)\right)_{\hat{\jmath}\in\hat{J}_R} \not\in CS^{TS}({1-\alpha;\beta,n})
    \right) \nonumber\\
    & \leq \limsup_{n\to\infty} \sup_{P\in\mathcal{P}}{\Pr}_P\left(
\left(
\mu_{Y,\hat{\jmath}}\left(P
\right)\right)_{\hat{\jmath}\in\hat{J}_R} \not\in CS^{TS}({1-\alpha;\beta,n}) \cap B^n
    \right) + {\Pr}_P((B^n) ^ c)\nonumber\\
    &\leq \limsup_{n\to\infty} \sup_{P\in\mathcal{P}}{\Pr}_P\left(
\left(
\mu_{Y,\hat{\jmath}}\left(P
\right)\right)_{\hat{\jmath}\in\hat{J}_R} \not\in CS^{TS}({1-\alpha;\beta,n}) \cap B^n
    \right)\nonumber +\limsup_{n\to\infty} \sup_{P\in\mathcal{P}}{\Pr}_P((B^n) ^ c)\nonumber\\
    & \leq \limsup_{n\to\infty} \sup_{P\in\mathcal{P}}{\Pr}_P\left(
\left(
\mu_{Y,\hat{\jmath}}\left(P
\right)\right)_{\hat{\jmath}\in\hat{J}_R} \not\in CS^{TS}({1-\alpha;\beta,n}) \cap B^n
    \right) + \beta \nonumber\\
& = \limsup_{n\to\infty} \sup_{P\in\mathcal{P}}{\Pr}_P\left(
\bigcup_{\hat{\jmath}\in\hat{J}_R} \left\{
\frac{\sqrt{n}\left|
\xi_{\widetilde{S}_Y^n,\hat{\jmath}}
\right|}{\sqrt{\hat{\Sigma}_{Y,\hat{\jmath}\hat{\jmath}}^n}}
> \rho_{1-\alpha+\beta}^n\left(L^n,U^n\right)\right\} \cap B^n
\right) + \beta\nonumber\\
& \leq \limsup_{n\to\infty} \sup_{P\in\mathcal{P}}{\Pr}_P\left(
\bigcup_{\hat{\jmath}\in\hat{J}_R} \left\{
\frac{\sqrt{n}\left|
\xi_{\widetilde{S}_Y^n,\hat{\jmath}}
\right|}{\sqrt{\hat{\Sigma}_{Y,\hat{\jmath}\hat{\jmath}}^n}} >\rho_{1-\alpha+\beta}^n\left(\Delta,\Delta\right)\right\}
\right) + \beta\nonumber\\
&\leq \alpha-\beta+\beta = \alpha \nonumber
\end{align}
where the third inequality follows by lemma \ref{lem:L_asymptotic_validity}, the fifth inequality holds since, on $B^n$, $\rho_{1-\alpha+\beta}^n\left(\Delta,\Delta\right) \leq \rho_{1-\alpha+\beta}^n\left(L^n,U^n\right)$, and the final inequality holds by lemma \ref{lem:rho_asymptotic_validity}. 
\end{proof}

\section{Comparisons to Projection}
\label{app:comp_to_projection}

In the following section, we will first provide an asymptotic counterpart to proposition 3 in proposition \ref{prop:clear_winner_two_step_best}. We will then provide a proof of proposition \ref{prop:clear_winner_two_step_best}. Using the same arguments, we will prove proposition 3.

Before stating and proving proposition \ref{prop:clear_winner_two_step_best}, we denote by $J_R(P)$ the set of true selections:
$$\left\{j : \sum_{j'\in J}\mathds{1}\left(\mu_{X,j'}(P) \leq \mu_{X,j}(P) \right) \in R\right\}~.$$
We may equivalently define $J_R(P)$ as $\hat{J}_R(\mu_X(P))$. Let $CS^{P}({1-\alpha;n})$ denote the projection confidence set with $n\hat\Sigma^n$ in lieu of $\Sigma(P)$. 

\begin{prop}
\label{prop:clear_winner_two_step_best}
Suppose that, for some fixed $P$ in $\mathcal{P}$, the set of true selections $J_R(P)$ is a proper subset of $J$ and that $\Sigma(P)$ is full rank. We also assume that assumption \ref{assu:Uniform_Asymptotic_Normality_Sufficient} holds on $\mathcal{P}$. Under these assumptions, the coverage probability of the two-step confidence set is pointwise, asymptotically smaller than that of the projection confidence set, in the sense that for any $\beta$ sufficiently small, we have:
\begin{equation*}
\resizebox{0.9\textwidth}{!}{$
    \lim_{n\to\infty}{\Pr}_P\left(\left(
\mu_{Y,\hat{\jmath}}(P)\right)_{\hat{\jmath}\in\hat{J}_{R;n}} \in CS^{TS}({1-\alpha;\beta,n})
    \right)
    \leq \lim_{n\to\infty}{\Pr}_P\left(\left(\mu_{Y,\hat{\jmath}}(P)\right)_{\hat{\jmath}\in\hat{J}_{R;n}} \in CS^{P}({1-\alpha;n})
    \right)$}
\end{equation*}
\end{prop}
Before providing a proof of proposition \ref{prop:clear_winner_two_step_best}, we provide the following technical lemma.
\begin{lem}
\label{lem:need_for_quantile_bounds}
    Let $X_n$ and $Y_n$ be random variables for $n\in\mathbb{N}$ with cumulative distribution functions $J_{X,n}$ and $J_{Y,n}$. Suppose also that:
    \begin{enumerate}
        \item $X_n\leq Y_n$ with probability approaching one.
        \item $Y_n$ has a weak limit $Y$ with a continuous cumulative distribution function $J_Y$ that admits a strictly positive density on $\mathbb{R}$.
    \end{enumerate} 
    For any $\alpha \in (0,1)$, the quantile functions satisfy:
    \begin{equation}
    \label{eq:quantile_functions_less}
         \limsup_{n\to\infty} J_{X,n}^{-1}(\alpha) - J_{Y,n}^{-1}(\alpha) \leq 0
    \end{equation}
\end{lem}
\begin{proof}
    Notice that the following holds:
    \begin{align*}
        & \liminf_{n\to\infty} \inf_{x\in\mathbb{R}} P(X_n\leq x) - P(Y_n\leq x)\\
        & = \liminf_{n\to\infty} \inf_{x\in\mathbb{R}} P(\{X_n\leq x\}\cap \{X_n\leq Y_n\}) - P(Y_n\leq x) + P(\{X_n\leq x\}\cap \{X_n> Y_n\})\\
         & \geq \liminf_{n\to\infty} \inf_{x\in\mathbb{R}} P(\{X_n\leq x\}\cap \{X_n\leq Y_n\}) - P(Y_n\leq x)\\
        & \geq \liminf_{n\to\infty} \inf_{x\in\mathbb{R}} P(\{X_n\leq x\}\cap \{X_n\leq Y_n\}) - P(\{Y_n\leq x\} \cap \{X_n\leq Y_n\}) - P(X_n>Y_n)\\
        & \geq \liminf_{n\to\infty}  - P(X_n > Y_n) = 0~.
    \end{align*}
    The final equality holds by condition 1. It follows that:
    \begin{equation*}
        \liminf_{n\to\infty} \inf_{x\in\mathbb{R}} J_{X,n}(x) - J_{Y,n}(x) \geq 0
    \end{equation*}
    and similarly, by Polya's theorem, that:
    \begin{equation*}
        \liminf_{n\to\infty} \inf_{x\in\mathbb{R}} J_{X,n}(x) - J_{Y}(x) \geq 0~.
    \end{equation*}
    Equivalently:
    \begin{equation*}
        \limsup_{n\to\infty} \sup_{x\in\mathbb{R}} J_{Y}(x) - J_{X,n}(x) \leq 0~.
    \end{equation*}
    Consequently, by part (ii) of lemma A.1 in \cite{RomanoShaikh_2012}, we obtain that for any $\varepsilon > 0$, there exists sufficiently large $n$ such that $J_{X,n}^{-1}(\alpha) \leq J_Y^{-1}(\alpha+\varepsilon)$. Because $Y$ admits a strictly positive density by condition 2, we obtain that $\limsup_{n\to\infty} J_{X,n}^{-1}(\alpha) \leq J_Y^{-1}(\alpha)$. Similarly, we know that $\lim_{n\to\infty} J_{Y,n}^{-1}(\alpha) = J_Y^{-1}(\alpha)$, finally implying (\ref{eq:quantile_functions_less}).
\end{proof}
Now, we prove proposition \ref{prop:clear_winner_two_step_best}. Intuitively, we show that $L^n$ and $U^n$ approach the true $\Delta(P)$ in probability. Given this, we show that our modeled, selected errors $f(L^n,U^n)$ can be bounded above by the largest absolute errors in $J_R(P)$ with probability approaching one. An application of lemma \ref{lem:need_for_quantile_bounds} concludes the proof.
\begin{proof}
\textbf{Proof of Proposition \ref{prop:clear_winner_two_step_best}.} Let us first fix some arbitrary $\beta>0$. We begin by recalling that: 
\begin{align*}
    L_{jj'}^n &= \widetilde{S}_{X,j}^n - \widetilde{S}_{X,j'}^n - d_{1-\beta\mbox{}}(\hat{\Sigma}^n)\sqrt{\widehat{\text{var}}^n_{jj'}}\\
    U_{jj'}^n &= \widetilde{S}_{X,j}^n - \widetilde{S}_{X,j'}^n + d_{1-\beta\mbox{}}(\hat{\Sigma}^n)\sqrt{\widehat{\text{var}}^n_{jj'}}
\end{align*}
We note that, by the law of large numbers and by the continuous mapping theorem, for fixed $P$ in $\mathcal{P}$, $\left\|n\hat{\Sigma}^n - \Sigma(P)\right\| = o_P(1)$. We also know that, for any fixed $\Sigma$, $d_{1-\beta\mbox{}}(\Sigma)$ is $O(1)$, meaning that $d_{1-\beta\mbox{}}(\hat\Sigma^n) = d_{1-\beta\mbox{}}(n\hat\Sigma^n)$ is $O_P(1)$. Finally, recalling that $\widehat{\text{var}}^n_{jj'} = \hat{\Sigma}^n_{jj}+\hat{\Sigma}^n_{j'j'}-2\hat{\Sigma}^n_{j'j}$, we notice that $\widehat{\text{var}}^n_{jj'} = o_P(1)$. Thus, by applying the weak law of large numbers to $\widetilde{S}_{X}^n$ and that fact that the above facts imply that $d_{1-\beta\mbox{}}(\hat{\Sigma}^n)\sqrt{\widehat{\text{var}}^n_{j'j}} = o_P(1)$:
\begin{equation}
\label{eq:L_a_s_convergence}
    \begin{pmatrix}
        L_{jj'}^n\\
        U_{jj'}^n
    \end{pmatrix} \xrightarrow[]{P} \begin{pmatrix}
        \Delta_{jj'}(P)\\
        \Delta_{jj'}(P)
    \end{pmatrix}~.
\end{equation}
In what follows, we will suppress the dependence of $\mu_X$, $\mu_Y$, and $\Delta$ on $P$ in our notation. We will also write $\xi_j^n := \xi_{\Tilde{S}_X^n,j}$, for notational simplicity. 

We define $\delta := \min_{j \in J_R(P), i\not\in J_R(P)}\left|\mu_{X,j} - \mu_{X,i}\right|$. Because $J_R(P)$ is a proper subset of $J$, we have that $\delta > 0$. It follows from (\ref{eq:L_a_s_convergence}) that, with $\|\cdot\|$ being the max norm, the event $A_n := \left\{\left\| L^n - \Delta \right\|\vee \left\|U^n - \Delta \right\| \leq \delta / 6 \right\}$ satisfies $Pr_P(A_n) \to 1$. We seek to study the behavior of $\rho_{1-\alpha+\beta}^n\left(
L^n,U^n
\right)$ on $A_n$. To do so, we first show that:
\begin{align*}
    & \left(\mathds{1}\left(\left[\sum_{j'\in J}\mathds{1}\left(
\xi_j^n \geq \xi_{j'}^n + U_{j'j}^n\right),
\sum_{j'\in J}\mathds{1}\left(
\xi_j^n \geq \xi_{j'}^n + L_{j'j}^n\right)
\right]\cap R\neq\emptyset \right)\right)_{j\in J}\\
& \leq \left(\mathds{1}\left(
j \in J_R(P)
\right)\right)_{j\in J}
\end{align*}
with probability approaching one, on $A_n$, with inequality interpreted elementwise. We let $\ell^n$ and $u^n$ be such that $\left|\ell^n - \Delta \right|,\ \left|u^n - \Delta \right| \leq \delta / 6$. We are interested in the following probability:
\begin{equation*}
        {\Pr}_P\left( \mathds{1}\left(\left[\sum_{j'\in J}\mathds{1}\left(
\xi_j^n \geq \xi_{j'}^n + u_{j'j}^n\right),
\sum_{j'\in J}\mathds{1}\left(
\xi_j^n \geq \xi_{j'}^n + \ell_{j'j}^n\right)
\right]\cap R\neq\emptyset \right)_{j\in J} \leq
    \mathds{1}\left(
j \in J_R(P)
\right)_{j\in J}
    \right)~.
\end{equation*}
We define $B_n$ to be correspond to the event that all $\xi_{\widetilde{S}_X^n,j}$ are within $\delta/ 6$ of zero. Formally, $B_n := \{\max_{j\in J}|\xi_j^n| <\delta/ 6\}$. Since $\xi_j^n \xrightarrow{P} 0$, it follows that $Pr_P(B_n)$ approaches one as $n\to\infty$. For such $\ell^n$ and $u^n$ as above, we obtain that:
\begin{align}
    &\lim_{n\to\infty} {\Pr}_P\left( \begin{aligned} & \left(\mathds{1}\left(\left[\sum_{j'\in J}\mathds{1}\left(
\xi_j^n \geq \xi_{j'}^n + u_{j'j}^n\right),
\sum_{j'\in J}\mathds{1}\left(
\xi_j^n \geq \xi_{j'}^n + \ell_{j'j}^n\right)
\right]\cap R\neq\emptyset \right)\right)_{j\in J} \\
& \quad \leq \left(\mathds{1}\left(
j \in J_R(P)
\right)
    \right)_{j\in J} \end{aligned}\right)\nonumber\\
    & = 1 - \lim_{n\to\infty} {\Pr}_P\left(\left\{
\exists j\not\in J_R(P) :\left[\sum_{j'\in J}\mathds{1}\left(
\xi_j^n \geq \xi_{j'}^n + u_{j'j}^n\right),
\sum_{j'\in J}\mathds{1}\left(
\xi_j^n \geq \xi_{j'}^n + \ell_{j'j}^n\right)
\right]\cap R\neq\emptyset\right\}
    \right) \nonumber\\
    & = 1 - \lim_{n\to\infty} {\Pr}_P\left(\left\{
\exists j\not\in J_R(P) :\left[\sum_{j'\in J}\mathds{1}\left(
\xi_j^n \geq \xi_{j'}^n + u_{j'j}^n\right),
\sum_{j'\in J}\mathds{1}\left(
\xi_j^n \geq \xi_{j'}^n + \ell_{j'j}^n\right)
\right]\cap R\neq\emptyset\right\}\cap B_n
    \right) \nonumber\\
    & \geq 1 - \lim_{n\to\infty} {\Pr}_P\left(\left\{
\exists j\not\in J_R(P) :\left[\sum_{j'\in J}\mathds{1}\left(
0\geq \Delta_{j'j} + \delta/2 \right),
\sum_{j'\in J}\mathds{1}\left(
0\geq \Delta_{j'j} - \delta/2\right)
\right]\cap R\neq\emptyset\right\}
    \right) \nonumber\\
    & = 1 - {\Pr}_P(\emptyset) = 1 \nonumber,
\end{align}
To see that the inequality holds, we notice that on $B_n$:
\begin{align*}
    &\left[\sum_{j'\in J}\mathds{1}\left(
\xi_j^n \geq \xi_{j'}^n + u_{j'j}^n\right),
\sum_{j'\in J}\mathds{1}\left(
\xi_j^n \geq \xi_{j'}^n + \ell_{j'j}^n\right)
\right]\\
& \subseteq \left[\sum_{j'\in J}\mathds{1}\left(
0\geq \Delta_{j'j} + \delta/2 \right),
\sum_{j'\in J}\mathds{1}\left(
0\geq \Delta_{j'j} - \delta/2\right)
\right]
\end{align*}
To see that the last equality holds, notice that $j\in J_R(P)$ if and only if:
\begin{equation*}
    \sum_{j'\in J}\mathds{1}\left(
0\geq \Delta_{j'j} \right)\in R~.
\end{equation*}
Suppose, by way of contradiction, that there exists $j'\not\in J_R(P)$ such that there exists $r\in R$ satisfying:
\begin{equation*}
    r \in \left[\sum_{j''\in J}\mathds{1}\left(
0\geq \Delta_{j''j'} + \delta/2 \right),
\sum_{j''\in J}\mathds{1}\left(
0\geq \Delta_{j''j'} - \delta/2\right)
\right]
\end{equation*}
Let $j_r(\mu_X)$ be some element in $J_R(P)$ such that $r = \sum_{j''\in J}\mathds{1}\left(
0\geq \Delta_{j''j_r(\mu_X)} \right)$. Assume that $j'$ is such that $\mu_{X,j'} < \mu_{X,j_r(\mu_X)}$. A symmetric argument holds when $\mu_{X,j'} > \mu_{X,j_r(\mu_X)}$. We have that:
\begin{equation*}
    \sum_{j''\in J}\mathds{1}\left(
0\geq \Delta_{j''j'} + \delta/2 \right)\leq
\sum_{j''\in J}\mathds{1}\left(
0\geq \Delta_{j''j_r(\mu_X)} \right) \leq
\sum_{j''\in J}\mathds{1}\left(
0\geq \Delta_{j''j'} - \delta/2\right)~.
\end{equation*}
Since the indices of the smallest $r$ elements of the vector $(\Delta_{j'' j_r(\mu_X)})_{j''\in J}$ coincide with those of $(\Delta_{j'' j'})_{j''\in J}$, these inequalities imply that $\mathds{1}\left(
0\geq \Delta_{j_r(\mu_X)j_r(\mu_X)} \right) \leq \mathds{1}\left(
0\geq \Delta_{j_r(\mu_X)j'}-\delta/2 \right)$. Of course, $\mathds{1}\left(
0\geq \Delta_{j_r(\mu_X)j_r(\mu_X)} \right) = 1$ since $\Delta_{j_r(\mu_X)j_r(\mu_X)} = 0$. It follows that $\Delta_{j_r(\mu_X)j'}\leq \delta/2$ and consequently that
$|\Delta_{j_r(\mu_X)j'}|\leq \delta / 2$, since we assumed that $\mu_{X,j'} < \mu_{X,j_r(\mu_X)}$. This yields a contradiction, since $j'\not\in J_R(P)$ and thus $|\Delta_{j'j_r(\mu_X)}|\geq\delta$, by our definition of $\delta$. Thus, the event:
\begin{equation*}
    \left\{
\exists j\not\in J_R(P) :\left[\sum_{j'\in J}\mathds{1}\left(
0\geq \Delta_{j'j} - \delta/2 \right),
\sum_{j'\in J}\mathds{1}\left(
0\geq \Delta_{j'j} + \delta/2\right)
\right]\cap R\neq\emptyset\right\}
\end{equation*}
is indeed the empty set. Consequently, whenever $\ell^n$ and $u^n$ are as above, $\rho_{1-\alpha+\beta}(\ell^n,u^n;\hat{\Sigma}^n) \leq c_{1-\alpha+\beta}(J_R(P)) + o(1)$ by lemma \ref{lem:need_for_quantile_bounds}. That condition 1 in the lemma holds is a consequence of the fact that, as established above:
\begin{equation*}
        {\Pr}_P\left( \mathds{1}\left(\left[\sum_{j'\in J}\mathds{1}\left(
\xi_j^n \geq \xi_{j'}^n + u_{j'j}^n\right),
\sum_{j'\in J}\mathds{1}\left(
\xi_j^n \geq \xi_{j'}^n + \ell_{j'j}^n\right)
\right]\cap R\neq\emptyset \right) \leq
    \mathds{1}\left(
j \in J_R(P)
\right)
    \right)\to 1
\end{equation*}
Letting $\xi_Y$ follow $\Phi_{\Sigma_Y(P)}$, we find that condition 2 of the lemma holds since, by the continuous mapping theorem and an application of lemma s.6.1 in \cite{RomanoShaikh_2012}, we obtain:
\begin{equation}
    \max_{j\in J_R(P)} \frac{|\xi_{\widetilde{S}_Y^n,j}|}{\sqrt{\hat{\Sigma}_{Y,jj}^n}} \xrightarrow{d}\max_{j\in J_R(P)} \frac{|\xi_{Y,j}|}{\sqrt{\Sigma_{Y,jj}(P)}} 
\end{equation}
where the right hand side admits a positive-everywhere, continuous density over $\mathbb{R}_+$, since we assumed that $\Sigma(P)$ is full rank. Similarly, because $Pr_P(A_n)\to 1$ and because $\|n\hat{\Sigma}^n-\Sigma(P)\| = o_p(1)$, we obtain $\rho^n_{1-\alpha+\beta}(L^n,U^n) \leq c_{1-\alpha+\beta}(J_R(P)) + o_P(1)$. Since $\Sigma(P)$ is full rank, there exists $\bar\beta$ sufficiently small such that for any $\beta \leq \bar\beta$, $c_{1-\alpha+\beta}(J_R(P)) < c_{1-\alpha}(J)$ and thus we obtain $\rho^n_{1-\alpha+\beta}(L^n,U^n) < c_{1-\alpha}(J) + o_p(1)$. The desired result now follows, simply by applying the law of total probability to the event $\left(\mu_{Y,\hat{\jmath}}(P)\right)_{\hat{\jmath}\in\hat{J}_{R;n}} \in CS^{P}({1-\alpha;n})$ with the event $\rho^n_{1-\alpha+\beta}(L^n,U^n) \geq c_{1-\alpha}(J)$. \end{proof}

Finally, we provide a proof of proposition 3 from the main text. We use the notation from the normal location model. 
\begin{proof}
    \textbf{Proof of Proposition 3.} 
    First, let us define $\text{var}_{jj'}^c = \text{Var}(\xi_{X,j} - \xi_{X,j'})$ to be the variance of the pairwise difference $\xi_{X,j} - \xi_{X,j'}$ under $c\Sigma$. We may note that $\text{var}_{jj'}^c = c\text{var}_{jj'}$. Consequently, we may write: 
\begin{align*}
    L_{jj'}^c &= X_j - X_{j'} - d_{1-\beta\mbox{}}(c\Sigma)\sqrt{c\text{var}_{jj'}}\\
    U_{jj'}^c &= X_j - X_{j'} + d_{1-\beta\mbox{}}(c\Sigma)\sqrt{c\text{var}_{jj'}}~.
\end{align*}
    Firstly, we notice that $d_{1-\beta\mbox{}}(c\Sigma)$ does not vary with $c$. As $c\downarrow 0$, we obtain that for any $\varepsilon>0$, $P_{\mu,c\Sigma}(\max_{j\in J}|X_j - \mu_{X,j}| > \varepsilon) \to 0$ and that $c\text{var}_{jj'} \to 0$. Consequently, we obtain that, for any $\varepsilon > 0$:
    \begin{equation}
        P_{\mu,c\Sigma}\left(
        \max_{j,j'\in J}\left\{\left|
L_{jj'}-\Delta_{jj'}
        \right|\vee \left|
U_{jj'}-\Delta_{jj'}
        \right|\right\} > \varepsilon
        \right)\to 0
    \end{equation}
    as $c\downarrow 0$. The remainder of our proof proceeds exactly as the proof of proposition \ref{prop:clear_winner_two_step_best}, modulo considerations of estimation error in $\Sigma$, and we therefore omit details.
\end{proof}

\section{Proofs of Other Theoretical Results}
\label{app:Proofs_of_other_theoretical_results_appendix}
First, we present the proof of our generalized polyhedral lemma. The proofs in this section take place in the normal location model. 
\begin{proof}
\textbf{Proof of Lemma \ref{lem:Polyhedral_lemma}.} Let $W := \begin{pmatrix}
        X' &
        Y'
    \end{pmatrix}'$.
Following the reasoning from the proof of Lemma 5.1 from \cite{Leeetal2016}, the following holds:
\begin{align*}
    \{AW\geq 0\} &= \{A(c(B'W) + Z) \geq 0\}\\
    & = \{(Ac)(B'W) \geq - AZ\}
\end{align*}
yielding a set of linear constraints on $B'W$, when conditioning on $Z$. Thus, because $Z$ is independent of $B'W$, we find that $B'W$, conditional on the selection event $AW\geq 0$ and sufficient statistic $Z$,\footnote{Here, $Z$ can be thought of as a sufficient statistic for the nuisance parameters $(\mu_{Y,j})_{j\not\in \hat{J}_R}$ in our model.} is distributed according to a multivariate normal with mean $\mu_B := B'\mu$ and variance-covariance $\Sigma_B := B'\Sigma B$, truncated to the polyhedron $\mathcal{Y}((i_l)_{l\in R},z)$. \end{proof}

In addition, we present a proof of the finite sample validity of conditional inference:
\begin{proof}
    \textbf{Proof of Proposition \ref{prop:conditional_validity}.}
    First, we notice that $\left\{\mu_{Y,i_l}\right\}_{l \in R} \in CS_{1-\alpha}^c$ if and only if our test $\phi\left(\ \cdot \ ;(i_l)_{l\in R}, z, (\mu_{Y,i_l})_{l\in R}\right)$ fails to reject. This test is a valid test at level $\alpha$ by construction, implying that (\ref{eq:conditional_validity_CS}) holds.
\end{proof}
Now, we present a proof of proposition 1 in the main text:

\begin{proof} \textbf{Proof of Proposition 1.}
    We recall that $P_{\mu,\Sigma}(B)\geq 1-\beta$, where $B := \{L \leq\Delta\leq U\}$. Moreover, on $B$, we have $\rho_{1-\alpha+\beta}(L,U) \geq \rho_{1-\alpha+\beta}(\Delta,\Delta)$. We write:
\begin{small}
\begin{align*}
& P_{\mu,\Sigma}\left(\bigcup_{\hat{\jmath}\in\hat{J}_R}\left\{
\frac{|\xi_{Y,\hat{\jmath}}|}
{\sqrt{\Sigma_{Y,\hat{\jmath}\hat{\jmath}}}} > \rho_{1-\alpha+\beta}(L,U)
\right\}
\right)\\
& \leq P_{\mu,\Sigma}\left(
\bigcup_{\hat{\jmath}\in\hat{J}_R}\left\{
\frac{|\xi_{Y,\hat{\jmath}}|}
{\sqrt{\Sigma_{Y,\hat{\jmath}\hat{\jmath}}}} > \rho_{1-\alpha+\beta}(L,U)
\right\}\cap B
\right) + P_{\mu,\Sigma}(B^c)\\
& \leq P_{\mu,\Sigma}\left(
\bigcup_{\hat{\jmath}\in\hat{J}_R}\left\{
\frac{|\xi_{Y,\hat{\jmath}}|}
{\sqrt{\Sigma_{Y,\hat{\jmath}\hat{\jmath}}}} > \rho_{1-\alpha+\beta}(\Delta,\Delta)
\right\}\cap B
\right) + \beta\\
& \leq P_{\mu,\Sigma}\left(
\bigcup_{\hat{\jmath}\in\hat{J}_R}\left\{
\frac{|\xi_{Y,\hat{\jmath}}|}
{\sqrt{\Sigma_{Y,\hat{\jmath}\hat{\jmath}}}} > \rho_{1-\alpha+\beta}(\Delta,\Delta)
\right\}
\right) + \beta\\
& \leq P_{\mu,\Sigma}\left(
\max_{j\in J} \frac{|\xi_{Y,j}|}{\sqrt{\Sigma_{Y,jj}}}\mathds{1}\left(\sum_{j'\in J}\mathds{1}\left(
\xi_{X,j} \geq \xi_{X,j'} + \Delta_{j'j}\right)
\in R \right)>\rho_{1-\alpha+\beta}(\Delta,\Delta)
\right)  +\beta\\
& = \alpha-\beta+\beta = \alpha
\end{align*}
\end{small}
for all $\mu$ and $\Sigma$. Since there exists $\hat\jmath \in \hat{J}_R$ such that $\mu_{Y,\hat\jmath}\not\in CS_{\hat\jmath}^{TS}(1-\alpha;\beta)$ if and only if there exists $\hat\jmath \in \hat{J}_R$ such that $\frac{|\xi_{Y,\hat{\jmath}}|}
{\sqrt{\Sigma_{Y,\hat{\jmath}\hat{\jmath}}}} > \rho_{1-\alpha+\beta}(L,U)$, the result holds.
\end{proof}

We now provide a proof of proposition \ref{prop:normal_validity_non_inferior}, which proceeds much as the proof of proposition 1.
\begin{proof}
    \textbf{Proof of Proposition \ref{prop:normal_validity_non_inferior}.} We divide our proof into cases. In the first case, we have $\rho_{1-\alpha+\beta}(\Delta,\Delta) \leq \widetilde{c}_{1-\alpha}$. In the second case, we have $\rho_{1-\alpha+\beta}(\Delta,\Delta) > \widetilde{c}_{1-\alpha}$. The proof of validity in the first case proceeds exactly as in proposition 1, so we omit details and focus on proving validity in the second case. As before, we take $B:=\{L\leq\Delta\leq U\}$. In the second case, we have the following:
    \begin{align*}
        & P_{\mu,\Sigma}\left(
        \mu_{Y,\hat{\jmath}}\in CS^{TS2}({1-\alpha;\beta})\quad\text{for all }\hat{\jmath}\in\hat{J}_R
        \right)\\
        & \geq P_{\mu,\Sigma}\left(\left\{
        \mu_{Y,\hat{\jmath}}\in CS^{TS2}({1-\alpha;\beta})\quad\text{for all }\hat{\jmath}\in\hat{J}_R \right\}\cap B
        \right)\\
        & = P_{\mu,\Sigma}\left(
        \left\{
        \mu_{Y,\hat{\jmath}}\in \left[
        Y_{\hat{\jmath}}\pm \left( \rho_{1-\alpha+\beta}(\Delta,\Delta)\wedge\widetilde{c}_{1-\alpha}\right)\sqrt{\Sigma_{Y,\hat{\jmath}\hat{\jmath}}}
        \right]\quad\text{for all }\hat{\jmath}\in\hat{J}_R
        \right\}\cap B
        \right)\\
        & = P_{\mu,\Sigma}\left(
        \left\{
        \mu_{Y,\hat{\jmath}}\in \left[
        Y_{\hat{\jmath}}\pm \widetilde{c}_{1-\alpha}\sqrt{\Sigma_{Y,\hat{\jmath}\hat{\jmath}}}
        \right]\quad\text{for all }\hat{\jmath}\in\hat{J}_R
        \right\}\cap B
        \right)\\
        & \geq P_{\mu,\Sigma}\left(
        \left\{
        \mu_{Y,j}\in \left[
        Y_j\pm \widetilde{c}_{1-\alpha}\sqrt{\Sigma_{Y,jj}}
        \right]\quad\text{for all }j \in J
        \right\}\cap B
        \right)\\
        & \geq P_{\mu,\Sigma}\left(
        \left\{
        \mu_{Y,j}\in \left[
        Y_j\pm \widetilde{c}_{1-\alpha}\sqrt{\Sigma_{Y,jj}}
        \right], \mu_{X,j}\in \left[
        X_j\pm \widetilde{c}_{1-\alpha}\sqrt{\Sigma_{X,jj}}
        \right]\quad\text{for all }j\in J
        \right\}\cap B
        \right)\\
        & = P_{\mu,\Sigma}\left(
        \mu_{Y,j}\in \left[
        Y_j\pm \widetilde{c}_{1-\alpha}\sqrt{\Sigma_{Y,jj}}
        \right], \mu_{X,j}\in \left[
        X_j\pm \widetilde{c}_{1-\alpha}\sqrt{\Sigma_{X,jj}}
        \right]\quad\text{for all }j\in J
        \right)\\
        & \geq 1-\alpha~.
    \end{align*}
    The final equality holds since $\widetilde{c}_{1-\alpha}(\sqrt{\Sigma_{X,jj}}+\sqrt{\Sigma_{X,j'j'}})\leq d_{1-\beta}(\Sigma)\sqrt{\text{var}_{jj'}}$ for all $j\neq j'$ implies that:
    \begin{equation*}
      \left\{
        \mu_{Y,j}\in \left[
        Y_j\pm \widetilde{c}_{1-\alpha}\sqrt{\Sigma_{Y,jj}}
        \right], \mu_{X,j}\in \left[
        X_j\pm \widetilde{c}_{1-\alpha}\sqrt{\Sigma_{X,jj}}
        \right]\quad\text{for all }j\in J
        \right\} \subseteq B~.  
    \end{equation*}
    Consequently, this casework gives that proposition \ref{prop:normal_validity_non_inferior} holds.
\end{proof}
Now, we provide a proof of proposition 6, which concerns the validity of locally simultaneous approaches to inference.
\begin{proof}\textbf{Proof of Proposition 6.} Our proof follows via application of theorem 1 of \cite{ZrnicFithian2024}. We can define the following set of plausible targets, given a realization $X=x$ and $Y=y$:
\begin{equation}
\label{eq:J_R+_defn}
    \hat{J}_R^+ = \bigcup_{\widetilde{x} : \sup_{j'\neq j} |(\widetilde{x}_{j'} - x_{j'})-(\widetilde{x}_j - x_j)|\leq 2\bar{d}_{1-\beta}(\Sigma)}\hat{J}_R(\widetilde{x})~.
\end{equation}
We define the acceptance region $A_\beta(\widetilde{\mu})$ as follows, for $\widetilde{\mu} = \begin{pmatrix}
    \widetilde{\mu}_X' & \widetilde{\mu}_Y'
\end{pmatrix}'$:
\begin{equation*}
    A_\beta(\widetilde{\mu}) = \left\{(\widetilde{y},\widetilde{x}): \sup_{1\leq j,j'\leq p} \left|\widetilde{x}_j-\widetilde{\mu}_{X,j}- (\widetilde{x}_{j'}-\widetilde{\mu}_{X,j'})\right|\leq \bar{d}_{1-\beta\mbox{}}(\Sigma)
    \right\}~.
\end{equation*}
We can invert $A_\beta(\widetilde{\mu})$ to obtain the confidence set $C_\beta((y,x))$. We notice that we may rewrite $\hat{J}_R^+$ as follows:
\begin{equation*}
    \hat{J}_R^+ = \bigcup_{\widetilde{\mu} \in C_\beta((y,x))}\bigcup_{(\widetilde{y},\widetilde{x})\in A_\beta(\widetilde{\mu})}\hat{J}_R(\widetilde{x})~.
\end{equation*}
This equivalence holds by double inclusion. We notice that for any $\widetilde{x}$ such that $\sup_{j'\neq j} |(\widetilde{x}_{j'} - x_{j'})-(\widetilde{x}_j - x_j)|\leq 2\bar{d}_{1-\beta}(\Sigma)$, we can choose $\widetilde{\mu}_X = (x + \widetilde{x})/2$ and $\widetilde{\mu_Y} = y$ such that $\widetilde{\mu} \in C_\beta((y,x))$ and such that $(y,\widetilde{x}) \in A_\beta(\widetilde{\mu})$. Similarly, if $\widetilde{x} \in A_\beta(\widetilde{\mu})$ for some $\widetilde{\mu} \in C_\beta((y,x))$, the triangle inequality implies that $\sup_{j'\neq j} |(\widetilde{x}_{j'} - x_{j'})-(\widetilde{x}_j - x_j)|\leq 2\bar{d}_{1-\beta}(\Sigma)$.

Following the reasoning of theorem 3 in \cite{ZrnicFithian2024}, we seek to find a most favorable choice of $\widetilde{x}$ for an index $j$ to be included in $\hat{J}_R(\widetilde{x})$, where $\widetilde{x}$ must satisfy:
\begin{equation}
\label{eq:x_prime_condn}
    \sup_{j'\neq j} |(\widetilde{x}_{j'} - x_{j'})-(\widetilde{x}_j - x_j)|\leq 2\bar{d}_{1-\beta}(\Sigma)~.
\end{equation}
To allow $j\in \hat{J}_R(\widetilde{x})$, let $j'$ be the element in $\hat{J}_R(x)$ that minimizes $|x_j - x_{j'}|$. Suppose, without loss, that $x_j < x_{j'}$. We can obtain the most favorable perturbation by taking $\widetilde{x}_j := x_j+ \bar{d}_{1-\beta\mbox{}}(\Sigma)$ and $\widetilde{x}_{j'} := x_{j'}-\bar{d}_{1-\beta\mbox{}}(\Sigma)$. For any $k\neq j,j'$, we may take $\widetilde{x}_k = x_k$. Consequently, since we have written an explicit $\widetilde{x}$ satisfying \eqref{eq:x_prime_condn} such that $j \in \hat{J}_R(x')$, we can write:
\begin{equation*}
    \left\{
j: \exists~ j'\in \hat{J}_R\text{ s.t. }|X_{j'}-X_j|\leq2\bar{d}_{1-\beta\mbox{}}(\Sigma)
    \right\} \subseteq \hat{J}_R^+~.
\end{equation*}
To see that the reverse inclusion holds, we will proceed by contradiction. Suppose that there exists $\widetilde{x}$ satisfying \eqref{eq:x_prime_condn} such that $j\in\hat{J}_R(\widetilde{x})$ and that for any $j' \in \hat{J}_R$, $|x_j - x_{j'}| >2 \bar{d}_{1-\beta}(\Sigma)$. To simplify our analysis, let us consider $j' \in \hat{J}_R$ such that $x_j < x_{j'}$.\footnote{If no such $j'$ exists, then we consider $j' \in \hat{J}_R$ such that $x_j > x_{j'}$, and the argument is symmetric.} For $j$ to be in $\hat{J}_R(\widetilde{x})$, it must be the case that $\widetilde{x}$ is such that $\widetilde{x}_j > \widetilde{x}_{j'}$, for one such $j'$. However, this implies that $|(\widetilde{x}_{j'} - x_{j'})-(\widetilde{x}_j - x_j)|> 2\bar{d}_{1-\beta}(\Sigma)$, providing a contradiction. Thus:
\begin{equation*}
    \left\{
j: \exists~ j'\in \hat{J}_R\text{ s.t. }|X_{j'}-X_j|\leq2\bar{d}_{1-\beta\mbox{}}(\Sigma)
    \right\} = \hat{J}_R^+
\end{equation*}
by double inclusion. Because we have show that $\hat{J}_R^+$ can be constructed in the spirit of equation (2) of \cite{ZrnicFithian2024}, we conclude the proof by appealing to theorem 1 of \cite{ZrnicFithian2024}.
\end{proof}
Finally, we provide a proof of proposition 7, which compares our two-step approach to inference to the locally simultaneous approach of \cite{ZrnicFithian2024}.
\begin{proof}\textbf{Proof of Proposition 7.} Notice that, whenever $\max_{j,j'\in J} |\mu_{X,j} - \mu_{X,j'}| \leq\bar{d}_{1-\beta}(\Sigma)$, the following series of implications holds, for arbitrary $j,j'\in J$:
\begin{align*}
    & |\xi_{X,j}-\xi_{X,j'}| \leq \bar{d}_{1-\beta}(\Sigma)\\
    & \implies |\xi_{X,j}-\xi_{X,j'}| + |\mu_{X,j} - \mu_{X,j'}| \leq 2\bar{d}_{1-\beta}(\Sigma)\\
    & \implies |X_j - X_{j'}| \leq 2\bar{d}_{1-\beta}(\Sigma)
\end{align*}
where the final implication is a consequence of the triangle inequality. Consequently, $P_{\mu,\Sigma}(\max_{j,j'\in J} |X_j - X_{j'}| \leq 2\bar{d}_{1-\beta}(\Sigma)) \geq 1-\beta$. On this event, it holds that $\hat{J}_R^+ = J$. Moreover, on the event that $|\xi_{X,j}-\xi_{X,j'}| \leq \bar{d}_{1-\beta}(\Sigma)$, it holds that $\bar{L}_{jj'} \geq -3\bar{d}_{1-\beta}(\Sigma)$ and $\bar{U}_{jj'} \leq 3\bar{d}_{1-\beta}(\Sigma)$, for all $j,j'$. Consequently, we notice that, with probability at least $1-\beta$, $\rho_{1-\alpha+\beta}(\bar{L},\bar{U})$ is bounded above by the $1-\alpha+\beta$ quantile of:
\begin{equation}
f(-3\bar{d}_{1-\beta}(\Sigma)\mathds{1}_p\mathds{1}_p', 3\bar{d}_{1-\beta}(\Sigma)\mathds{1}_p\mathds{1}_p') \leq \max_{j\in J}\frac{|\xi_{Y,j}|}{\sqrt{\Sigma_{Y,jj}}}
\end{equation}
where the inequality is strict with positive probability, since we assume that $\Sigma$ is full rank and that $R$ is a proper subset of $J$. It follows that $\rho_{1-\alpha+\beta}(\bar{L},\bar{U}) < c_{1-\alpha+\beta}(J) = c_{1-\alpha+\beta}\left(\hat{J}_R^+\right)$ with probability at least $1-\beta$.
\end{proof}
\section{Simulation Study Results}
\label{app:all_sims}
In this section, we present the results of a simulation study on our two-step approach to inference after ranking. We compare our two-step method to the existing approaches from section 4 and from appendix \ref{app:alternative_approaches}, namely the locally simultaneous approach of \cite{ZrnicFithian2024} and test inversion approach of \cite{ZrnicFithian2024WP}. Two-step inference performs well in simulations and relative to these methods, a finding which is robust to our choice of $\Sigma$. 

We consider the following designs:
\begin{itemize}
    \item \textbf{Design 1} $J = [5]$, $R=\{5\}$, $\mu_Y = 0$, $\mu_{X} = \left\{\arctan(i - 3)\right\}_{i=1}^5$
    \item \textbf{Design 2} $J = [10]$, $R=\{10\}$, $\mu_Y = 0$, $\mu_{X} = \left\{\arctan(i - 5)\right\}_{i=1}^{10}$
    \item \textbf{Design 3} $J = [5]$, $R=\{5\}$, $\mu_Y = 0$, $\mu_{X} = 0$
    \item \textbf{Design 4} $J = [5]$, $R=\{5\}$, $\mu_Y = 0$, $\mu_{X} = \{\mathds{1}(i=1)\}_{i=1}^5$
    \item \textbf{Design 5} $J = [5]$, $R=\{5\}$, $\mu_Y = 0$, $\mu_{X} = \{\mathds{1}(i\leq 2)\}_{i=1}^5$
    \item \textbf{Design 6} $J = [5]$, $R=\{5\}$, $\mu_Y = 0$, $\mu_{X} = \{\mathds{1}(i\leq 3)\}_{i=1}^5$
    \item \textbf{Design 7} $J = [5]$, $R=\{5\}$, $\mu_Y = 0$, $\mu_{X} = \{\mathds{1}(i\leq 4)\}_{i=1}^5$
\end{itemize}

In addition, we consider $\Sigma \in \{\Sigma_{simple},\Sigma_{low},\Sigma_{medium},\Sigma_{high}\}$. For $\Sigma_{simple}$, we take $\Sigma_{simple,X} = \Sigma_{simple,Y} = \Sigma_{simple,XY} = I_p$. For the remaining cases, we take:
\begin{align*}
    & \Sigma_{low,X} = \Sigma_{low,Y} = I_p,~\Sigma_{low,XY} = 0.5\cdot I_p\\
    & \Sigma_{medium,X} = \Sigma_{medium,Y} = 0.5\cdot I_p + 0.5 \mathds{1}_p\mathds{1}_p^\top,~\Sigma_{medium,XY} = 0.25\cdot I_p + 0.5 \mathds{1}_p\mathds{1}_p^\top\\
    & \Sigma_{high,X} = \Sigma_{high,Y} = 0.05\cdot I_p + 0.95 \mathds{1}_p\mathds{1}_p^\top,~ \Sigma_{high,XY} = 0.025\cdot I_p + 0.95 \mathds{1}_p\mathds{1}_p^\top~.
\end{align*}
For each of the seven designs, we take $\Sigma \in \{\Sigma_{simple},\Sigma_{low},\Sigma_{medium},\Sigma_{high}\}$, creating a total of 28 designs, and scale $\Sigma$ by $1/n$ for $n \in \{1,10,100,1000,10000\}$. 

We present results from these simulations below. We denote the two-step approach to inference in red, projection in black, the zoom test (based on the step-wise implementation) in light gray, and locally simultaneous inference in dark gray.

\newpage

\begin{figure}[H]
    \centering
    \includegraphics[width=1\linewidth]{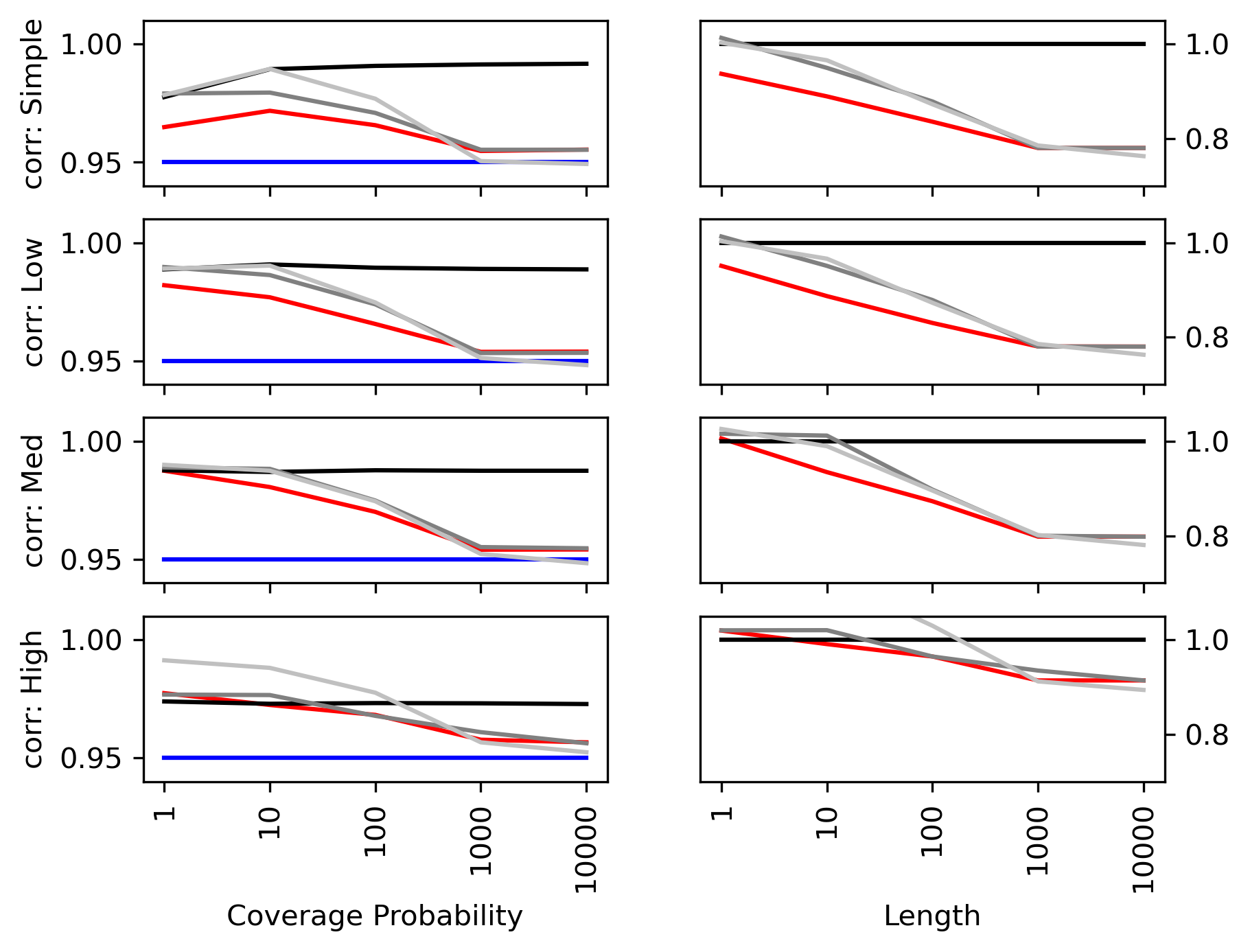}
    \caption{Coverage probability and length in design 1. CI lengths are presented as fractions of the projection CI length. We denote the two-step approach to inference in red, projection in black, the zoom test (based on the step-wise implementation) in light gray, and locally simultaneous inference in dark gray.}
    \label{fig:desn_1}
\end{figure}

\begin{figure}[H]
    \centering
    \includegraphics[width=1\linewidth]{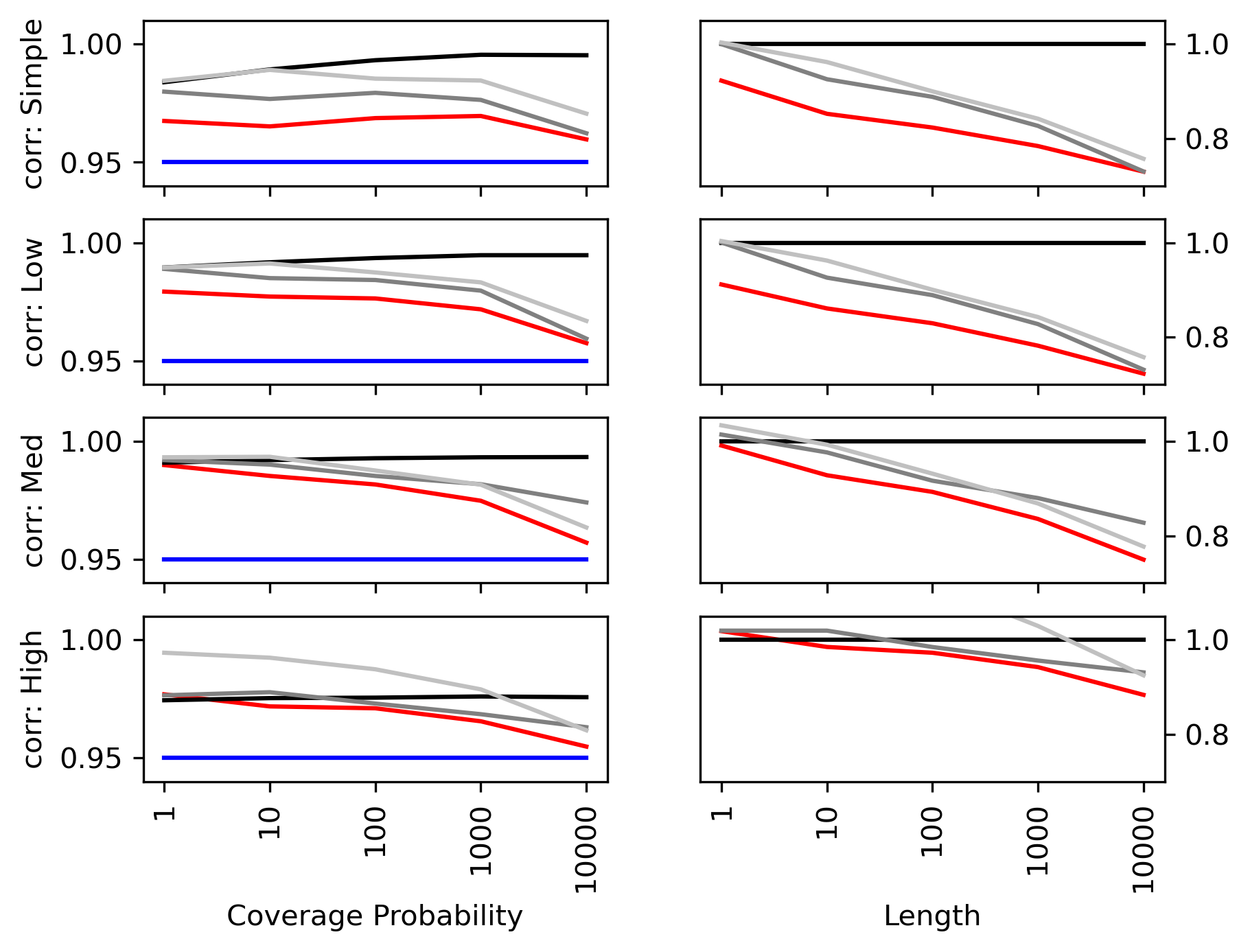}
    \caption{Coverage probability and length in design 2. CI lengths are presented as fractions of the projection CI length. We denote the two-step approach to inference in red, projection in black, the zoom test (based on the step-wise implementation) in light gray, and locally simultaneous inference in dark gray.}
    \label{fig:desn_1}
\end{figure}

\begin{figure}[H]
    \centering
    \includegraphics[width=1\linewidth]{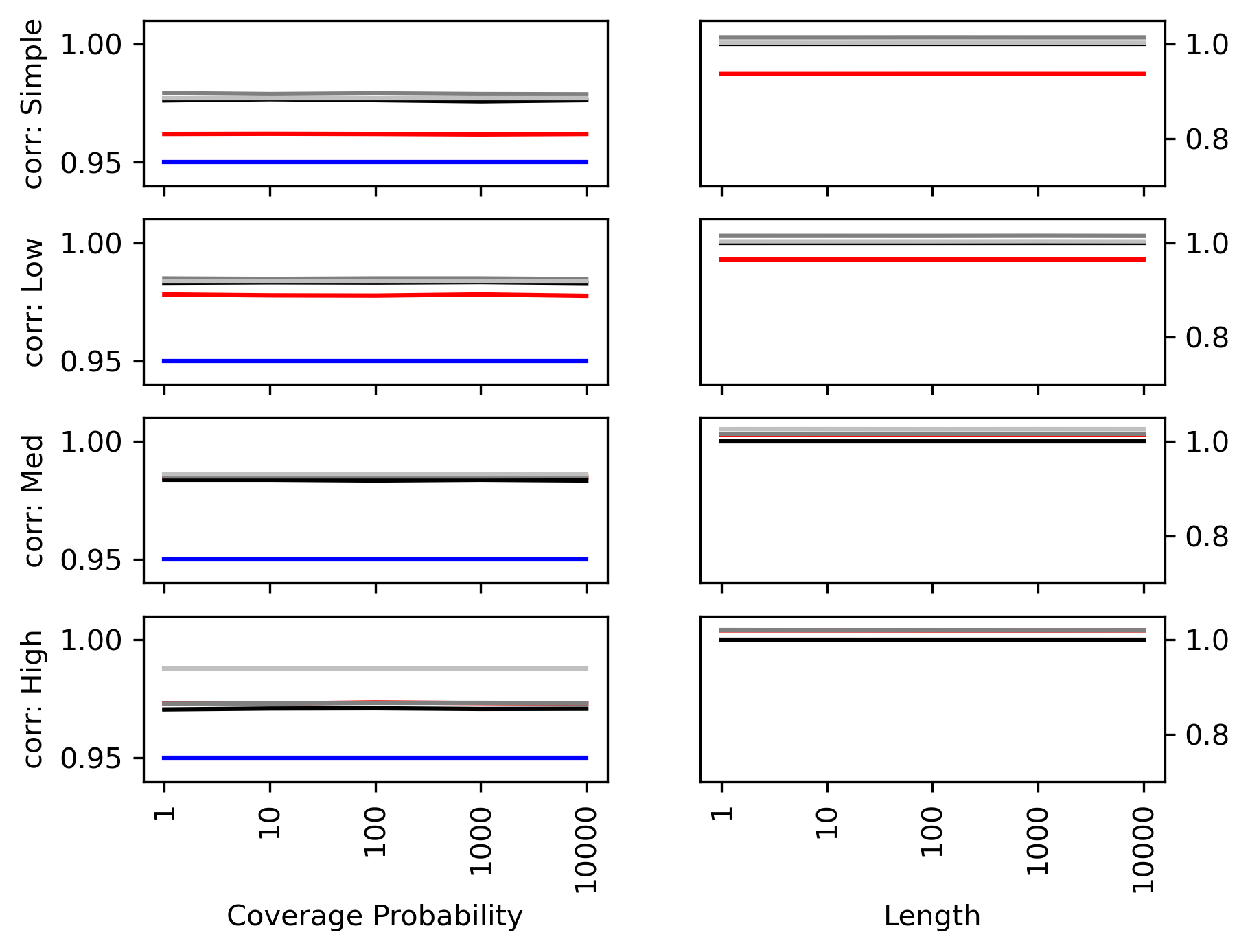}
    \caption{Coverage probability and length in design 3. CI lengths are presented as fractions of the projection CI length. We denote the two-step approach to inference in red, projection in black, the zoom test (based on the step-wise implementation) in light gray, and locally simultaneous inference in dark gray.}
    \label{fig:desn_1}
\end{figure}

\begin{figure}[H]
    \centering
    \includegraphics[width=1\linewidth]{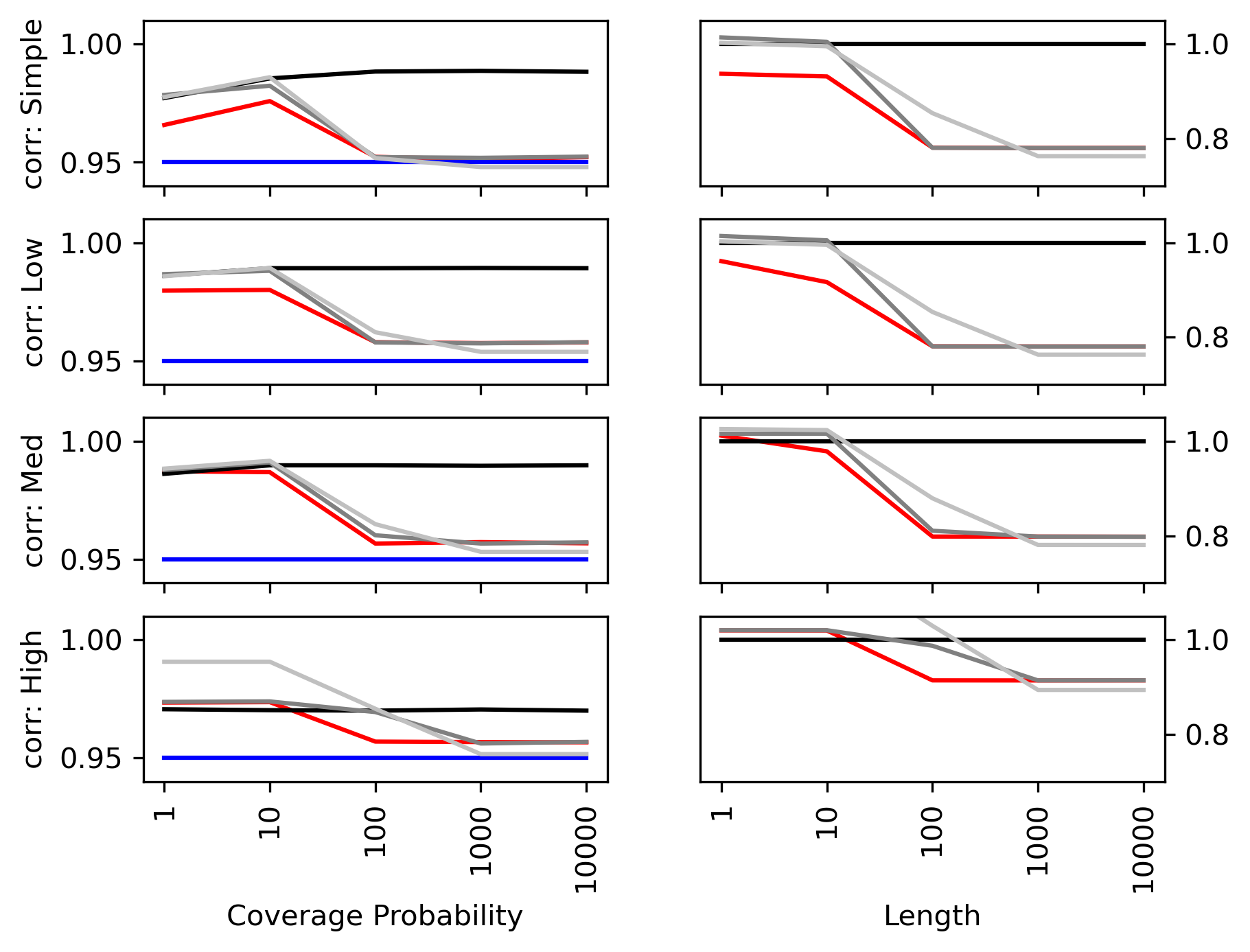}
    \caption{Coverage probability and length in design 4. CI lengths are presented as fractions of the projection CI length. We denote the two-step approach to inference in red, projection in black, the zoom test (based on the step-wise implementation) in light gray, and locally simultaneous inference in dark gray.}
    \label{fig:desn_1}
\end{figure}

\begin{figure}[H]
    \centering
    \includegraphics[width=1\linewidth]{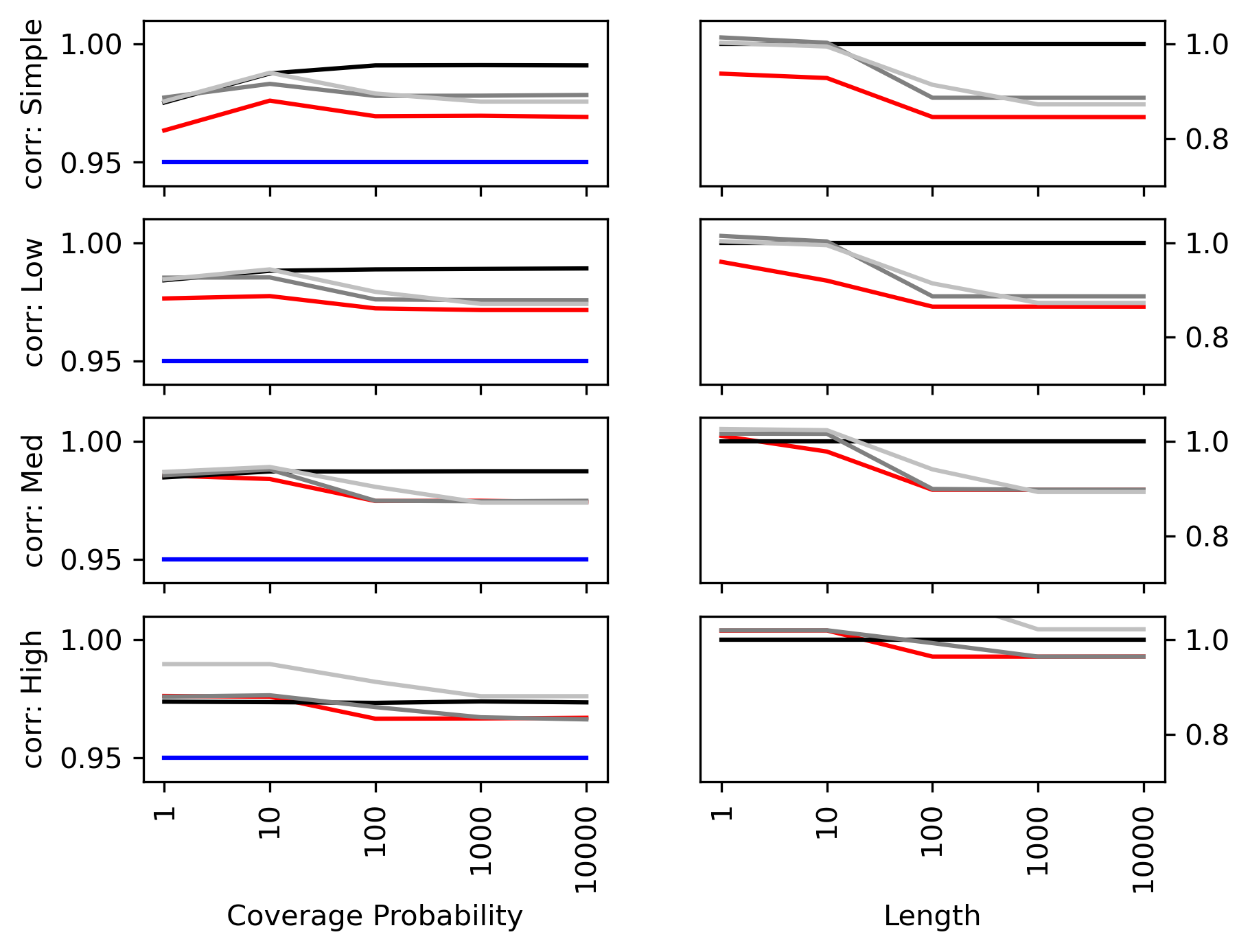}
    \caption{Coverage probability and length in design 5. CI lengths are presented as fractions of the projection CI length. We denote the two-step approach to inference in red, projection in black, the zoom test (based on the step-wise implementation) in light gray, and locally simultaneous inference in dark gray.}
    \label{fig:desn_1}
\end{figure}

\begin{figure}[H]
    \centering
    \includegraphics[width=1\linewidth]{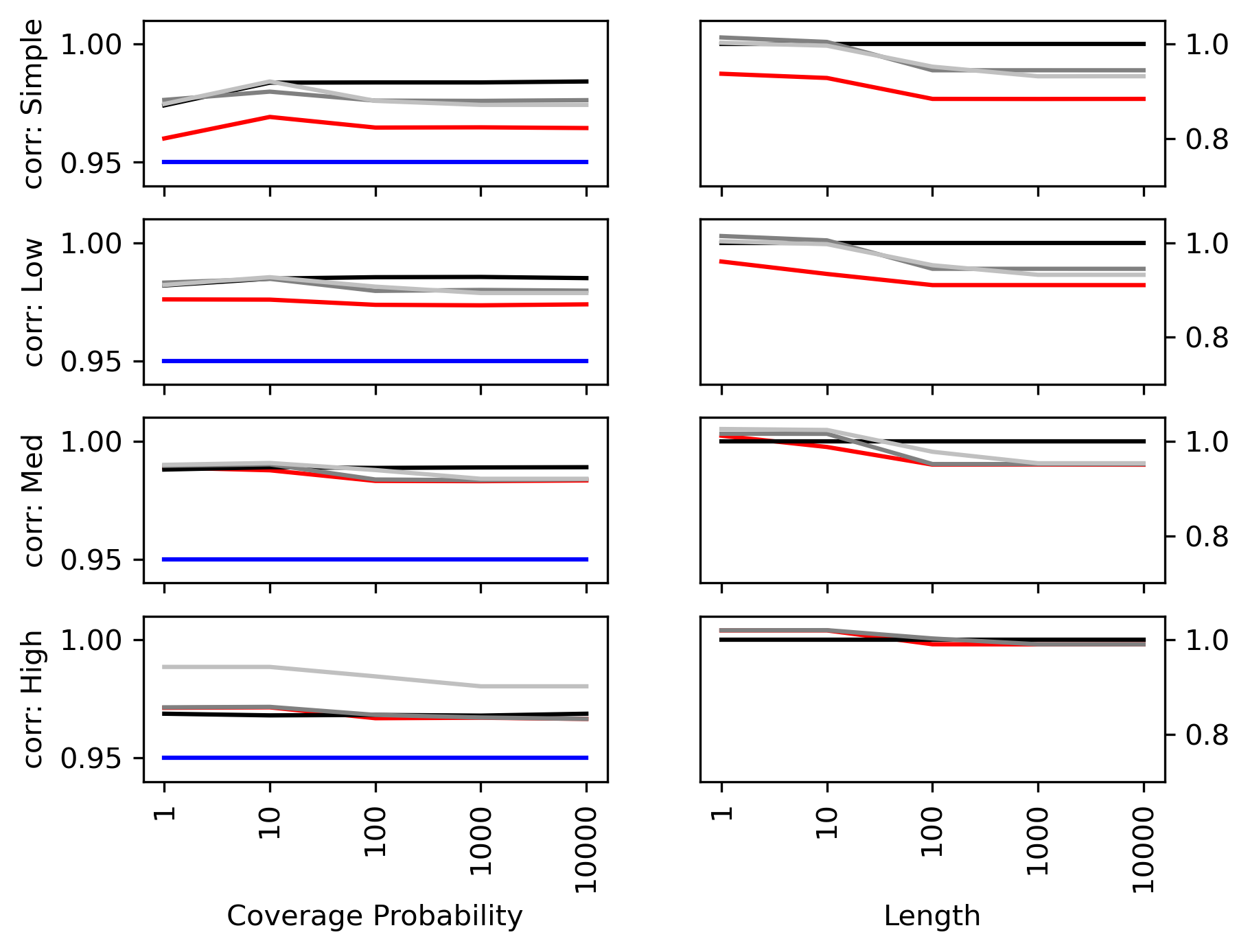}
    \caption{Coverage probability and length in design 6. CI lengths are presented as fractions of the projection CI length. We denote the two-step approach to inference in red, projection in black, the zoom test (based on the step-wise implementation) in light gray, and locally simultaneous inference in dark gray.}
    \label{fig:desn_1}
\end{figure}

\begin{figure}[H]
    \centering
    \includegraphics[width=1\linewidth]{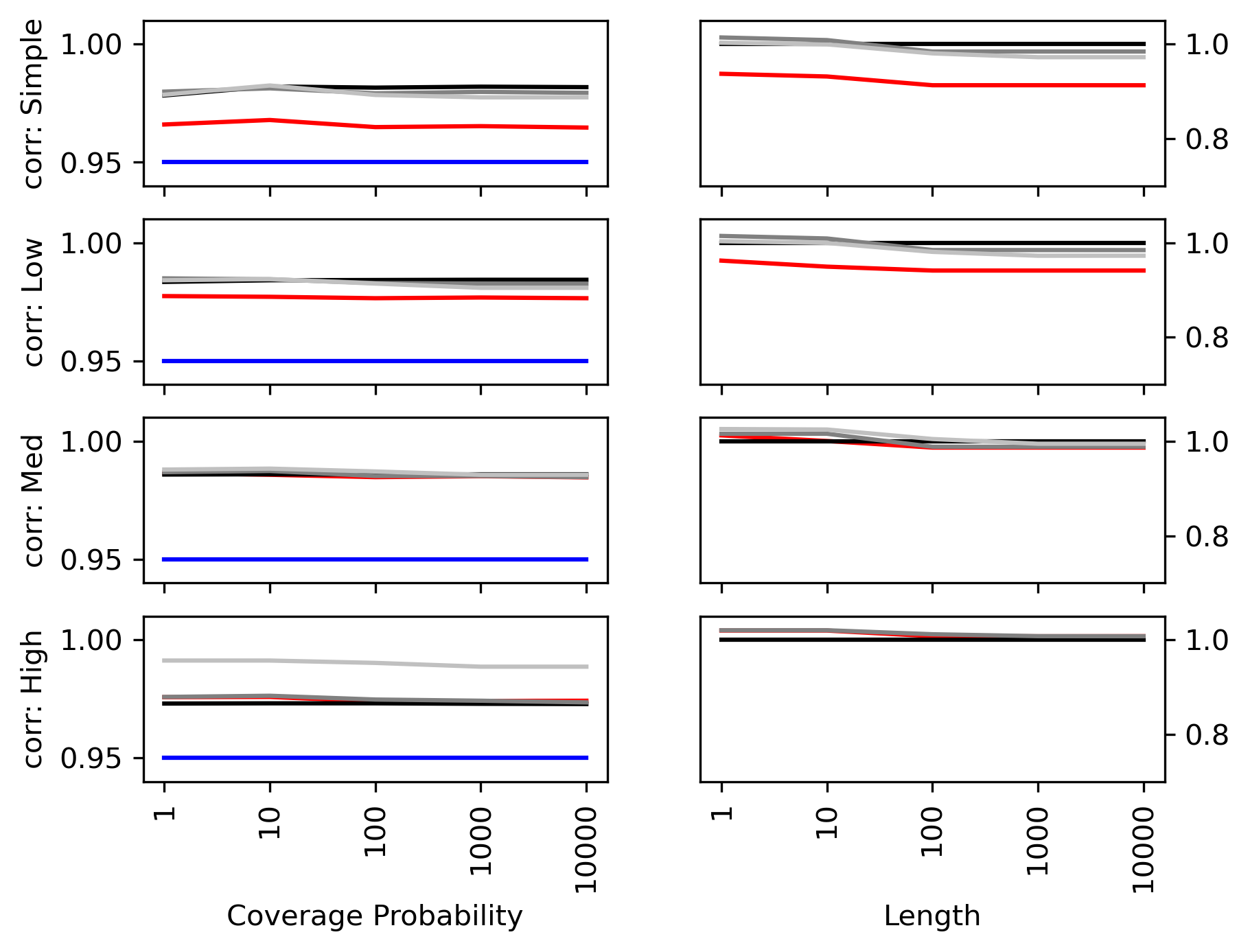}
    \caption{Coverage probability and length in design 7. CI lengths are presented as fractions of the projection CI length. We denote the two-step approach to inference in red, projection in black, the zoom test (based on the step-wise implementation) in light gray, and locally simultaneous inference in dark gray.}
    \label{fig:desn_1}
\end{figure}

\end{appendix}

\end{document}